\documentclass[acmsmall,screen,review=false,authorversion,nonacm]{acmart}

\colorlet{RED}{red}

\title[FPGA Innovation Research in the Netherlands]{FPGA Innovation Research in the Netherlands: \\Present Landscape and Future Outlook}

\AtBeginDocument{%
  \providecommand\BibTeX{{%
    \normalfont B\kern-0.5em{\scshape i\kern-0.25em b}\kern-0.8em\TeX}}}

\setcopyright{none} 
\makeatletter
\let\@authorsaddresses\@empty
\makeatother

\begin{document}


\author{Nikolaos Alachiotis}
\email{n.alachiotis@utwente.nl}
\orcid{0000-0001-8162-379}
\affiliation{%
  \institution{University of Twente}
  \streetaddress{}
  \city{}
  \state{}
  \country{The Netherlands}
  \postcode{}
}

\author{Sjoerd van den Belt}
\email{s.p.vandenbelt@utwente.nl}
\orcid{0009-0005-8812-9929}
\affiliation{%
  \institution{University of Twente}
  \streetaddress{}
  \city{}
  \state{}
  \country{the Netherlands}
  \postcode{}
}

\author{Steven van der Vlugt}
\email{vlugt@astron.nl}
\orcid{0000-0001-6834-4860}
\affiliation{%
  \institution{Netherlands Institute for Radio Astronomy (ASTRON)}
  \streetaddress{Oude Hoogeveensedijk 4}
  \city{Dwingeloo}
  \state{Drenthe}
  \country{the Netherlands}
  \postcode{7990 AA}
}

\author{Reinier van der Walle}
\email{walle@astron.nl}
\orcid{0009-0005-8097-6170}
\affiliation{%
  \institution{Netherlands Institute for Radio Astronomy (ASTRON)}
  \streetaddress{Oude Hoogeveensedijk 4}
  \city{Dwingeloo}
  \state{Drenthe}
  \country{the Netherlands}
  \postcode{7990 AA}
}

\author{Mohsen Safari}
\email{mohsen.safari@surf.nl}
\orcid{0000-0003-0839-3251}
\affiliation{%
  \institution{SURF}
  \streetaddress{}
  \city{}
  \state{}
  \country{the Netherlands}
  \postcode{}
}

\author{Bruno Endres Forlin}
\email{b.endresforlin@utwente.nl}
\orcid{0000-0003-4822-1841}
\affiliation{%
  \institution{University of Twente}
  \streetaddress{}
  \city{Enschede}
  \state{}
  \country{the Netherlands}
  \postcode{}
}

\author{Tiziano De Matteis}
\email{t.de.matteis@vu.nl}
\orcid{0000-0002-9158-6849}
\affiliation{%
  \institution{Vrije Universiteit Amsterdam}
  \streetaddress{}
  \city{Amsterdam}
  \state{}
  \country{The Netherlands}
  \postcode{}
}

\author{Zaid Al-Ars}
\email{z.al-ars@tudelft.nl}
\orcid{0000-0001-7670-8572}
\affiliation{%
  \institution{Delft University of Technology}
  \streetaddress{}
  \city{Delft}
  \state{}
  \country{the Netherlands}
  \postcode{}
}

\author{Roel Jordans}
  \email{r.jordans@tue.nl}
  \affiliation{
  \institution{Eindhoven University of Technology}
  \streetaddress{Den Dolech 2}
  \city{Eindhoven}
  \country{the Netherlands}
}

\additionalaffiliation{
  \institution{Radboud Radiolab, Institute for Mathemathics, Astrophysics and Particle Physics (IMAPP), Radboud University}
  \streetaddress{Heyendaalseweg 135}
  \city{Nijmegen}
  \country{the Netherlands}
}

\author{Ant\'onio J. Sousa de Almeida}
\email{ajsousal@gmail.com}
\orcid{0000-0002-7024-2262}
\affiliation{%
  \institution{University of Twente}
  \streetaddress{}
  \city{}
  \state{}
  \country{the Netherlands}
  \postcode{}
}

\author{Federico Corradi}
\email{f.corradi@tue.nl}
\orcid{0000-0002-5868-8077}
\affiliation{%
  \institution{Eindhoven University of Technology}
  \streetaddress{}
  \city{}
  \state{}
  \country{the Netherlands}
  \postcode{}
}

\author{Christiaan Baaij}
\email{christiaan@qbaylogic.com}
\affiliation{%
  \institution{QBayLogic B.V.}
  \streetaddress{}
  \city{}
  \state{}
  \country{the Netherlands}
  \postcode{}
}

\author{Ana-Lucia Varbanescu}
\email{a.l.varbanescu@utwente.nl}
\orcid{0000-0002-4932-1900}
\affiliation{%
  \institution{University of Twente}
  \streetaddress{}
  \city{}
  \state{}
  \country{The Netherlands}
  \postcode{}
}


\renewcommand{\shortauthors}{Alachiotis et al.}

\begin{abstract}

FPGAs have transformed digital design by enabling versatile and customizable solutions that balance performance and power efficiency, yielding them essential for today’s diverse computing challenges.
Research in the Netherlands, both in academia and industry, plays a major role in developing new innovative FPGA solutions. This survey presents the current landscape of FPGA
innovation research in the Netherlands by delving into ongoing projects, advancements, and
breakthroughs in the field. Focusing on recent research outcome (within the past 5 years), we have identified five key research areas: a) FPGA architecture, b) FPGA robustness, c) data center infrastructure and high-performance computing, d) programming models and tools, and e) applications. 
This survey provides in-depth insights beyond a mere snapshot of the current innovation research landscape by highlighting future research directions within each key area; these insights can serve as a foundational resource to inform potential national-level investments in FPGA technology.

\end{abstract}






\maketitle


\section{Introduction}

The current surge in computationally intensive workloads, such as artificial intelligence (AI) 
and high-performance computing (HPC), creates an unprecedented need for ever-increasing computational capacity. Most modern-day applications and technologies are data-driven and must process large volumes of data at high speeds, or operate under stringent time constraints. Emerging technologies, such as distributed sensor networks in radio astronomy \cite{big-data-radio-astronomy} and next generation sequencing in genetics \cite{genetic-big-data}, enable the accumulation of vast amounts of data, further intensifying the need for more 
computational power. 
Keeping up with the ever-increasing demand for computational capacity requires excessive amounts of energy, at a time when there is an urgent need to reduce unsustainable energy consumption. 
AI applications, powered by large language and generative models, are rapidly increasing in scale and complexity, further increasing their energy demand \cite{energy-llm}.
Given the increasing amounts of data to be processed and the high computational requirements of modern-day applications, the energy needed to sustain these applications can often not be supplied exclusively by renewable sources of energy \cite{green-data-centers, enegry-efficiency-cloud-dc}. This  
raises the need for energy-efficient hardware that is able to facilitate the next generation of data-driven applications.

To enable the deployment of modern applications that need to process large amounts of data faster, energy efficiently, and/or in real time, hardware acceleration is necessary. Various hardware technologies can be used for this purpose, such as GPUs, FPGAs, and ASICs. 
A GPU (Graphics Processing Unit) is a massively parallel architecture that comprises numerous small processing cores~\cite{gpuComputing}, making it well-suited for computationally intensive workloads such as graphics rendering and parallel computing, particularly for vector-processing operations. GPUs are widely deployed for accelerating AI/ML model training. 
%
An ASIC (Application-Specific Integrated Circuit) 
offers the highest performance and energy efficiency for an application by implementing in hardware only the logic that is actually needed by the target application. Designing and fabricating an ASIC, however, requires an extensive design process and is extremely costly \cite{asic-challenges}. These limitations can be alleviated by using an FPGA (Field Programmable Gate Array). FPGAs are frequently used instead of GPUs as well, because of their I/O capabilities, e.g., optical links or direct network connections.

FPGAs have revolutionized digital design and prototyping due to their versatility and adaptability to varying computational requirements. Unlike ASICs, FPGAs are reprogrammable and can be reconfigured after manufacturing multiple times, allowing for cost-effectively realizing highly efficient custom computer architectures. FPGA technology relies on an array of configurable logic blocks that are interconnected through programmable routing channels; the configuration of these elements results in the implementation of specialized digital circuits that are tailored to the specific requirements of a particular domain, application, or even workload. Eliminating unnecessary microarchitecture-level components and optimizing the design for a specific domain or application leads to considerably higher performance and energy efficiency than general-purpose processors without the prohibitively high costs for manufacturing an ASIC.

\begin{figure}[t]
    \centering
    \includegraphics[width=\textwidth]{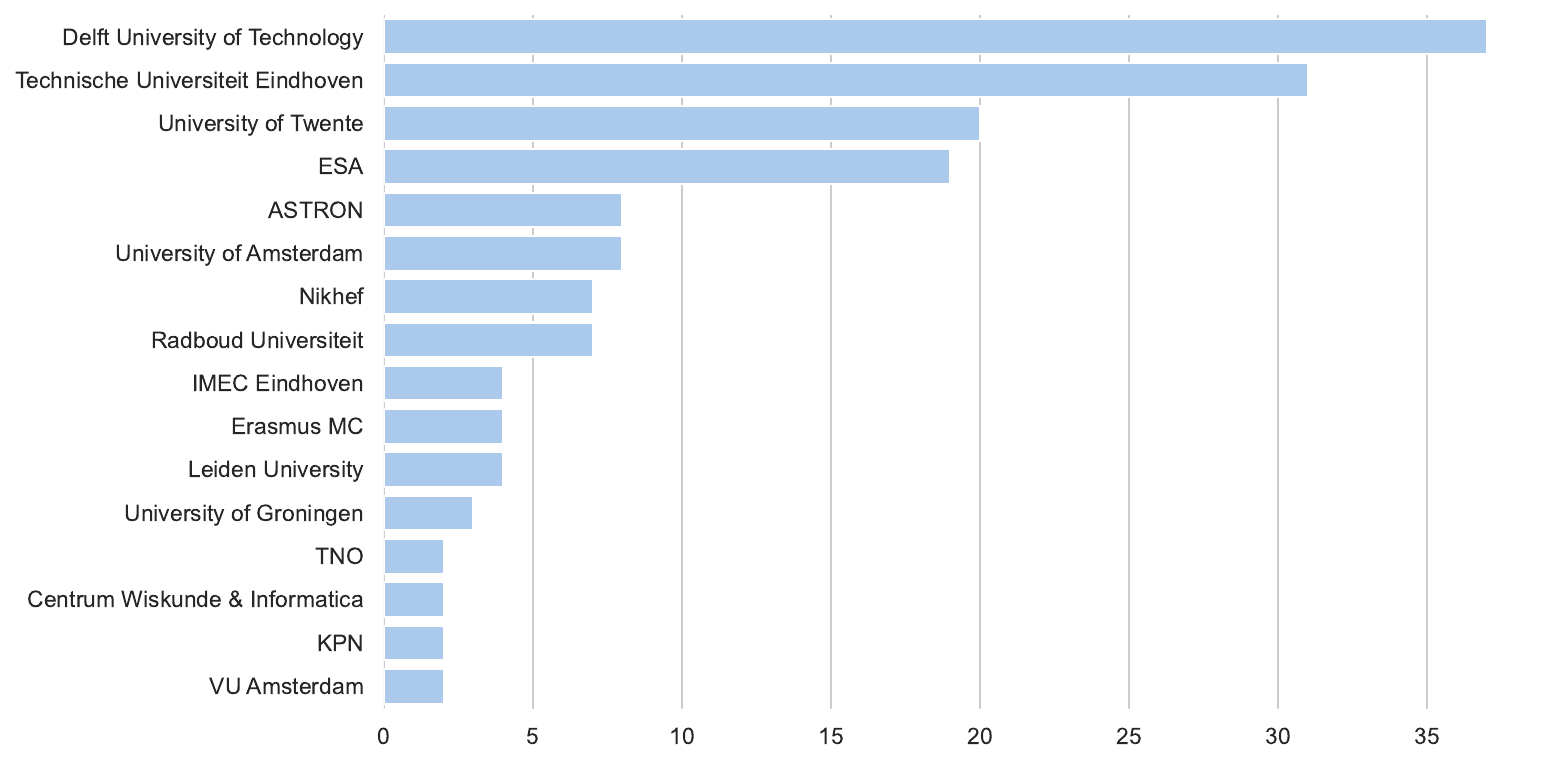}
    \caption{The number of FPGA related publications published by Dutch organizations within the past 5 years. Only organizations with at least two relevant publications are shown.}
    \label{fig:org-publications}
\end{figure}

\paragraph{Context} The unique ability of modern FPGA technology to balance performance and power efficiency through customization makes it a pivotal technology in addressing the diverse and evolving challenges of today's highly heterogeneous computing landscape. Research in the Netherlands plays a major role in developing new innovative FPGA technologies. The country has a strong reputation for research excellence, particularly in areas such as technology, engineering, agriculture, environmental science, and healthcare. Dutch universities and research institutions are actively involved in collaborative international projects, fostering a culture of knowledge exchange and cooperation, while Dutch companies contribute to global initiatives and advancements in fields like sustainable energy, water management, and digital technology. Besides universities and industry, public institutes such as the European Space Research and Technology Center (ESTEC), the National Institute for Nuclear and High energy physics (Nikhef), and the Netherlands Institute for Radio Astronomy (ASTRON) also conduct FPGA-related research. Figure~\ref{fig:org-publications} provides an overview of Dutch organizations that conduct research in this area, and the number of recent publications per organization. 
The figure shows that 
research by universities and ESTEC constitutes the majority of the relevant published work. Two private companies, IMEC NL and KPN, have also contributed with at least two publications, indicating that private organizations also play a role in the scientific development of FPGA technology. Overall, FPGA technology is an active field of research in the Dutch scientific community, with relevance in academia and industry. 

Advancing FPGA technology aligns with European goals, and strengthens the position of the Netherlands in a future where efficient large-scale computing will be increasingly important. 
In the past decade, the Netherlands has made significant scientific contributions to a large number of European research projects, accelerating the development of various technologies. Within the European Horizon 2020 funding program, the Netherlands has been 
successfully 
participating 
in scientific European projects 
when normalized to the number of scientific person-year effort in the country. Within Horizon 2020, the Netherlands has made a comparatively high contribution to the program pillar focusing on Excellent Science \cite{rathenauNederlandHorizon}. Current major strategic goals outlined by the European Commission are the development of autonomous technologies, such as artificial intelligence, and the development of technologies to combat climate change \cite{rathenauEuropeseWetenschap}. Both of these goals align with developing energy efficient hardware acceleration. The Netherlands also houses the European Space Research and Technology Center (ESTEC), which is the main center for research and development of the European Space Agency (ESA). Space applications are another specific field where FPGA technology has a prominent role. The flexibility and performance that FPGAs offer aligns with the needs of space applications. Recently, space-grade FPGA technology has been developed as a European research effort, strengthening the application of FPGAs in space projects \cite{EuEsaSpaceGradeFpgProject}.  Overall, Dutch contributions to research and technology are important on both European and global scale in various scientific and technological domains.

\paragraph{Survey focus} Harnessing the full potential of advanced FPGA-based systems demands more than just comprehending the inherent capabilities and limitations of FPGA technology. It necessitates a deep understanding of how these attributes align with and serve the diverse computational requirements across various domains and applications. To this end, we survey the present landscape of FPGA innovation research in the Netherlands.  
By delving into the ongoing projects, advancements, and breakthroughs in the field, this survey aims to provide valuable insights that go beyond a mere snapshot of the present landscape. More importantly, it aspires to serve as a foundational resource, possibly guiding future national-level investments in FPGA technology. By understanding the strengths, challenges, and emerging trends in FPGA research, policymakers, researchers, and industry stakeholders can make informed decisions about allocating resources and shaping strategies that will contribute to the continued growth and impact of FPGA technology. This survey endeavors to be a pivotal tool in fostering a forward-looking approach, ensuring that the nation remains at the forefront of FPGA innovation and leverages its potential for future technological advancements. Considering the relevance of FPGA technology in the European and Dutch context, there is ample reason for the Netherlands to invest in FPGA technology.

\section{Survey Method}

\label{select-lit-section}
We performed the literature collection in this survey study as a three-step process. First, we employed the online literature search tool Scopus \cite{elsevierScopus} to gather relevant literature from several publishers. Second, we performed a more in-depth search for articles by specific publishers. Third, a selection of conferences and journals were manually checked for relevant papers in proceedings and articles, respectively, to include publications that were possibly omitted in previous steps.  Finally, all duplicates were filtered out and we manually reviewed the resulting publications for relevance.

\paragraph{Step 1: Scopus-wide search}
We used Scopus to search for publications from over 7,000 publishers. We used the following search criteria: a) the word ``FPGA'' appears in the title, abstract, and/or keywords, b) at least one of the authors has a Dutch affiliation, and c) the publication year is 2019 or later. These translate into the following Scopus query:  
\texttt{TITLE-ABS-KEY (FPGA) AND AFFILCOUNTRY (Netherlands) AND PUBYEAR > 2019}.
The search was performed using this query on the October 30, 2023, which delivered a list of 186 papers. As of the publication date, this query is expected to yield an increasing number of papers, as more papers aligning with the search criteria are published. 
To put things in context, we also conducted a world-wide search for papers on FPGAs published in the last 5 years. The search resulted in a list of over 27,600 titles from 84 countries with at least 10 publications, indicating the Netherlands contributes roughly 1\% to the world-wide research on FPGAs, and ranks on position 22 in terms of output volume. In Europe, Dutch FPGA research ranks 10th in terms of output volume. 



\paragraph{Step 2: Select publishers}
In addition to the Scopus-based search, we conducted an in-depth search for relevant papers in proceedings and articles published by ACM and IEEE. Publications (co-)authored by Dutch institutes, published in 2019 or later, containing both the terms ``FPGA'' and ``HPC'' 
were queried for.  
Compared to the initial Scopus-based search, this new search is more specific in terms of keywords (by using one additional keyword), but more inclusive because it considers the entire text for keyword matches. 
We used advanced search tools provided by IEEE and ACM \cite{acm_advanced_search} \cite{ieee_advanced_search}, and obtained a list of 65 articles (including possible duplicates). 

\paragraph{Step 3: Select conferences and journals}
We conducted a final search within topic-relevant conferences and journals;  
we restricted the scope of this manual search to conference proceedings and journal issues published in the last 5 years. The following conferences and journals were considered in this step: 
\begin{itemize}
    \item International Symposium on Field-Programmable Gate Arrays (FPGA)
    \item International Symposium on Field-Programmable Custom Computing Machines (FCCM)
    \item International Conference on Field Programmable Logic and Applications (FPL)
    \item International Conference on Field-Programmable Technology (FPT)
    \item ACM Transactions on Reconfigurable Technology and Systems (TRETS)
\end{itemize}
This search delivered a list of 21 publications from Dutch-affiliated authors (including possible duplicates).

\paragraph{Final literature selection}

We verified all collected publications for relevance and excluded duplicates. This process yielded a total of 212 relevant publications, which we further classified into research themes. A significant number of these papers (120, over 55\%) focus on FPGA applications. Thus, we selected six applications for further review, based on the number of relevant papers and the significance of FPGAs in the application; the selection  resulted in a shortlist of 49 application papers. These applications are discussed in section \ref{sec:applications}. Other applications which are not reviewed in depth include Network Processing~\cite{Kundel2021OpenBNG:Hardware}, Cryptography~\cite{Massolino2020ASIKE}, Control Systems~\cite{Moonen2021Simulink-BasedSystems} and Weather Prediction~\cite{Singh2020NERO:Modeling}.  
Section~\ref{themes-section} introduces the research themes by which all reviewed papers were categorized. 


\section{Research Themes}
\label{themes-section}
Based on the literature found through the process outlined in Section \ref{select-lit-section}, articles with similar subjects or covering similar themes of research are grouped together. The themes are selected in such a way that most publications can be exclusively divided into one of the themes, i.e., the themes should not have significant overlap. Furthermore the themes should effectively separate domain specific research and more generally applicable research. 
Considering these requirements the following list of themes is selected:

\begin{enumerate}
    \item FPGA architecture
    \item Robustness of FPGAs
    \item Data center infrastructure \& HPC
    \item Programming models \& tools
    \item Applications
\end{enumerate}

\begin{figure}[!htbp]
    \centering
    \includegraphics[width=0.5\textwidth]{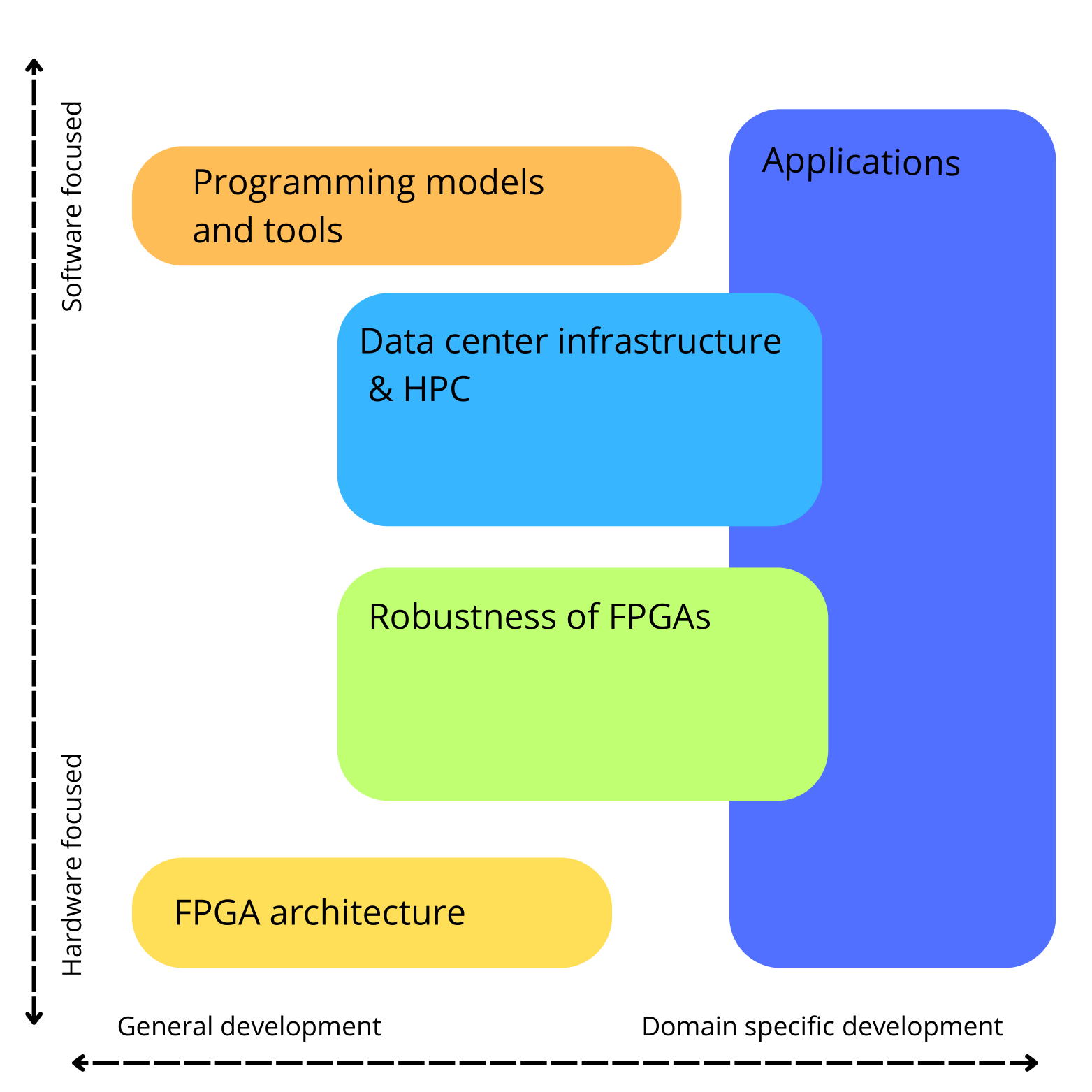}
    \caption{The themes that are selected can be differentiated based on their domain-specificity, ranging from very domain-focused to general purpose, and based on whether the main focus is on hardware or software. }
    \label{fig:theme-distribution}
\end{figure}

Figure~\ref{fig:theme-distribution} illustrates how these themes cover both general and domain-specific development, and shows whether a theme is hardware or software focused. The theme ``Applications'' focuses on research applying FPGA technology to domain specific problems. This can be in the form of hardware architectures for domain-specific applications, as well as software tools enabling FPGA technology in a specific domain of research. The themes ``Programming models and tools'' and ``FPGA architecture'' focus on research and development of solutions that are generally applicable in a wide range of domains, while the ``Robustness of FPGAs'' and ``Data center infrastructure \& HPC'' themes feature both hardware- and software-focused research. Moreover, these themes focus on a narrower selection of FPGA applications and can, therefore, be considered  domain-specific. 

Based on common subjects in each theme, the themes are further organized into 
subcategories. Table \ref{tab:overview-themes-most-cited} shows the subcategories and 
the prevalence of each  subject based on the number of published articles. 
Finally, the most influential articles, based on the highest number of citations within each category (Google Scholar), is shown. This selection excludes survey publications.

\begin{table}[!ht]
\centering
\caption{Overview of highly cited papers per category, with number of published articles per theme and category in parentheses.}
\label{tab:overview-themes-most-cited}
{\small
\begin{tabular}{lll}
\textbf{Theme and category} & \textbf{Highly cited publications} & \textbf{Dutch affiliation} \\ \hline
\textbf{\textbf{FPGA architecture (11)}} &  &  \\ \cline{1-1}
Near-memory processing (4) & \citet{Singh2021FPGA-BasedApplications} & Eindhoven University \\
Coarse-grained reconfigurable architecture (4) & \citet{Wijtvliet2019Blocks:Efficiency} & Eindhoven University \\
Network-on-Chip (3) & \citet{RibotGonzalez2020HopliteRT:FPGA} & Eindhoven University \\ \hline
\multicolumn{2}{l}{\textbf{\textbf{Data center infrastructure \& HPC (40)}}}  &  \\ \cline{1-1}
Big data processing and analytics (22) & \citet{Peltenburg2019Fletcher:Arrow} & Delft University of Tech. \\
Distributed computing (5) & \citet{Bielski2018DReDBox:Datacenter} & Sintecs B.V. \\
Optical hardware communication (9) & \citet{Yan2018HiFOST:Switches} & Eindhoven University \\
High performance computing (4) & \citet{Katevenis2018NextDevelopment} & MonetDB Solutions \\ \hline
\multicolumn{2}{l}{\textbf{\textbf{Programming models \& tools (15)}} }  &  \\ \cline{1-1}
Programming models and frameworks (8) & \citet{Peltenburg2020Tydi:Streams} & Delft University of Tech.  \\
Performance prediction (7) & \citet{Yasudo2018PerformancePlatforms} & University of Amsterdam \\ \hline
\textbf{\textbf{Robustness of FPGAs (26)}} &  &  \\ \cline{1-1}
Reliability (12) & \citet{Du2019UltrahighFPGA} & ESTEC \\
Hardware security (14) & \citet{Labafniya2020OnPrevention} & Delft University of Tech. \\ \hline
\textbf{\textbf{Applications (49)}} &  &  \\ \cline{1-1}
Machine learning (12) & \citet{Rocha2020BinaryWrist-PPG} & IMEC NL \\
Astronomy (11) & \citet{Ashton2020ATelescopes} & University of Amsterdam \\
Particle physics experiments (7) & \citet{FernandezPrieto2020PhaseExperiment} & Nikhef \\
Quantum computing (5) & \citet{Philips-nat-2022} & Delft University of Tech. \\
Space (9) & \citet{Barrios2020SHyLoCMissions} & ESTEC \\
Bioinformatics (5) & \citet{Malakonakis2020ExploringRAxML} & University of Twente \\ \hline
\end{tabular}
}
\end{table}

\begin{figure}[!htb]
    \centering
    \includegraphics[width=\textwidth]{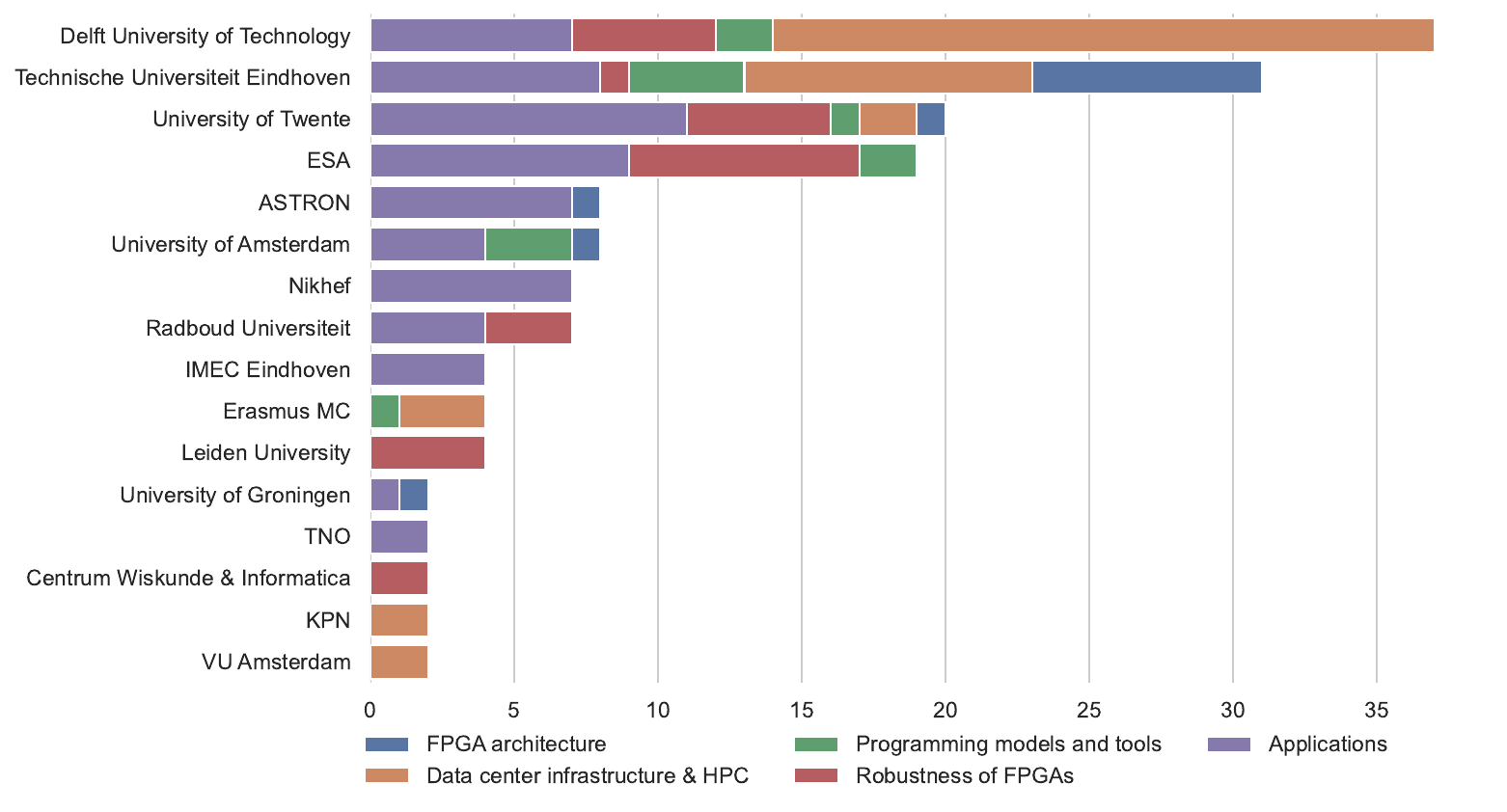}
    \caption{Number of publications per theme for each organization with more than one relevant publication.}
    \label{fig:org-publish-per-theme}
\end{figure}

Figure~\ref{fig:org-publish-per-theme} illustrates an overview of publications per theme for each organization with more than one publication. It is clear that most major contributors to FPGA research publish mostly application-specific research. Out of the major contributors, Delft University of Technology focuses more on the ``Data center \& infrastructure'' domain, while Eindhoven University of Technology is a larger contributor to the ``FPGA architecture'' theme. A brief description of each theme is provided below.

\begin{itemize}
    \item {\bf FPGA architecture}: This research theme covers literature related to the design of novel digital hardware architectures. Efficient architectures, fast on-chip memory access, coarse grained hardware design, and partially reconfigurable hardware are covered in this theme.
\item {\bf Data center infrastructure \& HPC}: This theme includes literature on FPGAs used in high-performance computing environments. This covers papers on the rapid processing of big data, and research towards distributed computing infrastructures deploying FPGAs. Furthermore, research focusing on employing FPGAs for processing communication between computing nodes using optical links is also covered here.
\item {\bf Programming models \& tools}:
This theme covers literature related to tools and models used to program FPGAs, ranging from research on high-level synthesis (HLS) tools to tools that enable accessible hardware acceleration of conventional software. This theme also features research efforts on tools for accurate performance prediction of synthesized 
FPGA solutions.
\item {\bf Robustness of FPGAs}:
This theme covers literature regarding the reliability and resilience of FPGAs to specific environments. Specifically, resilience to radiation in environments where this is prevalent is a common subject. Furthermore, this theme expands on the security of FPGAs with regards to cyberattacks.
\item {\bf Applications}: The literature on specific applications using FPGAs is more extensive than that of the other themes. This is expected since FPGAs can be applied in various fields, whereas the advancement of FPGA architectures and development tools is generally a more narrow area of research. 
Machine learning has been the main focus in recent years.
\end{itemize}

\section{FPGA Architecture}
\label{sec:archi}

This section discusses how FPGAs enable advanced computer architecture concepts that eliminate common bottlenecks in existing computer architectures. Here, we discuss three main architectural topics and their applications: Near-Memory Computing (NMC), Coarse-Grained Reconfigurable Arrays (CGRAs), and Network-on-Chip (NoC). 

\subsection{Near-Memory Computing (NMC)}

NMC is a promising approach to mitigate memory access bottlenecks in high-performance computing (HPC) systems. NMC bridges the widening gap between the computation capabilities of processors and the latency and bandwidth limitations of memory subsystems. This is specifically important for applications where conventional CPU architectures struggle due to complex data access patterns, limited data reusability, and low arithmetic intensity. These challenges are attributed to the inefficiencies in conventional memory hierarchies. FPGAs play an important role in enabling NMC due to their reconfigurable nature, allowing for the co-location of computation with memory. This co-location significantly reduces the data movement that typically limits performance and increases energy consumption in traditional architectures. FPGAs facilitate the implementation of customized, application-specific datapaths and processing units adjacent to memory, enabling more efficient data handling and processing. 

\subsubsection*{\bf{Research topics}}
There are multiple research topics being investigated by Dutch organizations that focus on harnessing NMC to enhance the performance and energy efficiency of applications with complex data access patterns, specifically through the use of FPGA-based accelerators.

\paragraph{Genomics} \citet{Singh2021FPGA-BasedApplications} 
propose improved computational solutions for genome analysis, which is an excessively expensive application in genomics. The study 
proposes a prealignment filter based on NMC to reduce the amount of data to be processed. 
\paragraph{Weather prediction} Solving large-scale weather prediction simulations suffer from limited performance and high energy consumption due to complex irregular memory access patterns and low arithmetic intensity. ~\citet{Singh2022AcceleratingFabric} 
propose 
an NMC solution with an FPGA and high-bandwidth memory (HBM) that is 5.3x faster than the baseline CPU solution. 
\paragraph{Phylogenetics} Reconstructing large evolutionary relationships among organisms is computationally expensive due to extensive calculation of probabilistic likelihood functions, which is a data-intensive, memory-bound operation. ~\citet{Alachiotis2021ScalableProcessing} 
describe an NMC solution that addresses the problem of workload distribution in a disaggregated datacenter architecture, and results in improved, scalable performance. 

\subsubsection*{\bf{Future directions}}
NMC is an emerging field that promises to have an impact on various applications domains where data access patterns hinder performance, including but not limited to, big data analytics, machine learning, and scientific simulations. Research continues to explore using FPGA-based accelerators for a broader range of applications. The integration of emerging memory technologies like high bandwidth memory with FPGAs is anticipated to further enhance bandwidth and reduce the energy footprint of such applications~\cite{Singh2019Near-memoryFuture}. Furthermore, investigations to improve the NMC architecture is on-going, by optimizing the granularity of data movement, refining memory hierarchies to match specific application patterns, and exploring the precision tolerance of various computational kernels for additional efficiency gains. The adaptability of FPGAs to various data access and processing patterns makes them attractive platforms for such optimizations~\cite{Singh2019Near-memoryFuture}. Moreover, as the NMC field progresses, there is a growing emphasis on software frameworks and tools that simplify the deployment of NMC solutions, making them accessible to a wider range of developers and applications~\cite{Abrahamse2022Memory-DisaggregatedApplications}.

\subsection{Coarse-Grained Reconfigurable Architecture (CGRA)}

CGRAs aim at improving the programming efficiency of 
FPGA platforms. Many applications do not require the bit-level reconfigurability that is provided by modern 
FPGA platforms. 
The trend toward coarser granularity is evident even in commercial FPGA architectures, as the number of Digital Signal Processing (DSP) blocks and other specialized hardware accelerators increases with each new generation.
%

\subsubsection*{\bf{Research topics}}
Multiple CGRA-related topics are explored 
by Dutch organizations, both at the level of application mapping and toward  the implementation of novel accelerator structures.



\paragraph{Application-mapping templates}\citet{Charitopoulos2021MC-DeF:Applications}
provide a template architecture that allows for device configuration at a coarser granularity (e.g., arithmetic operations), thereby  significantly reducing the architecture exploration space and significantly simplifying the work of 
application mapping tools. Due to the limited architecture exploration space, it becomes more likely that the application mapping process will find an efficient mapping. 

\paragraph{Accelerator architecture evaluation} Once a template architecture is designed, 
it can also be implemented directly on a chip, either as a standalone ASIC or integrated into a larger system as an accelerator.  This further reduces the reconfiguration overhead compared to 
FPGAs, and can result in overall improved performance \cite{Wijerathne2022HiMap:Abstraction,Wijtvliet2019Blocks:Efficiency,debruin2024rblocks}.  FPGAs, however, are still used in this context, mainly for prototyping, obtaining activity traces of applications for energy estimation, and exploring the template architecture design.

\subsubsection*{\bf{Future directions}}
Many FPGA platforms nowadays are part of a larger System-on-Chip (SoC).  These are increasingly introducing hardware accelerators for common tasks (e.g., machine learning).  CGRAs can play a key role in this context as they promise to maintain most of the high flexibility that an FPGA offers while providing  improved energy consumption and performance for many applications.
Furthermore, the more constrained architecture template of a CGRA alleviates 
the burden of application mapping, thereby benefiting 
human developers working on application mapping while 
simplifying the task of high-level synthesis tools.

\subsection{Network-on-Chip (NoC)}
NoCs play an important role in the development of complex, multi-core system on chips (SoCs), addressing the limitations of traditional bus architectures in scalability, bandwidth, and power efficiency. As embedded systems require advanced functionalities, leading to an increased number of processing elements (PEs) integrated into SoCs, the demand for efficient communication architectures has escalated. NoCs provide a scalable solution by facilitating parallel data transmission among multiple PEs through router-based packet switching networks. Complex hardware designs on FPGAs rely on 
NoCs to improve the efficiency of data communication. In addition, FPGAs serve as a popular platform in NoC development due to their reconfigurable nature, allowing for evaluating, prototyping and testing of various NoC architectures with different topologies and routing strategies. This flexibility is crucial in optimizing NoCs for specific applications, including real-time systems, where timely data delivery within predefined deadlines is essential.

\subsubsection*{\bf{Research topics}}
A number of papers have been published by Dutch organizations that focus on enhancing NoC architectures for real-time applications on FPGA platforms. 
These papers introduce novel NoC designs that aim to address the challenges of designing NoCs on FPGAs while minimizing hardware resources and power consumption. 
\paragraph{In-order NoCs} Traditional deflection-based NoCs, while efficient in resource usage, struggle with maintaining the order of flits, which is critical for various applications. IPDeN~\cite{IPDeN} addresses this challenge and ensures in-order flit delivery by incorporating a small, constant-size buffer in each router. This architecture  significantly reduces the hardware resources required compared to virtual-channel-based solutions.
\paragraph{FPGA optimized NoCs} To address the increasing communication volume between computation nodes integrated in FPGAs,  \citet{HopliteRT*} propose
HopliteRT*, 
a NoC architecture for real-time systems that is optimized for the constraints of FPGA platforms. HopliteRT* supports priority-based routing and a novel network topology to improve worst-case packet traversal time. Experiments show that the proposed NoC allows for at least 2\(\times\) improvement of traversal time of high priority packets for negligible additional hardware costs.
\paragraph{Predictable NoCs} Worst-case communication time analysis in NoCs is an important topic in real-time system design, but is rather challenging due to the unpredictable nature of NoC communication. nDimNoC~\cite{nDimNoC} is a new D-dimensional NoC that provides real-time guarantees for SoC systems. nDimNoC uses the properties of circulant topologies to ensure bounded worst-case communication delays. It requires 5x less silicon than routers that use virtual channels.

\subsubsection*{\bf{Future directions}} 
Future research on FPGA-based NoCs is expected to focus on further optimizations for real-time applications and efficient hardware utilization. New, more sophisticated routing policies and buffer management strategies can be explored to further enhance throughput and reduce latency. Commercial architectures such as the AMD Versal Adaptive SoC already contain hard NoC infrastructure. 
Given that data movement is a dominant challenge for many applications, it is likely that similar advancements will be adopted in other FPGA devices. Additionally, NoC architectures can be further developed to support dynamically changing workloads and applications with varying timing properties without necessitating network reconfiguration. This adaptability is important in addressing the evolving needs of real-time and embedded systems, where the workload can vary over time. Furthermore, intelligent routing algorithms, such as those based on machine learning, are a promising avenue for future NoC research. This approach could lead to more adaptive and efficient NoC designs that are capable of self-optimization based on current network conditions and application requirements.

\section{Data center infrastructure \& HPC}
\label{sec:HPC}
In this section, we overview research and activities related to FPGAs in HPC and data centers. The use of FPGAs in HPC and data centers raises a fundamental question: in which position FPGAs can play a significant role in the workflow of HPC and data centers?
One historical position of FPGAs in this ecosystem is in the network and communication. 
This is due to the direct I/O connection capabilities of these devices, allowing them to attach to network components (e.g., switches and routers) through a dedicated network stack directly implemented on FPGAs.
In the Microsoft Catapult project~\cite{caulfield2016cloud, PutnamAServices}, for instance, FPGAs are used in the Microsoft Bing search service as 
a re-configurable logic layer 
between network switches and servers. 

Another straightforward answer to this question is to deploy 
FPGAs as dedicated 
accelerators/co-processors. 
Due to their re-configurability and flexibility, 
FPGAs enable hardware-software co-design and implementation of domain specific applications. Moreover, FPGAs as accelerators facilitate 
spatial programming, 
e.g., dataflow implementations, to reduce data movement (from memory to compute units) compared to the traditional, control-based procedural programming~\cite{Licht2022PythonDesign}. 
As an instance of FPGAs as accelerators, Fugaku extends its supercomputer center with a scalable FPGA-cluster system~\cite{Sano2023ESSPER:Fugaku}. 
Another example is through the AMD university program~\cite{amd_hacc}, where some research institutes 
around the world deploy Heterogeneous Accelerated Compute Clusters (HACCs).  
These clusters support adaptive computing by incorporating FPGAs in their compute nodes 
to accelerate scientific applications.
Despite substantial efforts to improve the programmability of these devices for software developers and end users, achieving 
high performance through 
an optimized implementation of an algorithm remains a significant challenge in most cases.
Intel and Vmware, in collaboration with 
research institutes and universities, established the  Crossroads 3D-FPGA Academic Research Center~\cite{crossroads_fpga} to re-consider and find a permanent solution for this question. Their ambition is to define a fixed role for FPGAs as a central function in future data center servers. From 
their perspective, FPGAs will serve as 
the core 
of 
servers, acting as data movement and 
transformation engines between the network, traditional compute units, accelerators, and storage.

The aforementioned 
activities indicate the important, yet 
ambiguous role of FPGAs in the future of HPC ecosystems and data centers. ~\citet {Bobda2022TheCloud} provide 
an overview 
of existing academic and commercial efforts of employing 
in data centers. Among the commercial efforts, we observe that major data centers such as Microsoft, Alibaba, Amazon, Baidu, and Huawei 
benefit from FPGAs in their infrastructures~\cite{firestone2018azure,PutnamAServices,caulfield2016cloud,ernst2020competing,xilinx_alibaba}. 
Although this is still an open question, and various 
ad-hoc solutions have been proposed, 
one important factor will be the economic advantage; 
it will depend on whether these solutions can deliver more performance with less energy consumption and lower costs across a range of applications.
The rest of this section presents an 
overview of the FPGA research landscape 
in the Netherlands, organized into four categories: big data processing and analytics (\ref{sec:big-data-processing-analytics}), distributed computing (\ref{distcomp}), optical hardware communication (\ref{opthwcom}) and high performance computing (\ref{sec:high-performance-computing}).


\subsection{Big data processing and analytics} \label{sec:big-data-processing-analytics}

Several studies in Dutch academia 
 have assessed the domain of big data processing and analytics~\cite{Hoozemans2021FPGAOpportunities, Peltenburg2021GeneratingArrow, Rellermeyer2019TheProcessing, Fang2020In-memorySurvey}, identifying opportunities for FPGA accelerators 
 and describing the challenges faced in the wide adoption of FPGA technology. \citet{Peltenburg2021GeneratingArrow} identify 
 the programmability 
 of the accelerators, 
 the portability 
 of the implementation, 
 the interface design 
 to the data, and 
 the infrastructure 
 for data movement to/from the accelerator and across 
 kernels 
 running on 
 the accelerator, as the main challenges. Solutions 
 leveraging various existing technologies have been proposed, e.g., Apache Spark~\cite{ApacheSpark}, Apache Arrow Flight~\cite{ArrowFlight}, the IBM POWER architecture~\cite{7924241}, and OpenCAPI~\cite{OpenCAPI}, while applications 
of FPGA accelerators in this domain 
involve database search~\cite{Fang2020In-memorySurvey}, real-time data analysis~\cite{Chrysos2019DataNode}, graph-based processing~\cite{Iosup2023GraphContinuum, Prodan2022TowardsEurope}, high-frequency trading~\cite{Chen2021FPGAAlgorithm}, DNA analysis~\cite{Voicu2019SparkJNI:Spark}, and machine learning~\cite{Rellermeyer2019TheProcessing}.



\subsubsection*{\bf{Research topics}}
Several challenges in using FPGAs effectively as accelerators for big data processing and analytics have been addressed by the Dutch research community.
\paragraph{Interface design and infrastructure} Many data structures used in databases do 
not map well to the architecture of an FPGA, for example the alignment of data format or the method of data retrieval, 
thus making processing on an FPGA inefficient. Apache Arrow Flight~\cite{ArrowFlight} organizes data movement in a coherent and transparent way across various systems and applications. Fletcher~ \cite{Peltenburg2021GeneratingArrow, Ahmad2022BenchmarkingMicroservices} extends Apache Arrow Flight with FPGA support 
and defines inter-kernel infrastructure between 
processing kernels implemented in FPGA. Complementary work provides (on-line) data conversion from the widely 
used Parquet \cite{Peltenburg2020BattlingFPGA} and JSON \cite{Peltenburg2021TensAccelerators} formats to Arrow. 

\paragraph{Frameworks and tooling} 
Several frameworks have been 
developed to ease the programming of FPGA accelerators for big data processing and analytics. 
Fletcher 
integrates FPGA accelerators with tools and frameworks that use Apache Arrow as their back-end~\cite{Peltenburg2019Fletcher:Arrow}.
The open stream-oriented specification Tydi-spec \cite{Peltenburg2020Tydi:Streams} and language Tydi-lang \cite{Tian2022TydilangAL} are developed to specify and implement complex, dynamically sized data structures onto hardware streams.
SparkJNI 
enables heterogeneous CPU-FPGA systems based on the Apache Spark unified engine for large-scale data analytics \cite{Voicu2019SparkJNI:Spark}, while 
\citet{Abrahamse2022Memory-DisaggregatedApplications} extend the ThymesisFlow \cite{9252003} memory disaggregration system with a framework leveraging IBM POWER9 and FPGA accelerators.

\paragraph{Compression and decompression} 

Both the data storage size and the bandwidth required to move data from/to storage 
present significant challenges in efficiently deploying accelerators. Data compression is used to mitigate these challenges by reducing both storage size and bandwidth requirements. However, the processes of 
compression and decompression require considerable resources, and efforts have been undertaken to (de-)compress data on FPGAs, either to enable direct data processing on the FPGA itself, or to facilitate data transfers to another system component for further processing or storage. Solutions 
based on various (de-)compression algorithms have been presented, such as Snappy~\cite{Fang2019AModel}, LZ77~\cite{Fang2020AnLogic} and Zstd~\cite{Chen2021FPGAAlgorithm}. 
\citet{Hoozemans2021EnergyASIP} present an energy-efficient, FPGA-based co-processor that supports several 
decompression algorithms. 

\subsubsection*{\bf{Future directions}}

Research to develop frameworks that enable the efficient use of FPGA accelerators for big data processing and analytics is ongoing. By adopting high-level workflows tailored to these tasks, FPGA accelerators are becoming increasingly applicable within general data center infrastructures and applications. 
We see the work referred to in section \ref{sec:big-data-processing-analytics} being continued, 
as well as being extended with other partners in the industry 
\cite{10.1145/3624062.3624541, 10305451, Reukers2023AnIR, groet2024leveraging}.
Furthermore, one can not overstep the current rise of ML and AI, which, when applied to big data processing and analytics \cite{Rellermeyer2019TheProcessing} can benefit from FPGA acceleration \cite{10.1145/3613963}. The above listed technologies being developed in the Netherlands can enable the use of FPGAs as accelerators for ML and AI in big data analytics. 

\subsection{Distributed computing}
\label{distcomp}

Distributed computing involves the deployment of multiple computing nodes in parallel to increase performance and solve large computational problems. 
While 
research on distributed computing involving CPU and GPU nodes is well established, 
the emergence of FPGAs 
as a new type of computational resources and accelerators within data center infrastructures introduces a new and challenging area of research. 
The Dutch academia has mainly focused on applications that use distributed multi-FPGA systems for %
large-scale graph processing~\cite{Sahebi2023DistributedFPGAs} and deep neural networks (DNNs)~\cite{Alonso2021Elastic-DF:Partitioning}.  

\subsubsection*{\bf{Research topics}}
Several research topics have been 
investigated by Dutch researchers. 

\paragraph{Communication overhead} Reducing communication is a key factor in distributed computing, and in particular in multi-FPGA systems. By reducing communication overhead, computation time and latency reduces and efficiency increases. To reach this goal, researchers propose interconnection frameworks to establish flexible, reliable, efficient and custom communication protocols in multi-FPGA systems~\cite{salazar2020plasticnet,Salazar-Garcia2021PlasticNet+:Transceivers,Salazar-Garcia2022AApplications}. In addition to reducing latency, these 
frameworks are designed to work with various 
topology schemes and different FPGA technologies.

\paragraph{Partitioning and performance scaling} 
To increase performance in multi-FPGA systems, \citet{Alonso2021Elastic-DF:Partitioning} 
propose an open-source, distributed resource partitioning and allocator tool on FPGAs for data flow architectures targeting DNN inference; it 
works in conjunction with the FINN compiler~\cite{umuroglu2017finn}. The authors show that 
their methodology enables super-linear scaling of throughput by benefiting from model parallelism and direct FPGA-FPGA communication. 
\citet{Sahebi2023DistributedFPGAs} 
propose a (multi-FPGA) framework for 
large-scale graph processing. The framework uses an offline partitioning mechanism, and relies on 
Hadoop to map the graph into the underlying hardware. The authors show that graph partitioning using an FPGA architecture results in better performance on large graphs that include millions of vertices and billions of edges. Their results indicate 
significant speed-ups over 
state-of-the art CPU, GPU, and FPGA solutions.

\subsubsection*{\bf{Future directions}}

There are several 
challenges in distributed computing using multi-FPGA systems, thereby 
necessitating 
further research in this direction. For instance, overcoming communication barriers and designing protocols for FPGA-FPGA communication is an ongoing research domain. Moreover, at the application level, developing (standard) MPI-like collective communication libraries for multi-FPGA systems would be beneficial. 
Also, 
more case studies are needed in order to investigate and design efficient partitioning and workload distribution schemes 
for FPGA resources. Therefore, to bring ease-of-use and automation for distributed computing on FPGAs, developing libraries and tools  is crucial.

\subsection{Optical hardware communication}
\label{opthwcom}

Optical hardware communication is at the forefront of addressing the critical challenges faced by contemporary data center network (DCN) infrastructures, such as bandwidth limitations, latency issues, and scalability concerns. Optical communication is a viable alternative to conventional electrical data pathways, offering significant improvements in terms of efficiency and performance. The integration of FPGAs into optical communication systems has been a key development, providing the necessary flexibility and speed for dynamic network reconfiguration and management.

\subsubsection*{\bf{Research topics}}
The exploration of optical hardware communication utilizing FPGAs encompasses a variety of innovative research topics covered by Dutch organizations.
\paragraph{Optical wireless datacenter networks}
\citet{Zhang2022Low-LatencyRouter} have developed an optical wireless (OW)-DCN architecture that promises enhanced flexibility and scalability for DCNs, supporting high-speed optical packet-switching transmissions. FPGA-based switch schedulers are used for control of the implementation based on semiconductor optical amplifier (SOA)-based wavelength selectors and arrayed waveguide grating routers (AWGRs).

\paragraph{Disaggregated optical networks}
The DACON project~\cite{Guo2022DACON:Invited} introduces a Disaggregated, Application-Centric Optical Network that utilizes hybrid optical switches and FPGA-based controllers, resulting in improved application performance and reduced latency.

\paragraph{Low-latency edge networks}
The Electro-Optical Communication group at TU/e has proposed an edge data center network architecture that employs photonics and FPGA-based supervisory channels to achieve microsecond-time control and deterministic latency \cite{Santana2023SOA-BasedApplications}.

\paragraph{Nanosecond optical switching}
A novel optical switching and control system has been designed to address the bandwidth bottlenecks of electrical switching, featuring a distributed network architecture with optical label channels and the Optical Flow Control (OFC) protocol \cite{Xue2022NanosecondNetworks}.

\paragraph{Hybrid datacenter architectures} The HiFOST DCN architecture~\cite{Yan2018HiFOST:Switches} integrates flow-controlled fast optical switches with modified top-of-the-rack switches, offering substantial improvements in latency and cost efficiency.

\paragraph{Beyond 5G networks} \citet{Santana2022TransparentApplications} present a new Edge Cloud Network design 
that uses 
FPGA-based controllers for rapid reconfiguration of optical networks, catering to the low-latency requirements of 5G applications and beyond.  

\subsubsection*{\bf{Future directions}}

Through various ongoing developments, the field of optical hardware communication is poised for significant advancements. Efforts to mitigate LED nonlinearity have led to the development of a Legendre-polynomials-based post-compensator optimized for FPGA implementation, enhancing the bit rate efficiency of high-speed Visible Light Communications (VLC) systems \cite{Niu2021LEDCommunications}. The introduction of a real-time FPGA-based implementation of a nonlinear LED model and post-compensator marks a substantial contribution to VLC technology, enabling high data rates over bandwidth-limited LEDs \cite{Deng2022Physics-BasedImplementation}. 
A concurrency-aware mapping technique has been developed to reduce optical packet collisions in Architecture-On-Demand  (AoD) network infrastructures, improving buffer utilization and execution time degradation in HPC systems \cite{Meyer2018OpticalPerformance}.


\subsection{High performance computing}\label{sec:high-performance-computing}

Benefiting from FPGAs in 
HPC applications is an active research area. 
Even though GPUs remain 
the most prevalent 
accelerator technology in HPC, and AI-specific hardware is being  increasingly adopted, 
FPGAs are also 
increasingly employed in HPC centers. 

\subsubsection*{\bf{Research topics}}
Dutch institutes have been involved in European projects, e.g., 
ExaNeSt~\cite{Katevenis2018NextDevelopment} and MANGO~\cite{Flich2018ExploringApproach}, to 
design 
large-scale heterogeneous compute systems. We can observe 
the important role of FPGAs in these projects, facilitating network communication or accelerating execution. 

\paragraph{Architecture and system design}
    The 
    ExaNeSt European project~\cite{Katevenis2018NextDevelopment} deploys FPGAs 
    as accelerators in 
    a European 
    exascale supercomputer based on low-cost, low-power 
    ARM cores. They also employ an FPGA-based testbed for a low-latency, high bandwidth unified Remote Direct Memory Access (RDMA) interconnect, and present 
    a custom FPGA-based switch to support inner-cabinet communications.
   The MANGO project~\cite{Flich2018ExploringApproach} aims at addressing the PPP (Performance, Power, and Predictability) space in HPC 
   by exploring 
   customizabe and deeply heterogeneous accelerators. Their hardware concept consists of General-purpose compute Nodes (GNs) with 
   commercial accelerators such as Xeon Phi and NVIDIA GPUs, along with Heterogeneous Nodes (HNs). HNs are clusters of many-core chips coupled with customized heterogeneous computing resources, including high-capacity clusters of FPGAs.
    
\paragraph{Programming languages, tools, and applications}
    Within the ExaNeSt project, \citet{Katevenis2018NextDevelopment} 
    design a novel microarchitecture as Top-of-Rack switches. In one of their experiment, they port the OpenCL kernels of the molecular dynamics simulator LAMMPS~\cite{plimpton1995fast} to FPGAs using HLS tools. 
    They report that running the kernel on an FPGA requires 0.56 seconds while the 4 ARM cores requires 1.3 seconds. That is an improvement of more than a factor 2 in speed up.
    Within the MANGO project, ~\citet{Flich2018ExploringApproach} target three applications with significant QoS aspects: 1) online video transcoding, 2) rendering for medical imaging, and 3) error correcting codes in communication. The MANGO project relies on LLVM~\cite{lattner2004llvm} and their programming model is an extension of 
    existing languages and libraries (e.g., OpenCL~\cite{opencl}) for HPC by integrating the expression of new architectural features as well as QoS concerns and parameters. This is achieved 
    by augmenting the runtime library API with new functions, pragmas and keywords to the existing HPC languages (e.g., clang C/C++ frontend). 
    
\paragraph{Performance models}
    Combining the advantages of reconfigurability, dataflow computation, and heterogeneity results in 
    Reconfigurable Dataflow Platforms (RDPs) as a promising building block in 
    next-generation, large-scale high-performance machines. RDPs rely on 
    Reconfigurable Dataflow Accelerators (RDAs) to realize multiple streaming pipelines, each 
    comprising many parallel operations. Due to the 
    heterogeneous hierarchy, 
    performance prediction of RDPs is very challenging, in particular to detect bottlenecks within reasonable time and accuracy. 
\citet{Yasudo2021AnalyticalPlatforms,Yasudo2018PerformancePlatforms} 
propose a performance estimate framework for reconfigurable dataflow applications 
    named Performance Estimation for Reconfigurable Kernels and Systems (PERKS). 
    It 
    automatically extracts specific parameters from the application, the hardware, and the platform to calibrate the model. They use eight applications for their evaluation: AdPredictor (an online machine learning algorithm), N-body simulation, Monte Carlo simulation, sequence alignment, Asian option pricing, Jacobi solver, and Regression/regularisation solver. Their results show that PERKS achieves 
    accuracy of 91\% on these applications.

\subsubsection*{\bf{Future directions}}

Determining the role of FPGAs in HPC necessitates more research from both data center architecture design and FPGA programming model.
From a data center design perspective, the positioning of FPGAs in the architecture of HPC centers needs more investigation. This also depends on the targeted application workflow and how FPGA can impact the most.    
From a user perspective, the programmability of these devices is an important factor. Therefore, the gap between software developers and FPGA programming models and tools should be reduced further to use FPGA as mainstream HPC devices.

\section{Programming Models and Tools}
\label{sec:programming}

This section discusses approaches that boost developer productivity; Section~\ref{prog_models_frameworks} reviews programming models and frameworks that raise the abstraction level of describing hardware, while Section~\ref{perf_pred} presents 
techniques that predict performance of synthesized programs.

\subsection{Programming models and frameworks}
\label{prog_models_frameworks}

FPGA development is traditionally characterized by a steep learning curve, especially for non-experts. For this reason, High-Level Synthesis (HLS) tools and, more generally, high-level programming models and frameworks have been proposed to increase productivity by raising the abstraction level. 
HLS tools became commercial products in the early 2010s. Since then, they have been used in various application domains, including, but not limited to, deep learning, multimedia, graph processing, and genome sequencing ~\cite{Cong-2022}. HLS tools can reduce the average development time (up to two-thirds compared to RTL~\cite{Lahti-2019}). However, they still require considerable expertise to optimize the FPGA designs and achieve a Quality-of-Result that is on par with the one obtained through hardware description languages. For this reason, higher-level programming models and frameworks are now being proposed. They allow developers 
to describe hardware 
in a more convenient formalism (e.g., Clash~\cite{clash-2010,clash-website}, HeteroCL~\cite{heterocl-2019}, PyLog~\cite{pylog-2021}, and DaCe~\cite{dace-2021} which currently support Haskell, Python or a Python-embedded DSL), and 
automatically, or via user-provided hints, generate optimized 
HLS/HDL descriptions. 


\subsubsection*{\bf{Research topics}} Several papers have been published by Dutch organizations focusing on reducing development time for hardware design using HLS programming models and tools.

\paragraph{Abeto framework
}

\citet{Sanchez2022AbetoManagement} propose Abeto, a software tool for IP management and workflow automation. Historically, there has been no established standard for packing, documenting, and distributing IP core designs.  This prevents their re-usability, as each IP core has its unique learning curve and challenges for using them in an EDA toolchain. Abeto allows the user to operate in a unified manner with heterogeneous IP cores, and conveniently configure and launch the different stages of the IP workflow. To add an IP core, Abeto requires some auxiliary information to be provided: a database definition (containing information about the directory structure of the IP core) and a command dictionary (which includes the list of supported IP commands and how they must be executed). The tool has been validated against a subset of the ESA portfolio of IP cores\footnote{\url{https://www.esa.int/Enabling_Support/Space_Engineering_Technology/Microelectronics/ESA_HDL_IP_Cores_Portfolio_Overview}}, which constitutes a heterogeneous group of IP cores, demonstrating the tool's versatility.

\paragraph{
Design flow for Gowin FPGAs}
\citet{vos2020gowin} describe a method 
to create an open-source design flow for the Gowin LittleBee family of FPGAs.
The design flow is based on well-known open-source tools such as Yosys and nextpnr, as well as the newly developed bitstream generator.
Architectural details of the FPGA family were documented using input fuzzing and comparing results from the existing closed-source vendor tool flow.
While the created open-source flow is capable of synthesizing a full RISC-V core, many aspects, such as DSPs, RAMs, and PLLs are currently unsupported. The authors report that documenting the bitstream format for all of these features is 
the subject of future work.


\paragraph{AEx framework}

\citet{Hirvonen2023AEx:Co-Processors} propose 
AEx, a framework for automated High-Level synthesis of compiler programmable co-processors. AEx can be used to produce Application-Specific Instruction-Set (ASIP) architectures. ASIP processors have been proposed as a way to produce FPGA overlays starting from a software-programmable template. The program being executed can be easily changed, reducing design time and costs. The template being used by AEx is Transport Triggered Architecture (TTA).
AEx includes heuristics for design space exploration and pruning, aimed at finding the best architecture able to satisfy real-time execution time and clock frequency constraints. The user can then choose the results that better fit their need (e.g., minimum resource utilization). Evaluation 
shows how the tool is able to produce results in a reasonable amount of time, achieving 
performance close to that of the fixed-function implementations generated by HLS vendor tools such as AMD/Xilinx Vitis.


\paragraph{Synthesis from Simulink Models to FPGA for Aerospace Applications}


Reconfigurable hardware is becoming an attractive solution for aerospace applications, thanks to its power efficiency and capabilities of in-flight configuration. Algorithms are usually expressed in model-based programming frameworks, e.g., Matlab Simulink, but turning them into low-level hardware description languages can be cumbersome. \citet{Curzel2023ExplorationApplications} analyze solutions to automatically synthesize Simulink models. Matlab already provides an automated method (HDL coder) to translate part of Simulink models into Verilog/VHDL, but this still requires a certain level of expertise. Therefore, the authors propose to apply HLS on the code generated by Matlab's Embedded Coder tool, further automatizing the design process. Experiments with three 
benchmarks show that this solution is more efficient than relying on HDL coder, and it does not require specific hardware expertise. 

\paragraph{HLS optimizations for  post-quantum cryptography on Lattice FPGAs}

\citet{Guerrieri2022OptimizingSynthesis} discuss the process of porting Post-Quantum Cryptographic algorithms to an FPGA using HLS. While it can be reasonably straightforward 
to port an existing CPU implementation to an FPGA, 
performance can be low and resource utilization 
is not optimal. The authors discuss how, applying well-known HLS-specific optimization techniques, the code can be rewritten to leverage the capabilities of HLS tools and produce more efficient designs, reducing the computation latency of up to two orders of magnitudes in specific cases.

\paragraph{Optimizations for Heterogeneous HPC Systems (OPTIMA)}
\citet{Theodoropoulos2023optima} demonstrate the results of porting and optimizing industrial applications to two new heterogeneous HPC systems within the OPTIMA project. The results highlight the performance increase of using the available FPGA-based accelerators versus a pure software implementation running on the CPUs of the HPC system.


\paragraph{A Framework for Heterogeneous Cloud Applications (VINEYARD)}
\citet{Sidiropoulos2018vineyard} describe a framework to accelerate different parts of an application across different accelerators, like GPUs and FPGAs. They demonstrate the utility of this framework by creating a platform for computational neuroscience, called BrainFrame. The BrainFrame platform allows one to simulate spiking neural networks, and depending on the number of neurons and their interconnectivity, certain combinations of accelerators achieved the shortest simulation times.

\paragraph{Mapping data structures to hardware streams (Tydi)}
\citet{Peltenburg2020Tydi:Streams} describe a specification for mapping complex, dynamically sized data structures onto a fixed number of hardware streams.

\subsubsection*{\bf{Future directions}}
Traditional FPGA programming has been done using Hardware Description Languages, which have a steep learning curve that does not favor adopting reconfigurable devices in the scientific and industry community. To address this issue, there is a collective effort to increase the level of abstraction for FPGA designs without compromising performance. Achieving this goal requires a multidisciplinary approach involving programming languages, compilers, and optimization techniques. HLS tools and high-level approaches are being used in various application domains.
Although the current direction in this field in the Netherlands is unclear, our analysis has pinpointed specific domains of interest within local research communities, such as aerospace and accelerated big data processing, that could benefit from more accessible programming methods for FPGA devices.


\subsection{Performance prediction}
\label{perf_pred}

FPGA design and development processes are time-consuming activities due to, among others, the very fine granularity reconfigurability of FPGA designs, which translates into a large design space and long synthesis time. For this reason, it is crucial to enable quick performance prediction of synthesized programs to improve early-stage design analysis and exploration, and performance debugging.
We can distinguish between two main types of performance prediction models: analytical and ML-based. Analytical models (such as HLscope+~\cite{hlscope-2017} and COMBA~\cite{comba-2020}) analyze the source code and use mathematical modeling to estimate performance and resource utilization. They are able to produce quick estimates at the cost of reduced accuracy. ML-Based models~\cite{oneal-2018, ustun-2020, Sun-2021}, on the other hand, 
aim at improving prediction accuracy by considering device-specific features, but typically 
require long and expensive training procedures.


\subsubsection*{\bf{Research topics}} Several papers have been published by Dutch organizations focusing on performance prediction of synthesized codes.


\paragraph{System Performance Prediction via Transfer learning (LEAPER)}

\citet{Singha2022leaper} describe a method for predicting system performance and resource usage of FPGA accelerators using transfer learning.
They trained a performance predictor model for an edge/embedded FPGA, and used transfer learning so that the model can also be used for cloud/high-end FPGAs.
The method allows for design space exploration of mapping C/C++ programs to cloud/high-end FPGAs using HLS. The authors showed that it is 10x faster than the state of the art, achieving 85\% accuracy.

\paragraph{Modeling FPGA-Based Systems via Few-Shot Learning}
Machine learning based models are being proposed to provide fast and accurate performance predictions of FPGA-based designs. However, training these models is expensive, due to the time-consuming FPGA design cycle. 
\citet{Singh2021ModelingLearning} propose a transfer-learning-based approach for FPGA-based systems, to adapt an existing ML model, trained for a specific device, to a new, unknown environment, reducing the training costs.

\paragraph{
Energy and Area Estimation for Coarse-grained Reconfigurable Architectures}
Design space exploration is often required to achieve good Pareto points when creating reconfigurable architectures. 
\citet{Wijtvliet2021cgra} introduce the CGRA-EAM  model for energy and area estimation for CGRAs. It 
achieves a 15.5\% error for energy and 2.1\% error for area estimation for the Blocks~\cite{Wijtvliet2019Blocks:Efficiency} CGRA. 
The novelty of this work lies in its focus on CGRAs and its ability to handle multiple different applications running on a CGRA.

\paragraph{Analytical Performance Estimation for Large-Scale Reconfigurable Dataflow Platforms}

\citet{Yasudo2018PerformancePlatforms, Yasudo2021AnalyticalPlatforms} 
introduced and further expanded 
\emph{PERKS}, a performance estimation framework for reconfigurable dataflow platforms. The authors propose 
that reconfigurable accelerators, such as FPGAs, will play an important role in future exascale computing platforms and that such a framework is essential in the efficient deployment of applications on heterogenous platforms with reconfigurable accelerators. The PERKS framework uses parameters from the target platform and the application to build an analytical model to predict the performance of multi-accelerator systems. Experimental results with different reconfigurable dataflow applications are presented, showing that the framework can predict the performance of current workloads with high accuracy.


\paragraph{Memory and Communication Profiling for Accelerator-Based Platforms}
\citet{Ashraf2018MemoryPlatforms} present \emph{MCPROF}, an open-source memory-access and data-communication profiler. The tool provides a detailed profile of memory-access behavior for heterogeneous systems (CPU, GPU, and FPGA) for C/C++ applications, as well as data-communication-aware mapping of applications on these architectures. Comparison with the state of the art show that the proposed profiler has an order of magnitude, on average, lower overhead than state-of-the-art data-communication profilers over a wide range of benchmarks. A case study  with several image processing applications for heterogeneous multi-core platforms containing an FPGA and a GPU as accelerators was conducted. The authors also demonstrate that the tool can provide insights into whether 
a specific 
accelerator (GPU or FPGA) is a good fit 
for the application.

\paragraph{Delay Prediction for ASIC HLS}
The delay estimates of HLS tools can often deviate significantly from results obtained from logic synthesis. 
\citet{De2023hls} propose 
a hybrid model by incorporating graph-based learning models, which can infer structural features from a design, into traditional non-graph-based learning models for delay estimates. The hybrid model improves delay prediction by 93\% in comparison to the delay prediction reported 
by a 
commercial HLS tool. 

\subsubsection*{\bf{Future directions}}

Performance prediction for FPGAs and, more generally, reconfigurable dataflow devices is challenging. Traditionally, this was addressed with 
analytical approaches, but more recently, machine-learning-based approaches are becoming mainstream, and we can expect them to become a popular option given their successes in recent research. With CGRA-like architectures being commercialized, the interest in exploring these devices is increasing, and we 
expect future work to focus more on the predictability of the 
performance of such devices. Performance prediction is a device-specific task, yet only a 
few 
of the reviewed papers favor the reproducibility of the presented results, or publicly release the 
code used to generate these results. 
We should shift toward more reproducible and transparent research to foster 
collaborations and facilitate general progress. Thus, we advocate for the need for open-source practices 
in performance prediction as well.

\section{Robustness of FPGAs}
\label{sec:robustness}
This section discusses deployment challenges and ongoing research efforts in Dutch institutions toward achieving robust FPGA implementations. The robustness of FPGAs and the designs they support is a multi-dimensional challenge, affecting several stages of an FPGA-based systems' production chain. Here, we delve into two critical aspects: reliability (Sec.~\ref{reliabilitysec}) and security (Sec.~\ref{securitysec}). 
We discuss the intricacies, and present collaborative efforts required to address these challenges.

\subsection{Reliability}
\label{reliabilitysec}
This section delves into the reliability of FPGA devices 
in high-radiation environments, such as space and particle colliders---a topic that is at the forefront of research by ESTEC in the Netherlands.
%
%
FPGAs, particularly those based on SRAM, are vulnerable to charged or high-energy particles. The Configuration RAM (CRAM) of SRAM-based FPGAs is prone to Single Event Upsets (SEUs), which can alter bits within the configuration memory. These alterations can potentially change the logic functions and connections within the device, leading to Single Event Functional Interruptions (SEFIs). To mitigate these risks, protective measures such as Triple Modular Redundancy (TMR) or Error Correcting Code (ECC) scrubbers are often implemented. However, these solutions can complicate the design and add significant overhead. 
As an alternative, Flash-based FPGAs offer a robust option since their configuration memory is inherently immune to SEUs, although they face constraints in terms of computing resources and less mature toolchains. 



\subsubsection*{\bf{Research topics}}

FPGAs serve as an ideal test platform for advancing the maturity of RTL designs, especially in the context of Technology Readiness Levels (TRL) in space applications. 
FPGAs can be 
used to simulate early-stage prototypes, allowing for iterative refinement. In subsequent phases, they can be used to 
validate system components under simulated space-like conditions, ensuring RTL designs work as intended in the actual hardware configuration. In the final stages, FPGAs integrated with RTL designs are tested in operational environments, including actual space missions, demonstrating the system’s performance in real-world conditions. This approach 
provides a flexible and cost-effective method for testing and validating technology, and 
helps managing 
risks associated with deploying new technologies in space. Dutch research on FPGA devices focuses primarily on technology validation to assess and enhance their resilience to adverse conditions such as radiation, alongside the development of novel reliability techniques to mitigate vulnerabilities like high susceptibility to SEUs. Additionally, FPGAs are extensively employed for 
testing 
low TRL 
innovations and for realizing robust operational systems in various applications. The following list elaborates on these core research areas.


    \paragraph{Technology validation of FPGA devices}
    This topic assesses the resilience of FPGA devices under the effect of charged and high-energy particles. The Payload-XL project~\cite{Viel2023Payload-XL:FPGA} exemplifies practical deployment, validating the BRAVE FPGA's performance in orbit and demonstrating the system's real-world reliability under space conditions. 
    Studies that 
    use ultra-high-energy heavy ions~\cite{Vlagkoulis2021SingleIons,Du2019UltrahighFPGA} characterize and quantify how various FPGA technologies respond to environmental challenges, such as in space. \citet{Leon2021DevelopmentBenchmarks} employ 
    diverse benchmarks to evaluate the performance of new radiation hardened devices. 
    
    \paragraph{Advancements in FPGA device reliability}
    Research here develops innovative techniques aimed at enhancing FPGA reliability. 
    \citet{DeSio2023PyXEL:FPGAs} describe 
   PyXEL's 
   bitstream analysis tool for 
   enhancing FPGA robustness by automating reliability analysis and facilitating mitigation solutions.  \citet{Vlagkoulis2022ConfigurationTechnique} present 
   %
   configuration memory scrubbing methods that integrate mixed 2-D coding techniques, while 
\citet{Mousavi2023MTTRSensitivity}
   discuss scrubbing methods 
   that strive to reduce the mean time to repair, 
   thereby significantly improving error correction capabilities.
    
    \paragraph{
    FPGAs as research and development platforms}
    Various studies explore the role of FPGAs as both experimental test beds and final platforms for payload systems~\cite{Viel2023Payload-XL:FPGA,Forlin2023AnSounds,Bohmer2023NeutronFPGAs}. 
    \citet{Gambardella2022AcceleratedTraining} conducted accelerated radiation tests on quantized neural networks, while \citet{Anders2023AProcessors} and \citet{Hoozemans2018IncreasingProcessor} underline the capabilities of FPGAs to support the development of new processor architectures, such as RISC-V processors, and new computing paradigms, such as polymorphic VLIW processors, respectively.

   

\subsubsection*{\bf{Future directions}}

The future of reliability work on FPGAs in the Netherlands is increasingly shaped by the demand from the ESTEC ESA center and the expanding New Space economy for advanced space applications. Dutch companies, such as Technolution, with its FreNox RISC-V soft-core, play a pivotal role in building this high-reliability ecosystem. As new FPGA providers in Europe, like NanoExplore, emerge, the need for technology-agnostic designs that enable the seamless integration of external IPs becomes more critical. The primary focus moving forward will be on expanding this ecosystem by adapting and validating existing common IPs for high-reliability applications, while maintaining ease of integration across various platforms.

\subsection{Hardware security}
\label{securitysec}

This section explores the security challenges and innovative solutions in FPGA technology.
FPGAs present unique security risks due to their reconfigurability and wide deployment range, from cloud data centers to edge devices. Their flexibility and complex ecosystem makes them susceptible to various threats, including hardware Trojans, fault injection, and side-channel attacks.

\begin{figure}
    \centering
    \includegraphics[width=0.7\linewidth]{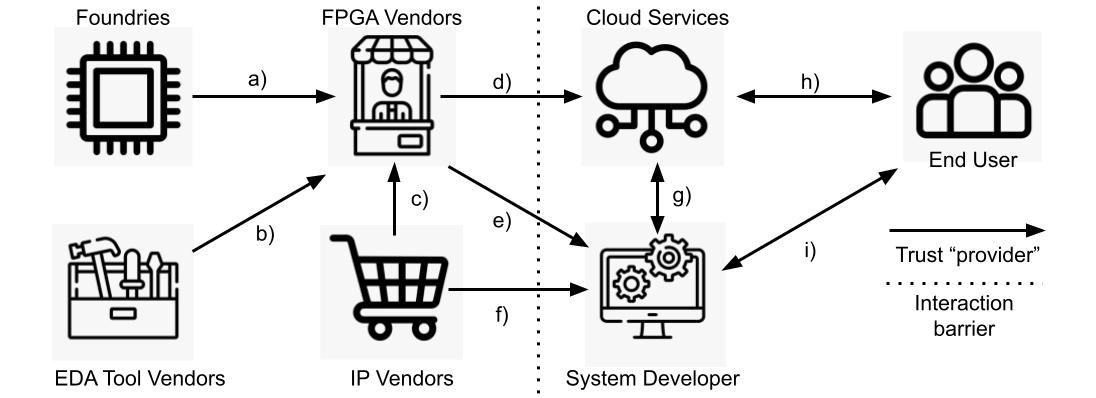}
    \caption{Trust chain demonstrated in the FPGA-based system market. The ``interaction barrier'' marks the point where the interactions become complex with trust being provided and required for the transactions. Adapted from~\cite{Zhang2014ASystems}.}
    \label{fig:FPGA-trust}
\end{figure}

Security, particularly in the realm of FPGAs, hinges on establishing and maintaining trust and isolation. This trust permeates the entire supply chain of the FPGA market, as detailed by \citet{Zhang2014ASystems}. Trust must exist between suppliers and consumers—for instance, between foundries and FPGA vendors—and is subsequently extended to downstream users. For example, system developers rely on FPGA vendors to ensure that their hardware and EDA tools are free from hardware Trojans, as discussed in various studies~\cite{Zeitouni2021TrustedFPGAs,Labafniya2020OnPrevention,Nikiema2023TowardsDevices,Palumbo2022IsAnswer}. The direct relationships of trust within this framework are illustrated in Figure~\ref{fig:FPGA-trust}. We observe that, at certain stages within the supply chain, the dynamics between service providers and consumers go beyond mere transactional interactions and become operational. This shift is represented by the dotted line in the figure. 
Typical examples include the relationship between end users and system developers, or between cloud service providers and end users. In these scenarios, interactions between the involved parties are more reciprocal, and the focus expands beyond trust. Critical issues then include ensuring isolation within cloud environments, protecting against malicious users, and implementing countermeasures to various attacks, as highlighted in recent research~\cite{Zeitouni2021TrustedFPGAs,Koylu2022ExploitingAttacks,Garaffa2021RevealingNeuron,Socha2020Side-channelHardware}.

\subsubsection*{\bf{Research topics}}

In the Netherlands, research on security spans multiple technical domains, employing a variety of techniques to ensure operational level guarantees.

\paragraph{Vendor-side:} Part of this research is devoted to assessing the implicit trust indicated by elements to the left-hand side of the dotted line in Figure~\ref{fig:FPGA-trust}. These efforts predominantly aim to detect and neutralize hardware trojans, as discussed in various studies~\cite{Zeitouni2021TrustedFPGAs,Labafniya2020OnPrevention,Nikiema2023TowardsDevices,Palumbo2022IsAnswer}. In cases involving hardware trojans, while FPGA vendors might not be directly accountable for the intrusion, they inherit a trust relationship from upstream entities such as foundries.

\paragraph{User-side:}
On the other hand, much of the research focus in the Netherlands is directed toward the interactions depicted on the right-hand side of the dotted line in Figure~\ref{fig:FPGA-trust}. This includes examining the connections among cloud services, system developers, and end users. Key focal areas include the development of hardware primitives like Programmable Unclonable Functions (PUFs)~\cite{Jin2022ProgrammableApplications,Jin2020ErasableDesign} and Roots-of-Trust ~\cite{Nikiema2023TowardsDevices}, as well as the evaluation of fault injection and protection mechanisms ~\cite{Miteloudi2022ROCKY:Data,Nikiema2023TowardsDevices,Koylu2022InstructionAttacks,Koylu2022ExploitingAttacks}. Additionally, significant attention is given to the detection and mitigation of side-channel attacks ~\cite{Lahr2020SideImplementation,Miteloudi2021EvaluatingLeakage, Socha2020Side-channelHardware, Garaffa2021RevealingNeuron} and microarchitectural attacks~\cite{Arikan2022ProcessorSketches,Nikiema2023TowardsDevices}.

\subsubsection*{\bf{Future directions}}

Current research in the Netherlands extensively utilizes FPGA systems across all stages of end development. However, there is relatively less focus on enhancing the resilience and security of FPGA production, likely due to the absence of commercial players developing these systems within the country. As a result, future research is expected to increasingly emphasize the integration of FPGAs into system-level solutions, addressing critical issues in security from an end-user perspective. The inherent flexibility of FPGA platforms also presents significant opportunities for innovative system design. They can serve as an initial gateway into a broader design ecosystem, where the development of intellectual properties (IPs) is validated for functional metrics such as side-channel leakage and protection mechanisms. This approach ensures that new designs meet stringent standards before deployment, fostering a robust and secure technological landscape.

\section{Applications}
\label{sec:applications}

FPGAs have emerged as powerful accelerators for a wide range of applications. In this section, we discuss FPGA-based solutions in machine learning (Section~\ref{sec:ml}), astronomy (Section~\ref{sec:astr}), particle physics experiments (Section~\ref{sec:phys}), quantum computing (Section~\ref{sec:quant}), space applications (Section~\ref{sec:space}), and bioinformatics (Section~\ref{sec:bio}).

\subsection{Machine learning}
\label{sec:ml}
In the field of machine learning, and in particular deep learning, hardware acceleration plays a vital role. GPUs are the predominant method for hardware acceleration due to their high parallelism, but FPGA research is showing promising results. FPGAs enable inference at greater speed and better power efficiency when compared to GPUs \cite{hw-efficiency-compare} by designing model-specific accelerated pipelines \cite{ml-energy-efficient-cnn}. Through the co-design of machine learning models and machine learning hardware on FPGAs, models are accelerated without compromising on performance metrics and utilizing limited FPGA resources. In addition, the flexibility of the FPGA's architecture enables the realization of unconventional deep learning technology, such as Spiking Neural Networks (SNNs). 

\paragraph{Hardware acceleration} Ample research on hardware acceleration focuses on accelerating existing neural network architectures. One common class of architectures is convolutional neural networks (CNNs), which learn image filters in order to identify abstract image features. CNNs are often deployed in embedded applications which require real-time image processing and low energy consumption, making FPGAs a suitable candidate for CNN acceleration. \citet{ml-energy-efficient-cnn} propose an implementation of the LeNet architecture using Vitis HLS, pipelining the CNN layers, and outperforms other FPGA based implementations at a processing time of $70 \mu s$. One downside to this approach is the inflexibility of designing a specific model architecture in HLS which can be resolved by using partial reconfiguration \cite{ml-cnn-acclr-part-reconf}. To increase CNN throughput, further parallelization can be exploited, and in combination with the use of the high bandwidth OpenCAPI interface, can achieve a latency of less than $10 \mu s$ on the LeNet-5 model, streaming data from an HDMI interface \cite{ml-FPQNet}. In each of these implementations, fully pipelined CNNs are possible due to the limited number of parameters in small CNNs. As larger pipelined networks are deployed on FPGAs, parallelization puts strain on the available resources, and in particular the amount of on-chip-memory becomes a bottleneck. A proposed solution to this is using Frequency Compensated Memory Packing \cite{ml-mem-efficient-df-inf}.

In addition to CNN acceleration, general neural network acceleration has been developed by means of a programmable Tensor Processing Unit (TPU) as an overlay on an FPGA accelerator \cite{ml-agile-tuned-tpu}. Deep learning acceleration using FPGAs is also relevant to space technology research. Since the reprogrammability of FPGAs make them a suitable contender for deployment on space missions, FPGA implementations of existing deep learning models are being benchmarked for space applications \cite{ml-myriad-2-space-cnn} \cite{ml-mem-efficient-df-inf}.

\paragraph{Spiking neural networks} Spiking Neural Networks (SNNs) are computational models formed using spiking neuronal units that operate in parallel and mimic the basic operational principles of biological systems. These features endow SNNs with potentially richer dynamics than traditional artificial neural network models based on the McCulloch-Pitts point neurons or simple ReLU activation functions that do not incorporate timing information. Thus, SNNs excel in handling temporal information streams and are well-suited for innovative non-von-Neumann computer architectures, which differ from traditional sequential processing systems. SNNs are particularly well-suited for implementation in FPGAs due to their massive parallelism and requirement for significant on-chip memories with high-memory bandwidth for storing neuron states and synaptic weights. Additionally, SNNs use sparse binary communication, which is beneficial for low-latency operations because both computing and memory updates are triggered by events. FPGAs' inherent flexibility allows for reprogramming and customization, which enable reprogrammable SNNs in FPGAs, resulting in flexible, efficient, and low-latency systems~\cite{Corradi2021Gyro:Analytics,Irmak2021ADesigns,SankaranAnInference}. \citet{corradi2024accelerated} demonstrated the application of a Spiking Convolutional Neural Network (SCNN) to population genomics. The SCNN architecture achieved comparable classification accuracy to state-of-the-art CNNs while processing only about 59.9\% of the input data, reaching 97.6\% of CNN accuracy for classifying selective-sweep and recombination-hotspot genomic regions. This was enabled by 
the SCNN's capability to temporize genetic information, allowing it to produce classification outputs without processing the entire genomic input sequence. Additionally, when implemented on FPGA hardware, the SCNN model exhibited over three times higher throughput and more than 100 times greater energy efficiency than a GPU implementation, markedly enhancing the processing of large-scale population genomics datasets.

\paragraph{Model/hardware co-design} Previous examples demonstrate that existing deep neural network models can be accelerated using FPGAs. Typically, research in this area focuses on designing an optimal hardware solution for an existing model. A more effective approach, however, is to co-design the model and the hardware accelerator simultaneously. However, simultaneous co-design of DNN models and accelerators is challenging. DNN designers often need more specialized knowledge to consider hardware constraints, while hardware designers may need help to maintain the quality and accuracy of DNN models. Furthermore, efficiently exploring the extensive co-design space is a significant challenge. This co-design methodology leads to better performance, leveraging FPGAs' flexibility and rapid prototyping capabilities. For example, \citet{Rocha2020BinaryWrist-PPG}, by co-designing the bCorNET framework, which combines binary CNNs and LSTMs, they were able to create an efficient hardware accelerator that processes HR estimation from PPG signals in real-time. The pipelined architecture and quantization strategies employed allowed for significant reductions in memory footprint and computational complexity, enabling real-time processing with low latency.

In SNNs, encoding information in spike streams is a crucial co-design aspect. SNNs primarily use two encoding strategies: rate-coding and time-to-first-spike (TTFS) coding. Rate coding is common in SNN models, encoding information based on the instantaneous frequency of spike streams. Higher spike frequencies result in higher precision but at the cost of increased energy consumption due to frequent spiking. While rate coding offers accuracy, it reduces sparsity. In FPGA implementations, rate coding is often used for its robustness, simplicity, ease of training through the conversion of analog neural networks to spiking neural networks, and practicality in multi-sensor data fusion, where it helps represent real values from various sensors (radars, cameras) even in the presence of jitter or imperfect synchronization~\cite{Corradi2021Gyro:Analytics}.
Conversely, TTFS coding has been demonstrated in SNNs implemented on FPGAs to enhance sparsity and has the potential of reducing energy consumption by encoding information in spike timing. For instance, Pes et al.~\cite{Pes2024ActiveNetworks} introduced a novel SNN model with active dendrites to address catastrophic forgetting in sequential learning tasks. Active dendrites enable the SNN to dynamically select different sub-networks for different tasks, improving continual learning and mitigating catastrophic forgetting. This model was implemented on a Xilinx Zynq-7020 SoC FPGA, demonstrating practical viability with a high accuracy of 80\% and an average inference time of 37.3 ms, indicating significant potential for real-world deployment in edge devices.

\subsection{Astronomy}
\label{sec:astr}

Astronomy is the study of everything in the universe beyond our Earth's atmosphere. Observations are done at different modalities and wavelengths, such as detection of a range of different particles (e.g., Cherenkov detector based systems such as KM3NeT \cite{KM3NeT:2009xxi}), gravitational waves, optical observations, gamma and x-ray observations and radio (e.g., WSRT \cite{van_Cappellen_2022}, LOFAR \cite{van_Haarlem_2013}, SKA \cite{book-SKA}). Observations can be done from space or from earth; in this section, we limit the scope to ground-based astronomy. A common denominator for instruments required for observation of the different modalities and different wave lengths is that the systems need to be very sensitive in order to observe very faint signals from outside the Earth's atmosphere. Instruments are typically large and/or distributed over a large area 
to achieve 
good sensitivity and resolution. Different modalities and wavelengths require distinct types of sensors, cameras, or antennas to convert observed phenomena into electrical signals. Each system is tailored to its specific modality and wavelength, necessitating specialized components to accurately capture and translate the data. 
At a certain stage in the signal chain, the electrical signal is converted into the digital domain, where it undergoes multiple processing stages. This processed signal ultimately results in an end product that can be utilized by scientists for analysis and research purposes.
Systems can roughly be split into two parts, a front-end and a back-end. The front-end requires interfacing with and processing of data from the sensor; electronics commonly deployed in the front-end are constrained in space (size), temperature, power, cost, RFI, environmental conditions and serviceability. The back-end processes data produced by the front-end(s) either in an online or offline fashion, which is usually 
done with server infrastructure in a data center. 
In the back-end, the main challenges are the high data bandwidth and large data size coming from the front-ends. Although the environment is more flexible, systems are still constrained in space, power, and cost.


FPGAs have been used in astronomy instrumentation for quite some time, as they 
are 
efficient in interfacing with Analog to Digital Converters (ADCs), and well suited to the conditions faced in instrumentation front-ends (e.g. NCLE \cite{karapakula2024ncle}). Moreover, FPGA are also used further down the processing stages for various signal processing operations, both in the front-ends (e.g., Uniboard2 in LOFAR \cite{doi:10.1142/S225117171950003X}) as well as in the back-ends of systems (e.g., MeerKAT \cite{2022JATIS...8a1006V} and SKA \cite{SKA-CBF}). GPUs represent a good alternative in back-end processing (e.g., LOFAR's system COBALT \cite{Broekema_2018}) as well. The work by Veenboer et al. \cite{10.1007/978-3-030-29400-7_36} describes a trade-off between using a GPU and an FPGA accelerator in the implementation of an image processing operation in a radio telescope back-end.

\paragraph{Hardware Development for the Radio Neutrino Observatory in Greenland (RNO-G)}
The RNO-G \cite{Smith2022HardwareRNO-G} is a radio detection array for neutrinos. It consists of 35 autonomous stations deployed over a $5 \times 6$ km grid near the NSF Summit Station
in Greenland. Each station includes an FPGA-based phased trigger. The station has to operate in a 25 W power envelope. The implementation on FPGA seems to be favorable due to environmental conditions and operation constraints.

\paragraph{Implementation of a Correlator onto a Hardware Beam-Former to Calculate Beam-Weights}
The Apertif Phased Array Feed (PAF) \cite{van_Cappellen_2022} is a radio telescope front-end used in the WSRT system in the Netherlands. FPGAs are used for antenna read out as well as signal processing close to the antenna. Schoonderbeek et al. \cite{Schoonderbeek2020ImplementationBeam-Weights} describe the transformation and implementation of a beamformer algorithm on FPGA in order to build a more efficient system.

\paragraph{Near Memory Acceleration and Reduced-Precision Acceleration for High Resolution Radio Astronomy Imaging}
\citet{Corda2020NearImaging} describe the implementation of a 2D FFT on FPGA, leveraging Near-Memory Computing. The 2D FFT is applied to an image processing implementation on FPGA in the back-end of a radio telescope and compared with implementations on CPU and GPU. \citet{Corda2022Reduced-PrecisionHardware} explore 
reduced-precision computation on an FPGA 
for the same image processing application. 

\paragraph{The MUSCAT Readout Electronics Backend: Design and Pre-deployment Performance}
The MUSCAT is a large single dish radio telescope with 1458 receives in the focal plane. The system uses FPGA based electronics to read out and pre-process the data from the receivers \cite{Rowe2023ThePerformance}. 


\paragraph{Cherenkov Telescope Arrays}
Three different contributions have been made to three different Cherenkov based Telescope Arrays.
Ashton et al.~\cite{Ashton2020ATelescopes} describe a system for the High Energy Stereoscopic System (H.E.S.S.) where a custom board with ARM CPU and an FPGA is used to read out and pre-process a custom designed NECTAr digitizer chip in the front-end of the system. After pre-processing, the data is distributed to a back-end over Ethernet.
Sánchez-Garrido et al.~\cite{Sanchez-Garrido2021AArray} present the design of a Zynq FPGA SoC based platform for White Rabbit time synchronization in the ZEN-CTA telescope array front-ends. Data captured and pre-processed at the front-ends is distributed over Ethernet to the back-end including the time stamp.
Aiello et al.~\cite{Aiello2021ArchitectureFirmware} outline the architecture and performance of the KM3Net front-end firmware. The KM3NeT telescope consists of two deep-sea three-dimensional sensor grids being deployed in the Mediterranean Sea. A central logic board with FPGA in the front-end serves as a Time to Digital Converter to record events and time at the sensors; the data is transmitted and further processed in a back-end on shore.


\vspace{0.4cm}
FPGA are mainly used for front-end sensor interfacing and pre-processing. \citet{Corda2020NearImaging, Corda2022Reduced-PrecisionHardware} underline that FPGAs are also still relevant in the back-end, providing improved performance over CPU and on-par performance with GPU accelerators. FPGA are expected to remain the dominant choice for platforms in astronomy instrumentation front-ends due to the strong interfacing capabilities and the adaptability and suitability to the constraints imposed by instrumentation front-ends. In the back-end, FPGAs provide a viable solution to application acceleration, but will have to compete with other accelerator architectures, e.g., GPUs~\cite{10.1007/978-3-030-29400-7_36}. 
An emerging new technology are the Artificial Intelligence Engines in the AMD Versal Adaptive SoC. The work from \citet{Versal-ACAP} evaluated the AI Engines for a signal processing application in radio astronomy. The flexibility and programmability of the AI Engines, combined with the interfacing capabilities of the FPGA can lead to a powerful platform for telescope front-ends.

\subsection{Particle physics experiments}
\label{sec:phys}
The Large Hadron Collider (LHC) features various particle accelerators to facilitate particle physics experiments. Experiments performed using particle accelerators can produce massive amounts of data that needs to be propagated and preprocessed at high speeds before the reduced relevant data is recorded for offline storage. FPGAs are widely employed throughout systems LHC particle accelerators, such as ATLAS and LHCb, for their high-bandwidth capabilities, and the flexibility that reconfigurable hardware offers without requiring hardware alterations to the system. Recently both the ATLAS and LHCb particle accelerators have been commissioned for upgrades. The Dutch Institute for Subatomic Physics (Nikhef) is one of the collaborating institutes working on the LHC accelerators.


LHCb is a particle accelerator that specializes in experiments that study the bottom quark. Major upgrades to the LHCb that enable handling a higher collision rate require new front-end and back-end electronics. To facilitate the back-end of the upgrade, the LHCb implements the custom PCIe40 board, which features an Intel Arria 10 FPGA. Four PCIe40 boards are dedicated for controlling part of the LHCb system, and 52 PCIe40 boards are used to read out each of the detector’s slices, producing an aggregated data rate of 2.85 Tb/s \cite{FernandezPrieto2020PhaseExperiment}.

ATLAS is one of the general particle accelerators of the LHC. ATLAS uses two trigger stages in order to record only the particle interactions of interest. In an upgrade to the ATLAS accelerator, ASIC-based calorimeter trigger preprocessor boards are replaced by FPGA-based hardware. Using FPGAs for this purpose allows implementing enhanced signal processing methods \cite{Aad2020PerformanceTrigger}. After the two trigger stages, FPGAs are deployed to process the triggers for tracking particles \cite{Aad2021TheSystem}.

Ongoing upgrades to the LHC particle accelerators, referred to as High Luminosity LHC (HL-LHC), will facilitate higher energy collisions. HL-LHC will produce increased background rates. To reduce false triggers due to background, the New Small Wheel checks for coinciding hits. Each trigger processor features Virtex-7, Kintex Ultrascale and Zynq FPGAs \cite{Iakovidis2023TheElectronics}. Interaction to and from front-end hardware is done through Front-End Link eXchange (FELiX) boards. As part of the HL-LHC upgrades, each FELiX board must facilitate a maximum throughput of 200 Gbps. To enable this, Remote Direct Memory Access (RDMA) over Converged Ethernet (RoCE) as part of the FELiX FPGA system is proposed \cite{Vasile2022FPGALHC, Vasile2023IntegrationLHC}. The performance of the FELiX upgrade in combination with an upgraded Software ReadOut Driver (SW ROD) satisfies the data transfer requirements of the upgraded ATLAS system \cite{Gottardo2020FEliXSystem}.

\subsection{Quantum computing}
\label{sec:quant}
Quantum computing promises to help solving many global challenges of our time such as quantum chemistry problems to design new medicines, the prediction of material properties for efficient energy storage, and the handling of big data needed for complex climate physics~\cite{Gibney-nat-2014}. The most promising quantum algorithms demand systems comprising thousands to millions of quantum bits~\cite{Meter-2013}, the quantum counterpart of a classical bit. A quantum processor comprising up to 50 qubits has been realized using solid-state superconducting qubits~\cite{Arute-nat-2019}, but its operation requires a combination of cryogenic temperatures below ~100 mK and hundreds of coaxial lines for qubit control and readout. 
While in systems with a few qubits, this can be controlled using off-the-shelf electronic equipment, such approach becomes infeasible when scaling qubit systems toward thousands or millions of qubits that are required for a practical quantum computer. 


A means to tackle the foreseeable bottleneck in scaling the operation of qubit systems is to integrate FPGA technology in the control and readout of solid-state qubits. FPGAs have been used to generate highly-stable waveforms suitable for the control of quantum bits with latency significantly lower than software alternatives~\cite{Ireland-2020}. In systems of semiconductor spin qubits, FPGAs have provided in-hardware syncing of quantum dot control voltages with the signal acquisition and buffering and thus enabled the observation of real-time charge-tunneling events~\cite{Hartman-2023}. FPGAs have also been used to configure and synchronize a cryo-controller with an arbitrary waveform generator required to generate complex pulse shapes and perform quantum operations~\cite{Xue-nat-2021}. Such setup has enabled the demonstration of universal control of a quantum processor hosting six semiconductor spin qubits~\cite{Philips-nat-2022}. FPGAs have proven to be essential for implementing quantum error correction algorithms, which are critical for mitigating the effects of dephasing and decoherence in solid-state qubits. 
In qubit systems based on superconducting quantum circuits, the first efficient demonstration of quantum error correction was made possible by a FPGA-controlled data acquisition system which provided dynamic real-time feedback on the evolution of the quantum system~\cite{Ofek-nat-2016}. It has been further predicted that FPGA can enable highly-efficient quantum error correction based on neural-network decoders~\cite{Overwater-2022}.


FPGA technology has proven invaluable in the development of the emerging research field of quantum computing.
However, the complexity of programming FPGA circuits hinders their implementation in quantum computing systems. Commercial efforts have been done toward providing graphical tools for designing FPGA programs, namely the Quantum Researchers Toolkit by Keysight Technologies and the FPGA-based multi-instrument platform Moku-Pro developed by Liquid Instruments. These tools are essential for implementing customized algorithms without the need for dedicated expertise in hardware description languages. Future research is also needed in integrating FPGAs in cryogenic platforms required to operate qubit systems. Such capability has already been demonstrated; commercial FPGAs can operate at temperatures below 4 K and be integrated in a cryogenic platform for qubit control~\cite{Homulle-2017}. These efforts provide evidence that FPGA technology is of great interest for enabling a scalable and practically applicable quantum computer.

\subsection{Space}
\label{sec:space}
The flexibility of FPGA technology makes it a suitable platform for many applications on-board space missions. The European Space Research and Technology Centre (ESTEC), as part of the European Space Agency (ESA) actively explores FPGA technology for space applications, and has an extensive portfolio of FPGA Intellectual Property (IP) Cores~\cite{esa_ip}.


FPGAs can flexibly route its input and output ports, and can be configured to support many different communication protocols. This makes FPGAs good contenders as devices that communicate with the various hardware platforms and sensors on a space mission. FPGAs and have been implemented as interface devices in novel on-board machine learning and digital signal processing  implementations~\cite{Leon2021ImprovingSoC, Leon2021FPGABenchmarks, karapakula2024ncle}. 

An on-board task for which FPGAs are used is hyperspectral imaging. This type of on-board imaging produces vast amounts of data. To reduce transmission bandwidth requirements when transmitting the sensory data to earth, real-time on-board compression handling high data rates is required. FPGAs are well-suited for such tasks, and research has been done on using space-grade radiation-hardened FPGAs \cite{Barrios2020SHyLoCMissions} as well as commercial off-the-shelf (COTS) FPGAs \cite{Rodriguez2019ScalableCompression} for on-board hyperspectral image compression. COTS devices are generally cheaper than space-grade devices, but the higher susceptibility of these devices to radiation-induced effects makes them challenging to employ.

Communication between on-board systems often requires high data-rates and is susceptible to radiation induced effects. To deal with the unique constraints of space applications, dedicated communication protocols such as SpaceWire, and its successor, SpaceFibre have been developed. These protocols are available as FPGA IP implementations, and testing environments of SpaceFibre have been developed \cite{MystkowskaSimulationSpaceFibre, AnSection}. SpaceWire can interface with the common AXI4 protocol using a dedicated bridge \cite{RubattuASystems}, enabling its integration with SpaceWire interfaces. Direct Memory Access (DMA) allows peripherals to transfer data to and from an FPGA without going through a CPU. The application of DMA in space is being investigated, however its application as of now is limited since DMA is susceptibility to radiation-induced effects \cite{Portaluri2022Radiation-inducedDevices}.

\subsection{Bioinformatics}
\label{sec:bio}
FPGA technology has been extensively explored for accelerating Bioinformatics kernels. Bioinformatics is an interdisciplinary scientific field that combines biology, computer science, mathematics, and statistics to analyze and interpret biological data. The field primarily focuses on the development and application of methods, algorithms, and tools to handle, process, and analyze large sets of biological data, such as DNA sequences, protein structures, and gene expression patterns.Continuous advances in DNA sequencing technologies~\cite{hu2021next} have led to the rapid accumulation of biological data, creating an urgent need for high-performance computational solutions capable of efficiently managing increasingly larger datasets.

\citet{Shahroodi2022KrakenOnMem:Profiling} describe a hardware/software co-designed framework to accelerate and improve energy consumption of taxonomic profiling. In metagenomics, the main goal is to understand the role of each organism in our environment in order to
improve our quality of life, and taxonomic profiling involves the identification and categorization of the various types of organisms present in a biological sample by analyzing DNA or protein sequences from the sample to determine which species or taxa are represented. The study focuses on boosting performance of table lookup, which is the primary bottleneck in taxonomic profilers, by proposing a processing-in-memory hardware accelerator. Using large-scale simulations, the authors report an average of 63.1\% faster execution and orders of magnitude higher energy efficiency than the  widely used metagenomic analysis tool Kraken2~\cite{wood2019improved} executed on a 128-core server with AMD EPYC 7742 processors  operating at 2.25 GHz. An FPGA was used for prototyping and emulation purposes.

\citet{Corts2022AcceleratedFPGAs} employ FPGAs to accelerate the detection of traces of positive natural selection in genomes. The authors designed a hardware accelerator for the $\omega$ statistic~\cite{kim2004linkage}, which is extensively used in population genetics as an indicator of positive selection. In comparison with a single CPU core,
the FPGA accelerator can deliver up to $57.1\times$ faster
computation of the $\omega$ statistic, using the OmegaPlus~\cite{alachiotis2012omegaplus} software implementation as reference.


\citet{Malakonakis2020ExploringRAxML} use FPGAs to accelerate the widely used phylogenetics software tool RAxML~\cite{stamatakis2014raxml}. The study implements the Phylogenetic Likelihood Function (PLF), which is used for evaluating phylogenetic trees, on a Xilinx ZCU102 development board and a cloud-based Amazon AWS EC2 F1 instance. The first system (ZCU102) can deploy two accelerator instances, operating at 250MHz, and delivers up to $7.7\times$ faster executions than sequential software execution on a AWS EC2 F1 instance. 
The AWS-based accelerated system is $5.2\times$ faster than the same software implementation. 

\citet{Alachiotis2021AcceleratingCloud} also target the PLF implementation in RAxML, and propose an optimization method for data movement in PCI-attached accelerators using tree-search algorithms. They developed a software cache controller that leverages data dependencies between consecutive PLF calls to cache data on the accelerator card. In combination with double buffering over PCIe, this approach led to nearly $4\times$ improvement in the performance of an FPGA-based PLF accelerator. Executing the complete RAxML algorithm on an AWS EC2 F1 instance, the authors observed up to $9.2\times$ faster processing of protein data than a $2.7$ GHz Xeon processor in the same cloud environment.

With genomic datasets continuing to expand, bioinformatics analyses are likely to increasingly rely on cloud computing in the future. This shift will be supported by new programming models and frameworks designed to address the data-movement challenges posed by cloud-based hardware accelerators. These accelerators, such as FPGAs and GPUs, need data transfers from the host processor, which can significantly impact execution times and negate gains from computation improvements. Fortunately, similar data-movement concerns exist for both FPGAs and GPUs, and ongoing engineering efforts are likely to converge on common solutions~\cite{Corts2023AGenetics}. This will help bring optimized, hardware-accelerate processing techniques into more widespread use among computational biologists and bioinformaticians in the future.

\section{Research and Development in Industry}
\label{sec:industry}


To understand the alignment between FPGA research in Dutch academia and industry, a questionnaire was initially distributed to Dutch companies working in FPGA technology. Nine companies responded, representing a broad range of FPGA applications, from custom hardware architecture development to programming models, tools, and application-specific designs. Thereafter, we conducted interviews with several industry representatives in the Netherlands, reviewed the initial survey findings with them, and integrated their feedback into this analysis. The survey posed nine questions, addressing the following subjects:
\begin{enumerate}
    \item The themes and applications where companies apply FPGAs (question 1 and 2)
    \item The reasons for companies to choose FPGA technology (question 3)
    \item To what extend the companies create custom FPGA designs (question 4)
    \item The satisfaction of companies with respect to the available FPGA technology on the market (question 5 and 6)
    \item The most beneficial advances in FPGA technology to companies (question 7)
    \item The involvement of companies in academic research (question 8 and 9)
\end{enumerate}
The companies involved and consented to their names being published are ProDrive Technologies, HDL Works, Sioux Technologies, Technolution, QBayLogic, Demcon and Core-Vision\footnote{\url{https://prodrive-technologies.com}, \url{https://hdlworks.com}, \url{https://sioux.eu}, \url{https://technolution.com}, \url{https://qbaylogic.com}, \url{https://demcon.com}, \url{https://www.core-vision.nl}}. Other companies choose to remain anonymous. 

\begin{table}[ht]
\caption{Most cited applications where FPGAs are used, and reasons for preferring a design involving FPGAs}
\label{tab:survey_results}
\resizebox{\columnwidth}{!}{%
\begin{tabular}{lc|lc}
\textbf{Applications for FPGA design} & \textbf{Companies} & \textbf{What makes FPGAs the preferred technology} & \textbf{Companies} \\ \hline
Control systems                       & 7                       & High degree of parallelism                         & 4                       \\
Near memory processing                & 3                       & High bandwidth and throughput                      & 3                       \\
Reliability and resilient design      & 3                       & Flexibility of development                         & 3                       \\
Machine learning                      & 2                       & Custom hardware interfacing                        & 2                       \\
Distributed computing                 & 2                       & Low-latency                                        & 2                       \\
Coarse grained architecture design    & 2                       & Affordable compared to ASIC design             & 2                      
\end{tabular}%
}
\end{table}

The most frequently cited applications for FPGAs are control systems, near-memory processing and resilient designs. Table \ref{tab:survey_results} shows an overview of the most cited applications (each company could submit multiple applications). Most companies also reported that they use existing FPGA technology and software to develop custom IP designs for their products, while a fraction (three companies) also highlighted their focus on custom software development to create FPGA solutions.

Companies cite a variety of reasons for choosing FPGAs over GPUs or CPUs in their solutions. The most frequently mentioned motivations are shown in table \ref{tab:survey_results} (there can be more than one motivation per company). These motivations align with the typical advantages of FPGAs, including high parallelism, high throughput, and substantial processing bandwidth. Additionally, companies appreciate the flexibility in hardware design that FPGAs offer, allowing for adjustments throughout a product's lifetime without major redesigns. Several companies also highlight the deterministic behavior and precise timing of FPGAs as key factors in their decision-making. Furthermore, the cost-effectiveness of FPGAs for implementing custom hardware designs—especially for low-volume products, compared to developing a custom ASIC—plays a significant role in their preference. Other advantages mentioned include flexible hardware interfacing, low latency, low power consumption, and sufficient I/O bandwidth.

Most companies express general satisfaction with the current state of accessible FPGA technology, with only two reporting dissatisfaction. However, they identify several developments that could enhance their FPGA designs, such as on-chip memory for smaller FPGA devices, improved quality of IP cores, better software for managing HDL dependencies and packages, and high-level languages that retain the details of hardware design. Many companies suggest that FPGA technology would be more attractive if high-level design efforts shifted focus from making FPGAs easier for software developers to enhancing the capability of hardware designers to work at a higher level of abstraction. Overall, tooling is regarded as a critical area for improvement. Frequent changes in vendor tooling often lead companies to develop custom solutions. Additionally, vendor-specific tools create varying workflows, making it challenging to work with devices from different manufacturers. While open-source tools exist, they are generally not considered suitable for professional use.

In conclusion, companies employ FPGA technology when their solutions require the specific advantages that FPGAs provide, such as low latency, high processing bandwidth, deterministic behavior, and flexible interfacing. The accessibility and adoption of FPGAs in the industry would significantly improve with the development of vendor-agnostic, open-source toolchains that are suitable for professional use. Moreover, creating a high-level programming language tailored for hardware designers, instead of focusing on accessibility for software developers, would greatly enhance FPGA design in industrial applications.


\section{Conclusions}
This paper presents an overview of the current research landscape, applications, and future potential of FPGA technology in the Netherlands. It highlights how academia and industry in the Netherlands play an important role in developing new and innovative FPGA-based technologies and solutions to address various important societal challenges ranging from healthcare to power efficiency. 

\subsection{Summary and main findings}
We selected a total of 
212 relevant FPGA-related papers published in the past 5 years. Most contributions to the national FPGA research effort stem from 16 organizations, including major Dutch universities, research institutes, and 
industry. Many of these publications are collaborations of national and international partners, mostly European. We note that the survey covers about 1\% of the relevant world-wide FPGA publications (from the same period), with the Netherlands ranking 22nd in the world and 10th in Europe in relevant FPGA-research output. 

We classified the selected papers into five major themes: a) FPGA architecture, with 11 published papers, b) data center infrastructure \& HPC, with 40 papers, c) programming models \& tools, with 15 papers, d) robustness of FPGAs, with 26 papers, and e) applications, with 120 papers. We paid specific attention and reviewed in depth popular FPGA applications, i.e, those applications with a significant number of relevant publications where FPGAs play an important role. We found 49 application papers over 6 subjects; these publications indicate that FPGAs have emerged as powerful accelerators for a wide range of applications, such as machine learning, astronomy, particle physics experiments, quantum computing, space applications, and bioinformatics. 

Our survey revealed ample future directions across all themes. In terms of architecture (see Section ~\ref{sec:archi}), specialized FPGA architectures (e.g., for in-memory computing) as well as embedding FPGA technology as part of (complex) computing systems beyond the computation (e.g., for on-chip and off-chip communication or for memory systems) are  promising paths towards reducing the memory gap of traditional systems. For datacenters and HPC infrastructure (see Section~\ref{sec:HPC}), we identify a promising research direction in FPGA-based acceleration in general, and tooling to enable its seamless integration into large-scale systems in particular; a second research direction should focus on supporting novel communication protocols and technologies, to further improve data movement in supercomputers, datacenters, and the computing continuum. For programming models and tools (see Section~\ref{sec:programming}), the main research developments target high-level tooling for easier development and deployment, accessible for more educated users than HDL experts, and tooling support for performance analysis, modeling, and prediction, which will help solution feasibility analysis and fair comparison against competing technologies. For robustness of FPGAs (see Section~\ref{sec:robustness}), reliability and security are key aspects where more research is needed in architecture and tooling for moving from proof-of-concept to complex (eco)systems, where FPGAs can guarantee such essential non-functional requirements in the most adverse conditions (e.g., for deployment in space). Finally, for applications, future research should focus on the adoption of FPGAs in more application domains where energy-efficient acceleration is essential; we suggest further research towards more efficient developments in AI (for both training acceleration and latency-sensitive deployment), as well as for various computing continuum layers, fast/specialized communication, and HPC kernels. We further expect more developments towards successful acceleration in simulation and digital twin applications, where FPGAs could play an interesting role in a simulation-emulation continuum.     

\subsection{Limitations}

The literature was collected by requiring at least one author to be affiliated with a Dutch organization. However, this criterion does not provide insights into the degree of involvement of the Dutch-affiliated author in the research. It is possible that a co-author with a Dutch affiliation made only a minor contribution to the overall work presented in the paper. Given the limitations of the available data, determining the specific contributions of individual authors is not feasible; hence, this survey does not include an analysis of the contributions of individual authors for each paper.

This survey categorized FPGA literature by subject and provided insights into the content of FPGA research by Dutch organizations. Moreover, literature categories were grouped per Dutch organization, identifying the primary FPGA-related research direction of each Dutch organization. The survey does not identify which research groups in each organization are focusing on a specific domain of FPGA research. As future work, mapping out these research groups, and identifying the lines of research connecting them, can give further insights into the directions of FPGA research in the Netherlands. Moreover, this can aid in identifying which FPGA-related lines of research are primarily established in the Netherlands.

\subsection{Future Work}
   
Our detailed analysis of the last half-decade of Dutch FPGA research does already emphasize that, while the technology holds tremendous potential, there are ongoing challenges, particularly in terms of development tools, performance predictability, and hardware security. Future research will need to focus on improving these aspects, especially with regards to user-friendly programming models and more robust performance estimation frameworks. The continued investment in FPGA technology will be vital for maintaining the Netherlands' competitive edge and addressing the growing demand for energy-efficient, high-performance computing solutions. Furthermore, fostering open-source tools, as well as deeper industry-academia collaboration, will be essential for sustaining long-term growth and innovation in this rapidly evolving field.

To further enlarge the scope of this analysis, we welcome a broader community-driven research, to survey the developments and needs of FPGA research and innovation at European or even world-wide level. Such analysis will reveal additional collaboration opportunities, cross-theme and cross-domain, across multiple layers of modern computing systems. While we recognize the scale and scope of such an effort are significantly larger, we argue that it is an effective way to map the current landscape of FPGA research, and further focus on relevant research challenges and opportunities.  

\bibliographystyle{ACM-Reference-Format-num}
\bibliography{references,mendeley}


\begin{thebibliography}{210}


\ifx \showCODEN    \undefined \def \showCODEN     #1{\unskip}     \fi
\ifx \showDOI      \undefined \def \showDOI       #1{#1}\fi
\ifx \showISBNx    \undefined \def \showISBNx     #1{\unskip}     \fi
\ifx \showISBNxiii \undefined \def \showISBNxiii  #1{\unskip}     \fi
\ifx \showISSN     \undefined \def \showISSN      #1{\unskip}     \fi
\ifx \showLCCN     \undefined \def \showLCCN      #1{\unskip}     \fi
\ifx \shownote     \undefined \def \shownote      #1{#1}          \fi
\ifx \showarticletitle \undefined \def \showarticletitle #1{#1}   \fi
\ifx \showURL      \undefined \def \showURL       {\relax}        \fi
\providecommand\bibfield[2]{#2}
\providecommand\bibinfo[2]{#2}
\providecommand\natexlab[1]{#1}
\providecommand\showeprint[2][]{arXiv:#2}

\bibitem[\protect\citeauthoryear{??}{Arr}{2019}]%
        {ArrowFlight}
 \bibinfo{year}{2019}\natexlab{}.
\newblock \bibinfo{title}{Introducing Apache Arrow Flight}.
\newblock
\newblock
\urldef\tempurl%
\url{https://arrow.apache.org/blog/2019/10/13/introducing-arrow-flight/}
\showURL{%
\tempurl}


\bibitem[\protect\citeauthoryear{??}{acm}{2024}]%
        {acm_advanced_search}
 \bibinfo{year}{2024}\natexlab{}.
\newblock \bibinfo{title}{{ACM} {Advanced} {Search}}.
\newblock
\newblock
\urldef\tempurl%
\url{https://dl.acm.org/search/advanced}
\showURL{%
\tempurl}


\bibitem[\protect\citeauthoryear{??}{amd}{2024}]%
        {amd_hacc}
 \bibinfo{year}{2024}\natexlab{}.
\newblock \bibinfo{title}{{AMD} {HACC} {Program}}.
\newblock
\newblock
\urldef\tempurl%
\url{https://www.amd-haccs.io/index.html}
\showURL{%
\tempurl}


\bibitem[\protect\citeauthoryear{??}{Apa}{2024}]%
        {ApacheSpark}
 \bibinfo{year}{2024}\natexlab{}.
\newblock \bibinfo{title}{Apache Spark}.
\newblock
\newblock
\urldef\tempurl%
\url{https://spark.apache.org/}
\showURL{%
\tempurl}


\bibitem[\protect\citeauthoryear{??}{cla}{2024}]%
        {clash-website}
 \bibinfo{year}{2024}\natexlab{}.
\newblock \bibinfo{title}{Clash: a modern, functional, hardware description language}.
\newblock \bibinfo{howpublished}{\url{https://clash-lang.org/}}.
\newblock
\newblock
\shownote{[Accessed 15-05-2024].}


\bibitem[\protect\citeauthoryear{??}{cro}{2024}]%
        {crossroads_fpga}
 \bibinfo{year}{2024}\natexlab{}.
\newblock \bibinfo{title}{{Crossroads} {3D-FPGA} {Academic} {Research} {Center}}.
\newblock
\newblock
\urldef\tempurl%
\url{https://www.crossroadsfpga.org/}
\showURL{%
\tempurl}


\bibitem[\protect\citeauthoryear{??}{esa}{2024}]%
        {esa_ip}
 \bibinfo{year}{2024}\natexlab{}.
\newblock \bibinfo{title}{ESA HDL IP Cores Portfolio Overview}.
\newblock \bibinfo{howpublished}{\url{https://www.esa.int/Enabling_Support/Space_Engineering_Technology/Microelectronics/ESA_HDL_IP_Cores_Portfolio_Overview}}.
\newblock
\newblock
\shownote{[Accessed 03-06-2024].}


\bibitem[\protect\citeauthoryear{??}{iee}{2024}]%
        {ieee_advanced_search}
 \bibinfo{year}{2024}\natexlab{}.
\newblock \bibinfo{title}{{IEEE} {Xplore} {Help}}.
\newblock
\newblock
\urldef\tempurl%
\url{https://ieeexplore.ieee.org/Xplorehelp/searching-ieee-xplore/advanced-search}
\showURL{%
\tempurl}


\bibitem[\protect\citeauthoryear{??}{rat}{2024}]%
        {rathenauNederlandHorizon}
 \bibinfo{year}{2024}\natexlab{}.
\newblock \bibinfo{title}{{N}ederland en {H}orizon 2020 | {R}athenau {I}nstituut}.
\newblock \bibinfo{howpublished}{\url{https://www.rathenau.nl/nl/wetenschap-cijfers/nederland-en-horizon-2020}}.
\newblock
\newblock
\shownote{[Accessed 19-02-2024].}


\bibitem[\protect\citeauthoryear{??}{Ope}{2024}]%
        {OpenCAPI}
 \bibinfo{year}{2024}\natexlab{}.
\newblock \bibinfo{title}{OpenCAPI Consortium}.
\newblock
\newblock
\urldef\tempurl%
\url{https://opencapi.org}
\showURL{%
\tempurl}


\bibitem[\protect\citeauthoryear{??}{els}{2024}]%
        {elsevierScopus}
 \bibinfo{year}{2024}\natexlab{}.
\newblock \bibinfo{title}{{S}copus | {A}bstract and citation database | {E}lsevier --- elsevier.com}.
\newblock \bibinfo{howpublished}{\url{https://www.elsevier.com/products/scopus}}.
\newblock
\newblock
\shownote{[Accessed 10-11-2023].}


\bibitem[\protect\citeauthoryear{??}{EuE}{2024}]%
        {EuEsaSpaceGradeFpgProject}
 \bibinfo{year}{2024}\natexlab{}.
\newblock \bibinfo{title}{{S}pace {Q}ualification and {V}alidation of {H}igh {P}erformance {E}uropean {R}ad-{H}ard {F}{P}{G}{A} | {O}{P}{E}{R}{A} {P}roject | {F}act {S}heet | {H}2020 | {C}{O}{R}{D}{I}{S} | {E}uropean {C}ommission --- cordis.europa.eu}.
\newblock \bibinfo{howpublished}{\url{https://cordis.europa.eu/project/id/821969}}.
\newblock
\newblock
\shownote{[Accessed 19-02-2024].}


\bibitem[\protect\citeauthoryear{??}{xil}{2024}]%
        {xilinx_alibaba}
 \bibinfo{year}{2024}\natexlab{}.
\newblock \bibinfo{title}{{Xilinx} {Case} {Study}. Xilinx Powers Alibaba Cloud FaaS with AI Acceleration Solution for E-Commerce Business}.
\newblock
\newblock
\urldef\tempurl%
\url{https://www.xilinx.com/publications/powered-by-xilinx/xilinx-alibaba-case-study.pdf}
\showURL{%
\tempurl}


\bibitem[\protect\citeauthoryear{Aad, Abbott, Abbott, Abed~Abud, Abeling, et~al\mbox{.}}{Aad et~al\mbox{.}}{2020}]%
        {Aad2020PerformanceTrigger}
\bibfield{author}{\bibinfo{person}{G. Aad}, \bibinfo{person}{B. Abbott}, \bibinfo{person}{D.~C. Abbott}, \bibinfo{person}{A. Abed~Abud}, \bibinfo{person}{K. Abeling}, {et~al\mbox{.}}} \bibinfo{year}{2020}\natexlab{}.
\newblock \showarticletitle{{Performance of the upgraded PreProcessor of the ATLAS Level-1 Calorimeter Trigger}}.
\newblock \bibinfo{journal}{\emph{Journal of Instrumentation}} \bibinfo{volume}{15}, \bibinfo{number}{11} (\bibinfo{date}{11} \bibinfo{year}{2020}), \bibinfo{pages}{P11016}.
\newblock
\showISSN{1748-0221}
\urldef\tempurl%
\url{https://doi.org/10.1088/1748-0221/15/11/P11016}
\showDOI{\tempurl}


\bibitem[\protect\citeauthoryear{Aad, Abbott, Abbott, Abud, Abeling, et~al\mbox{.}}{Aad et~al\mbox{.}}{2021}]%
        {Aad2021TheSystem}
\bibfield{author}{\bibinfo{person}{G. Aad}, \bibinfo{person}{B. Abbott}, \bibinfo{person}{D.~C. Abbott}, \bibinfo{person}{A.~Abed Abud}, \bibinfo{person}{K. Abeling}, {et~al\mbox{.}}} \bibinfo{year}{2021}\natexlab{}.
\newblock \showarticletitle{{The ATLAS Fast TracKer system}}.
\newblock \bibinfo{journal}{\emph{Journal of Instrumentation}} \bibinfo{volume}{16}, \bibinfo{number}{07} (\bibinfo{date}{7} \bibinfo{year}{2021}), \bibinfo{pages}{P07006}.
\newblock
\showISSN{1748-0221}
\urldef\tempurl%
\url{https://doi.org/10.1088/1748-0221/16/07/P07006}
\showDOI{\tempurl}


\bibitem[\protect\citeauthoryear{Abrahamse, Hadnagy, and Al-Ars}{Abrahamse et~al\mbox{.}}{2022}]%
        {Abrahamse2022Memory-DisaggregatedApplications}
\bibfield{author}{\bibinfo{person}{Robin Abrahamse}, \bibinfo{person}{Akos Hadnagy}, {and} \bibinfo{person}{Zaid Al-Ars}.} \bibinfo{year}{2022}\natexlab{}.
\newblock \showarticletitle{{Memory-Disaggregated In-Memory Object Store Framework for Big Data Applications}}.
\newblock \bibinfo{journal}{\emph{Proceedings - 2022 IEEE 36th International Parallel and Distributed Processing Symposium Workshops, IPDPSW 2022}} (\bibinfo{year}{2022}), \bibinfo{pages}{1228--1234}.
\newblock
\showISBNx{9781665497473}
\urldef\tempurl%
\url{https://doi.org/10.1109/IPDPSW55747.2022.00211}
\showDOI{\tempurl}


\bibitem[\protect\citeauthoryear{Ahmad, Al-Ars, and Hofstee}{Ahmad et~al\mbox{.}}{2022}]%
        {Ahmad2022BenchmarkingMicroservices}
\bibfield{author}{\bibinfo{person}{Tanveer Ahmad}, \bibinfo{person}{Zaid Al-Ars}, {and} \bibinfo{person}{H.~Peter Hofstee}.} \bibinfo{year}{2022}\natexlab{}.
\newblock \showarticletitle{{Benchmarking Apache Arrow Flight - A wire-speed protocol for data transfer, querying and microservices}}.
\newblock \bibinfo{journal}{\emph{Proceedings of the 4th Benchmarking in the Datacenter: Expanding to the Cloud 2022, BID 2022}}  \bibinfo{volume}{22} (\bibinfo{date}{4} \bibinfo{year}{2022}).
\newblock
\showISBNx{9781450393249}
\urldef\tempurl%
\url{https://doi.org/10.1145/3527199.3527264}
\showDOI{\tempurl}


\bibitem[\protect\citeauthoryear{Aiello, Albert, Garre, Aly, Ameli, et~al\mbox{.}}{Aiello et~al\mbox{.}}{2021}]%
        {Aiello2021ArchitectureFirmware}
\bibfield{author}{\bibinfo{person}{Sebastiano Aiello}, \bibinfo{person}{Arnauld Albert}, \bibinfo{person}{Sergio~Alves Garre}, \bibinfo{person}{Zineb Aly}, \bibinfo{person}{Fabrizio Ameli}, {et~al\mbox{.}}} \bibinfo{year}{2021}\natexlab{}.
\newblock \showarticletitle{{Architecture and performance of the KM3NeT front-end firmware}}.
\newblock \bibinfo{journal}{\emph{https://doi.org/10.1117/1.JATIS.7.1.016001}} \bibinfo{volume}{7}, \bibinfo{number}{1} (\bibinfo{date}{1} \bibinfo{year}{2021}), \bibinfo{pages}{016001}.
\newblock
\showISSN{2329-4124}
\urldef\tempurl%
\url{https://doi.org/10.1117/1.JATIS.7.1.016001}
\showDOI{\tempurl}


\bibitem[\protect\citeauthoryear{Al-Ars, Petri-Koenig, Hoozemans, Dierick, and Hofstee}{Al-Ars et~al\mbox{.}}{2023}]%
        {10.1145/3624062.3624541}
\bibfield{author}{\bibinfo{person}{Zaid Al-Ars}, \bibinfo{person}{Jakoba Petri-Koenig}, \bibinfo{person}{Joost Hoozemans}, \bibinfo{person}{Luc Dierick}, {and} \bibinfo{person}{H.~Peter Hofstee}.} \bibinfo{year}{2023}\natexlab{}.
\newblock \showarticletitle{OctoRay: Framework for Scalable FPGA Cluster Acceleration of Python Big Data Applications}. In \bibinfo{booktitle}{\emph{Proceedings of the SC '23 Workshops of The International Conference on High Performance Computing, Network, Storage, and Analysis}} \emph{(\bibinfo{series}{SC-W '23})}. \bibinfo{publisher}{Association for Computing Machinery}, \bibinfo{address}{New York, NY, USA}, \bibinfo{pages}{539–546}.
\newblock
\showISBNx{9798400707858}
\urldef\tempurl%
\url{https://doi.org/10.1145/3624062.3624541}
\showDOI{\tempurl}


\bibitem[\protect\citeauthoryear{Alachiotis, Brokalakis, Amourgianos, Ioannidis, Malakonakis, et~al\mbox{.}}{Alachiotis et~al\mbox{.}}{2021a}]%
        {Alachiotis2021AcceleratingCloud}
\bibfield{author}{\bibinfo{person}{Nikolaos Alachiotis}, \bibinfo{person}{Andreas Brokalakis}, \bibinfo{person}{Vasilis Amourgianos}, \bibinfo{person}{Sotiris Ioannidis}, \bibinfo{person}{Pavlos Malakonakis}, {et~al\mbox{.}}} \bibinfo{year}{2021}\natexlab{a}.
\newblock \showarticletitle{{Accelerating Phylogenetics Using FPGAs in the Cloud}}.
\newblock \bibinfo{journal}{\emph{IEEE Micro}} \bibinfo{volume}{41}, \bibinfo{number}{4} (\bibinfo{date}{7} \bibinfo{year}{2021}), \bibinfo{pages}{24--30}.
\newblock
\showISSN{19374143}
\urldef\tempurl%
\url{https://doi.org/10.1109/MM.2021.3075848}
\showDOI{\tempurl}


\bibitem[\protect\citeauthoryear{Alachiotis, Skrimponis, Pissadakis, and Pnevmatikatos}{Alachiotis et~al\mbox{.}}{2021b}]%
        {Alachiotis2021ScalableProcessing}
\bibfield{author}{\bibinfo{person}{Nikolaos Alachiotis}, \bibinfo{person}{Panagiotis Skrimponis}, \bibinfo{person}{Manolis Pissadakis}, {and} \bibinfo{person}{Dionisios Pnevmatikatos}.} \bibinfo{year}{2021}\natexlab{b}.
\newblock \showarticletitle{{Scalable Phylogeny Reconstruction with Disaggregated Near-memory Processing}}.
\newblock \bibinfo{journal}{\emph{ACM Transactions on Reconfigurable Technology and Systems (TRETS)}} \bibinfo{volume}{15}, \bibinfo{number}{3} (\bibinfo{date}{12} \bibinfo{year}{2021}).
\newblock
\showISSN{19367414}
\urldef\tempurl%
\url{https://doi.org/10.1145/3484983}
\showDOI{\tempurl}


\bibitem[\protect\citeauthoryear{Alachiotis, Stamatakis, and Pavlidis}{Alachiotis et~al\mbox{.}}{2012}]%
        {alachiotis2012omegaplus}
\bibfield{author}{\bibinfo{person}{Nikolaos Alachiotis}, \bibinfo{person}{Alexandros Stamatakis}, {and} \bibinfo{person}{Pavlos Pavlidis}.} \bibinfo{year}{2012}\natexlab{}.
\newblock \showarticletitle{OmegaPlus: a scalable tool for rapid detection of selective sweeps in whole-genome datasets}.
\newblock \bibinfo{journal}{\emph{Bioinformatics}} \bibinfo{volume}{28}, \bibinfo{number}{17} (\bibinfo{year}{2012}), \bibinfo{pages}{2274--2275}.
\newblock


\bibitem[\protect\citeauthoryear{Alonso, Petrica, Ruiz, Petri-Koenig, Umuroglu, et~al\mbox{.}}{Alonso et~al\mbox{.}}{2021}]%
        {Alonso2021Elastic-DF:Partitioning}
\bibfield{author}{\bibinfo{person}{Tobias Alonso}, \bibinfo{person}{Lucian Petrica}, \bibinfo{person}{Mario Ruiz}, \bibinfo{person}{Jakoba Petri-Koenig}, \bibinfo{person}{Yaman Umuroglu}, {et~al\mbox{.}}} \bibinfo{year}{2021}\natexlab{}.
\newblock \showarticletitle{{Elastic-DF: Scaling Performance of DNN Inference in FPGA Clouds through Automatic Partitioning}}.
\newblock \bibinfo{journal}{\emph{ACM Transactions on Reconfigurable Technology and Systems (TRETS)}} \bibinfo{volume}{15}, \bibinfo{number}{2} (\bibinfo{date}{12} \bibinfo{year}{2021}).
\newblock
\showISSN{19367414}
\urldef\tempurl%
\url{https://doi.org/10.1145/3470567}
\showDOI{\tempurl}


\bibitem[\protect\citeauthoryear{Anders, Andreu, Becker, Becker, Cantoro, et~al\mbox{.}}{Anders et~al\mbox{.}}{2023}]%
        {Anders2023AProcessors}
\bibfield{author}{\bibinfo{person}{Jens Anders}, \bibinfo{person}{Pablo Andreu}, \bibinfo{person}{Bernd Becker}, \bibinfo{person}{Steffen Becker}, \bibinfo{person}{Riccardo Cantoro}, {et~al\mbox{.}}} \bibinfo{year}{2023}\natexlab{}.
\newblock \showarticletitle{{A Survey of Recent Developments in Testability, Safety and Security of RISC-V Processors}}.
\newblock \bibinfo{journal}{\emph{Proceedings of the European Test Workshop}}  \bibinfo{volume}{2023-May} (\bibinfo{year}{2023}).
\newblock
\showISBNx{9798350336344}
\showISSN{15581780}
\urldef\tempurl%
\url{https://doi.org/10.1109/ETS56758.2023.10174099}
\showDOI{\tempurl}


\bibitem[\protect\citeauthoryear{Arikan, Palumbo, Cassano, Reviriego, Pontarelli, et~al\mbox{.}}{Arikan et~al\mbox{.}}{2022}]%
        {Arikan2022ProcessorSketches}
\bibfield{author}{\bibinfo{person}{Kerem Arikan}, \bibinfo{person}{Alessandro Palumbo}, \bibinfo{person}{Luca Cassano}, \bibinfo{person}{Pedro Reviriego}, \bibinfo{person}{Salvatore Pontarelli}, {et~al\mbox{.}}} \bibinfo{year}{2022}\natexlab{}.
\newblock \showarticletitle{{Processor Security: Detecting Microarchitectural Attacks via Count-Min Sketches}}.
\newblock \bibinfo{journal}{\emph{IEEE Transactions on Very Large Scale Integration (VLSI) Systems}} \bibinfo{volume}{30}, \bibinfo{number}{7} (\bibinfo{date}{7} \bibinfo{year}{2022}), \bibinfo{pages}{938--951}.
\newblock
\showISSN{15579999}
\urldef\tempurl%
\url{https://doi.org/10.1109/TVLSI.2022.3171810}
\showDOI{\tempurl}


\bibitem[\protect\citeauthoryear{Arute, Arya, Babbush, Bacon, Bardin, et~al\mbox{.}}{Arute et~al\mbox{.}}{2019}]%
        {Arute-nat-2019}
\bibfield{author}{\bibinfo{person}{Frank Arute}, \bibinfo{person}{Kunal Arya}, \bibinfo{person}{Ryan Babbush}, \bibinfo{person}{Dave Bacon}, \bibinfo{person}{Joseph~C. Bardin}, {et~al\mbox{.}}} \bibinfo{year}{2019}\natexlab{}.
\newblock \showarticletitle{Quantum supremacy using a programmable superconducting processor}.
\newblock \bibinfo{journal}{\emph{Nature}}  \bibinfo{volume}{574} (\bibinfo{year}{2019}), \bibinfo{pages}{505--510}.
\newblock


\bibitem[\protect\citeauthoryear{Ashraf, Khammassi, Taouil, and Bertels}{Ashraf et~al\mbox{.}}{2018}]%
        {Ashraf2018MemoryPlatforms}
\bibfield{author}{\bibinfo{person}{Imran Ashraf}, \bibinfo{person}{Nader Khammassi}, \bibinfo{person}{Mottaqiallah Taouil}, {and} \bibinfo{person}{Koen Bertels}.} \bibinfo{year}{2018}\natexlab{}.
\newblock \showarticletitle{{Memory and Communication Profiling for Accelerator-Based Platforms}}.
\newblock \bibinfo{journal}{\emph{IEEE Trans. Comput.}} \bibinfo{volume}{67}, \bibinfo{number}{7} (\bibinfo{date}{7} \bibinfo{year}{2018}), \bibinfo{pages}{934--948}.
\newblock
\showISSN{15579956}
\urldef\tempurl%
\url{https://doi.org/10.1109/TC.2017.2785225}
\showDOI{\tempurl}


\bibitem[\protect\citeauthoryear{Ashton, Backes, Balzer, Berge, Bolmont, et~al\mbox{.}}{Ashton et~al\mbox{.}}{2020}]%
        {Ashton2020ATelescopes}
\bibfield{author}{\bibinfo{person}{T. Ashton}, \bibinfo{person}{M. Backes}, \bibinfo{person}{A. Balzer}, \bibinfo{person}{D. Berge}, \bibinfo{person}{J. Bolmont}, {et~al\mbox{.}}} \bibinfo{year}{2020}\natexlab{}.
\newblock \showarticletitle{{A NECTAr-based upgrade for the Cherenkov cameras of the H.E.S.S. 12-meter telescopes}}.
\newblock \bibinfo{journal}{\emph{Astroparticle Physics}}  \bibinfo{volume}{118} (\bibinfo{date}{3} \bibinfo{year}{2020}), \bibinfo{pages}{102425}.
\newblock
\showISSN{0927-6505}
\urldef\tempurl%
\url{https://doi.org/10.1016/J.ASTROPARTPHYS.2019.102425}
\showDOI{\tempurl}


\bibitem[\protect\citeauthoryear{Baaij, Kooijman, Kuper, Boeijink, and Gerards}{Baaij et~al\mbox{.}}{2010}]%
        {clash-2010}
\bibfield{author}{\bibinfo{person}{C.P.R. Baaij}, \bibinfo{person}{Matthijs Kooijman}, \bibinfo{person}{Jan Kuper}, \bibinfo{person}{W.A. Boeijink}, {and} \bibinfo{person}{M.E.T. Gerards}.} \bibinfo{year}{2010}\natexlab{}.
\newblock \showarticletitle{ClaSH: Structural Descriptions of Synchronous Hardware using Haskell}. In \bibinfo{booktitle}{\emph{Proceedings of the 13th EUROMICRO Conference on Digital System Design: Architectures, Methods and Tools}}. \bibinfo{publisher}{IEEE}, \bibinfo{pages}{714--721}.
\newblock
\showISBNx{978-0-7695-4171-6}
\urldef\tempurl%
\url{https://doi.org/10.1109/DSD.2010.21}
\showDOI{\tempurl}


\bibitem[\protect\citeauthoryear{Bagley et~al\mbox{.}}{Bagley et~al\mbox{.}}{2009}]%
        {KM3NeT:2009xxi}
\bibfield{author}{\bibinfo{person}{P. Bagley} {et~al\mbox{.}}} \bibinfo{year}{2009}\natexlab{}.
\newblock \showarticletitle{{KM3NeT: Technical Design Report for a Deep-Sea Research Infrastructure in the Mediterranean Sea Incorporating a Very Large Volume Neutrino Telescope}}.
\newblock  (\bibinfo{year}{2009}).
\newblock


\bibitem[\protect\citeauthoryear{Bargholz, Rönninger, and Siegle}{Bargholz et~al\mbox{.}}{2022}]%
        {AnSection}
\bibfield{author}{\bibinfo{person}{Malte Bargholz}, \bibinfo{person}{Wolfgang Rönninger}, {and} \bibinfo{person}{Felix Siegle}.} \bibinfo{year}{2022}\natexlab{}.
\newblock \showarticletitle{An Optical SpaceFibre Testbed for Transceiver Evaluation and Validation of a Design-Time Configurable Router}. In \bibinfo{booktitle}{\emph{2022 International SpaceWire \& SpaceFibre Conference (ISC)}}. \bibinfo{pages}{01--08}.
\newblock


\bibitem[\protect\citeauthoryear{Barrios, Sanchez, Santos, and Sarmiento}{Barrios et~al\mbox{.}}{2020}]%
        {Barrios2020SHyLoCMissions}
\bibfield{author}{\bibinfo{person}{Yubal Barrios}, \bibinfo{person}{Antonio~J. Sanchez}, \bibinfo{person}{Lucana Santos}, {and} \bibinfo{person}{Roberto Sarmiento}.} \bibinfo{year}{2020}\natexlab{}.
\newblock \showarticletitle{{SHyLoC 2.0: A Versatile Hardware Solution for On-Board Data and Hyperspectral Image Compression on Future Space Missions}}.
\newblock \bibinfo{journal}{\emph{IEEE Access}}  \bibinfo{volume}{8} (\bibinfo{year}{2020}), \bibinfo{pages}{54269--54287}.
\newblock
\showISSN{21693536}
\urldef\tempurl%
\url{https://doi.org/10.1109/ACCESS.2020.2980767}
\showDOI{\tempurl}


\bibitem[\protect\citeauthoryear{Bhattacharya and Qin}{Bhattacharya and Qin}{2020}]%
        {green-data-centers}
\bibfield{author}{\bibinfo{person}{Tathagata Bhattacharya} {and} \bibinfo{person}{Xiao Qin}.} \bibinfo{year}{2020}\natexlab{}.
\newblock \showarticletitle{Modeling Energy Efficiency of Future Green Data centers}. In \bibinfo{booktitle}{\emph{2020 11th International Green and Sustainable Computing Workshops (IGSC)}}. \bibinfo{pages}{1--3}.
\newblock
\urldef\tempurl%
\url{https://doi.org/10.1109/IGSC51522.2020.9291049}
\showDOI{\tempurl}


\bibitem[\protect\citeauthoryear{Bielski, Syrigos, Katrinis, Syrivelis, Reale, et~al\mbox{.}}{Bielski et~al\mbox{.}}{2018}]%
        {Bielski2018DReDBox:Datacenter}
\bibfield{author}{\bibinfo{person}{M. Bielski}, \bibinfo{person}{I. Syrigos}, \bibinfo{person}{K. Katrinis}, \bibinfo{person}{D. Syrivelis}, \bibinfo{person}{A. Reale}, {et~al\mbox{.}}} \bibinfo{year}{2018}\natexlab{}.
\newblock \showarticletitle{{DReDBox: Materializing a full-stack rack-scale system prototype of a next-generation disaggregated datacenter}}.
\newblock \bibinfo{journal}{\emph{Proceedings of the 2018 Design, Automation and Test in Europe Conference and Exhibition, DATE 2018}}  \bibinfo{volume}{2018-January} (\bibinfo{date}{4} \bibinfo{year}{2018}), \bibinfo{pages}{1093--1098}.
\newblock
\showISBNx{9783981926316}
\urldef\tempurl%
\url{https://doi.org/10.23919/DATE.2018.8342174}
\showDOI{\tempurl}


\bibitem[\protect\citeauthoryear{Bobda, Mbongue, Chow, Ewais, Tarafdar, et~al\mbox{.}}{Bobda et~al\mbox{.}}{2022}]%
        {Bobda2022TheCloud}
\bibfield{author}{\bibinfo{person}{Christophe Bobda}, \bibinfo{person}{Joel~Mandebi Mbongue}, \bibinfo{person}{Paul Chow}, \bibinfo{person}{Mohammad Ewais}, \bibinfo{person}{Naif Tarafdar}, {et~al\mbox{.}}} \bibinfo{year}{2022}\natexlab{}.
\newblock \bibinfo{title}{{The Future of FPGA Acceleration in Datacenters and the Cloud}}.
\newblock
\newblock
\showISSN{19367414}
\urldef\tempurl%
\url{https://doi.org/10.1145/3506713}
\showDOI{\tempurl}


\bibitem[\protect\citeauthoryear{Bohmer, Forlin, Cazzaniga, Rech, Furano, et~al\mbox{.}}{Bohmer et~al\mbox{.}}{2023}]%
        {Bohmer2023NeutronFPGAs}
\bibfield{author}{\bibinfo{person}{Kevin Bohmer}, \bibinfo{person}{Bruno Forlin}, \bibinfo{person}{Carlo Cazzaniga}, \bibinfo{person}{Paolo Rech}, \bibinfo{person}{Gianluca Furano}, {et~al\mbox{.}}} \bibinfo{year}{2023}\natexlab{}.
\newblock \showarticletitle{{Neutron Radiation Tests of the NEORV32 RISC-V SoC on Flash-Based FPGAs}}.
\newblock \bibinfo{journal}{\emph{Proceedings - IEEE International Symposium on Defect and Fault Tolerance in VLSI and Nanotechnology Systems, DFT}} (\bibinfo{year}{2023}).
\newblock
\showISBNx{9798350315004}
\showISSN{2765933X}
\urldef\tempurl%
\url{https://doi.org/10.1109/DFT59622.2023.10313556}
\showDOI{\tempurl}


\bibitem[\protect\citeauthoryear{Broekema, Mol, Nijboer, van Amesfoort, Brentjens, et~al\mbox{.}}{Broekema et~al\mbox{.}}{2018}]%
        {Broekema_2018}
\bibfield{author}{\bibinfo{person}{P.~Chris Broekema}, \bibinfo{person}{J.~Jan~David Mol}, \bibinfo{person}{R. Nijboer}, \bibinfo{person}{A.S. van Amesfoort}, \bibinfo{person}{M.A. Brentjens}, {et~al\mbox{.}}} \bibinfo{year}{2018}\natexlab{}.
\newblock \showarticletitle{Cobalt: A GPU-based correlator and beamformer for LOFAR}.
\newblock \bibinfo{journal}{\emph{Astronomy and Computing}}  \bibinfo{volume}{23} (\bibinfo{date}{April} \bibinfo{year}{2018}), \bibinfo{pages}{180–192}.
\newblock
\showISSN{2213-1337}
\urldef\tempurl%
\url{https://doi.org/10.1016/j.ascom.2018.04.006}
\showDOI{\tempurl}


\bibitem[\protect\citeauthoryear{Caulfield, Chung, Putnam, Angepat, Fowers, et~al\mbox{.}}{Caulfield et~al\mbox{.}}{2016}]%
        {caulfield2016cloud}
\bibfield{author}{\bibinfo{person}{Adrian~M Caulfield}, \bibinfo{person}{Eric~S Chung}, \bibinfo{person}{Andrew Putnam}, \bibinfo{person}{Hari Angepat}, \bibinfo{person}{Jeremy Fowers}, {et~al\mbox{.}}} \bibinfo{year}{2016}\natexlab{}.
\newblock \showarticletitle{A cloud-scale acceleration architecture}. In \bibinfo{booktitle}{\emph{2016 49th Annual IEEE/ACM international symposium on microarchitecture (MICRO)}}. IEEE, \bibinfo{pages}{1--13}.
\newblock


\bibitem[\protect\citeauthoryear{Charitopoulos, Pnevmatikatos, and Gaydadjiev}{Charitopoulos et~al\mbox{.}}{2021}]%
        {Charitopoulos2021MC-DeF:Applications}
\bibfield{author}{\bibinfo{person}{George Charitopoulos}, \bibinfo{person}{Dionisios~N. Pnevmatikatos}, {and} \bibinfo{person}{Georgi Gaydadjiev}.} \bibinfo{year}{2021}\natexlab{}.
\newblock \showarticletitle{{MC-DeF: Creating Customized CGRAs for Dataflow Applications}}.
\newblock \bibinfo{journal}{\emph{ACM Transactions on Architecture and Code Optimization (TACO)}} \bibinfo{volume}{18}, \bibinfo{number}{3} (\bibinfo{date}{4} \bibinfo{year}{2021}).
\newblock
\showISSN{15443973}
\urldef\tempurl%
\url{https://doi.org/10.1145/3447970}
\showDOI{\tempurl}


\bibitem[\protect\citeauthoryear{Chen, Daverveldt, and Al-Ars}{Chen et~al\mbox{.}}{2021}]%
        {Chen2021FPGAAlgorithm}
\bibfield{author}{\bibinfo{person}{Jianyu Chen}, \bibinfo{person}{Maurice Daverveldt}, {and} \bibinfo{person}{Zaid Al-Ars}.} \bibinfo{year}{2021}\natexlab{}.
\newblock \showarticletitle{{FPGA Acceleration of Zstd Compression Algorithm}}.
\newblock \bibinfo{journal}{\emph{2021 IEEE International Parallel and Distributed Processing Symposium Workshops, IPDPSW 2021 - In conjunction with IEEE IPDPS 2021}} (\bibinfo{date}{6} \bibinfo{year}{2021}), \bibinfo{pages}{188--191}.
\newblock
\showISBNx{9781665435772}
\urldef\tempurl%
\url{https://doi.org/10.1109/IPDPSW52791.2021.00035}
\showDOI{\tempurl}


\bibitem[\protect\citeauthoryear{Choi, Zhang, Li, and Cong}{Choi et~al\mbox{.}}{2017}]%
        {hlscope-2017}
\bibfield{author}{\bibinfo{person}{Young-kyu Choi}, \bibinfo{person}{Peng Zhang}, \bibinfo{person}{Peng Li}, {and} \bibinfo{person}{Jason Cong}.} \bibinfo{year}{2017}\natexlab{}.
\newblock \showarticletitle{HLscope+: fast and accurate performance estimation for FPGA HLS}. In \bibinfo{booktitle}{\emph{Proceedings of the 36th International Conference on Computer-Aided Design}} \emph{(\bibinfo{series}{ICCAD '17})}. \bibinfo{publisher}{IEEE Press}, \bibinfo{pages}{691–698}.
\newblock


\bibitem[\protect\citeauthoryear{Chrysos, Papapetrou, Pnevmatikatos, Dollas, and Garofalakis}{Chrysos et~al\mbox{.}}{2019}]%
        {Chrysos2019DataNode}
\bibfield{author}{\bibinfo{person}{Grigorios Chrysos}, \bibinfo{person}{Odysseas Papapetrou}, \bibinfo{person}{Dionisios Pnevmatikatos}, \bibinfo{person}{Apostolos Dollas}, {and} \bibinfo{person}{Minos Garofalakis}.} \bibinfo{year}{2019}\natexlab{}.
\newblock \showarticletitle{{Data stream statistics over sliding windows: How to summarize 150 million updates per second on a single node}}.
\newblock \bibinfo{journal}{\emph{Proceedings - 29th International Conference on Field-Programmable Logic and Applications, FPL 2019}} (\bibinfo{date}{9} \bibinfo{year}{2019}), \bibinfo{pages}{278--285}.
\newblock
\showISBNx{9781728148847}
\urldef\tempurl%
\url{https://doi.org/10.1109/FPL.2019.00052}
\showDOI{\tempurl}


\bibitem[\protect\citeauthoryear{Cong, Lau, Liu, Neuendorffer, Pan, et~al\mbox{.}}{Cong et~al\mbox{.}}{2022}]%
        {Cong-2022}
\bibfield{author}{\bibinfo{person}{Jason Cong}, \bibinfo{person}{Jason Lau}, \bibinfo{person}{Gai Liu}, \bibinfo{person}{Stephen Neuendorffer}, \bibinfo{person}{Peichen Pan}, {et~al\mbox{.}}} \bibinfo{year}{2022}\natexlab{}.
\newblock \showarticletitle{FPGA HLS Today: Successes, Challenges, and Opportunities}.
\newblock \bibinfo{journal}{\emph{ACM Trans. Reconfigurable Technol. Syst.}} \bibinfo{volume}{15}, \bibinfo{number}{4}, Article \bibinfo{articleno}{51} (\bibinfo{date}{aug} \bibinfo{year}{2022}), \bibinfo{numpages}{42}~pages.
\newblock
\showISSN{1936-7406}
\urldef\tempurl%
\url{https://doi.org/10.1145/3530775}
\showDOI{\tempurl}


\bibitem[\protect\citeauthoryear{Corda, Veenboer, Awan, Kumar, Jordans, et~al\mbox{.}}{Corda et~al\mbox{.}}{2020}]%
        {Corda2020NearImaging}
\bibfield{author}{\bibinfo{person}{Stefano Corda}, \bibinfo{person}{Bram Veenboer}, \bibinfo{person}{Ahsan~Javed Awan}, \bibinfo{person}{Akash Kumar}, \bibinfo{person}{Roel Jordans}, {et~al\mbox{.}}} \bibinfo{year}{2020}\natexlab{}.
\newblock \showarticletitle{{Near Memory Acceleration on High Resolution Radio Astronomy Imaging}}.
\newblock \bibinfo{journal}{\emph{2020 9th Mediterranean Conference on Embedded Computing, MECO 2020}} (\bibinfo{date}{6} \bibinfo{year}{2020}).
\newblock
\showISBNx{9781728169477}
\urldef\tempurl%
\url{https://doi.org/10.1109/MECO49872.2020.9134089}
\showDOI{\tempurl}


\bibitem[\protect\citeauthoryear{Corda, Veenboer, Awan, Romein, Jordans, et~al\mbox{.}}{Corda et~al\mbox{.}}{2022}]%
        {Corda2022Reduced-PrecisionHardware}
\bibfield{author}{\bibinfo{person}{Stefano Corda}, \bibinfo{person}{Bram Veenboer}, \bibinfo{person}{Ahsan~Javed Awan}, \bibinfo{person}{John~W. Romein}, \bibinfo{person}{Roel Jordans}, {et~al\mbox{.}}} \bibinfo{year}{2022}\natexlab{}.
\newblock \showarticletitle{{Reduced-Precision Acceleration of Radio-Astronomical Imaging on Reconfigurable Hardware}}.
\newblock \bibinfo{journal}{\emph{IEEE Access}}  \bibinfo{volume}{10} (\bibinfo{year}{2022}), \bibinfo{pages}{22819--22843}.
\newblock
\showISSN{21693536}
\urldef\tempurl%
\url{https://doi.org/10.1109/ACCESS.2022.3150861}
\showDOI{\tempurl}


\bibitem[\protect\citeauthoryear{Corradi, Adriaans, and Stuijk}{Corradi et~al\mbox{.}}{2021}]%
        {Corradi2021Gyro:Analytics}
\bibfield{author}{\bibinfo{person}{Federico Corradi}, \bibinfo{person}{Guido Adriaans}, {and} \bibinfo{person}{Sander Stuijk}.} \bibinfo{year}{2021}\natexlab{}.
\newblock \showarticletitle{{Gyro: A digital spiking neural network architecture for multi-sensory data analytics}}.
\newblock \bibinfo{journal}{\emph{ACM International Conference Proceeding Series}} (\bibinfo{date}{1} \bibinfo{year}{2021}), \bibinfo{pages}{9--15}.
\newblock
\showISBNx{9781450389525}
\urldef\tempurl%
\url{https://doi.org/10.1145/3444950.3444951}
\showDOI{\tempurl}


\bibitem[\protect\citeauthoryear{Corradi, Shen, Zhao, and Alachiotis}{Corradi et~al\mbox{.}}{2024}]%
        {corradi2024accelerated}
\bibfield{author}{\bibinfo{person}{Federico Corradi}, \bibinfo{person}{Zhanbo Shen}, \bibinfo{person}{Hanqing Zhao}, {and} \bibinfo{person}{Nikolaos Alachiotis}.} \bibinfo{year}{2024}\natexlab{}.
\newblock \showarticletitle{Accelerated Spiking Convolutional Neural Networks for Scalable Population Genomics}. In \bibinfo{booktitle}{\emph{Proceedings of the 14th International Symposium on Highly Efficient Accelerators and Reconfigurable Technologies}}. \bibinfo{pages}{53--62}.
\newblock


\bibitem[\protect\citeauthoryear{Corts and Alachiotis}{Corts and Alachiotis}{2023}]%
        {Corts2023AGenetics}
\bibfield{author}{\bibinfo{person}{Reinout Corts} {and} \bibinfo{person}{Nikolaos Alachiotis}.} \bibinfo{year}{2023}\natexlab{}.
\newblock \showarticletitle{{A Survey of Processing Systems for Phylogenetics and Population Genetics}}.
\newblock \bibinfo{journal}{\emph{ACM Transactions on Reconfigurable Technology and Systems}} \bibinfo{volume}{16}, \bibinfo{number}{3} (\bibinfo{date}{6} \bibinfo{year}{2023}).
\newblock
\showISSN{19367414}
\urldef\tempurl%
\url{https://doi.org/10.1145/3588033}
\showDOI{\tempurl}


\bibitem[\protect\citeauthoryear{Corts, Sterenborg, and Alachiotis}{Corts et~al\mbox{.}}{2022}]%
        {Corts2022AcceleratedFPGAs}
\bibfield{author}{\bibinfo{person}{Reinout Corts}, \bibinfo{person}{Niek Sterenborg}, {and} \bibinfo{person}{Nikolaos Alachiotis}.} \bibinfo{year}{2022}\natexlab{}.
\newblock \showarticletitle{{Accelerated LD-based selective sweep detection using GPUs and FPGAs}}.
\newblock \bibinfo{journal}{\emph{Proceedings - 2022 IEEE 36th International Parallel and Distributed Processing Symposium Workshops, IPDPSW 2022}} (\bibinfo{year}{2022}), \bibinfo{pages}{196--205}.
\newblock
\showISBNx{9781665497473}
\urldef\tempurl%
\url{https://doi.org/10.1109/IPDPSW55747.2022.00044}
\showDOI{\tempurl}


\bibitem[\protect\citeauthoryear{Cromjongh, Tian, Hofstee, and Al-Ars}{Cromjongh et~al\mbox{.}}{2023}]%
        {10305451}
\bibfield{author}{\bibinfo{person}{Casper Cromjongh}, \bibinfo{person}{Yongding Tian}, \bibinfo{person}{Peter Hofstee}, {and} \bibinfo{person}{Zaid Al-Ars}.} \bibinfo{year}{2023}\natexlab{}.
\newblock \showarticletitle{Tydi-Chisel: Collaborative and Interface-Driven Data-Streaming Accelerators}. In \bibinfo{booktitle}{\emph{2023 IEEE Nordic Circuits and Systems Conference (NorCAS)}}. \bibinfo{pages}{1--7}.
\newblock
\urldef\tempurl%
\url{https://doi.org/10.1109/NorCAS58970.2023.10305451}
\showDOI{\tempurl}


\bibitem[\protect\citeauthoryear{Curzel, Fiorito, Cueva, Jorge, Tsiodras, et~al\mbox{.}}{Curzel et~al\mbox{.}}{2023}]%
        {Curzel2023ExplorationApplications}
\bibfield{author}{\bibinfo{person}{Serena Curzel}, \bibinfo{person}{Michele Fiorito}, \bibinfo{person}{Patricia~Lopez Cueva}, \bibinfo{person}{Tiago Jorge}, \bibinfo{person}{Thanassis Tsiodras}, {et~al\mbox{.}}} \bibinfo{year}{2023}\natexlab{}.
\newblock \showarticletitle{{Exploration of Synthesis Methods from Simulink Models to FPGA for Aerospace Applications}}.
\newblock \bibinfo{journal}{\emph{Proceedings of the 20th ACM International Conference on Computing Frontiers 2023, CF 2023}}  \bibinfo{volume}{7} (\bibinfo{date}{5} \bibinfo{year}{2023}), \bibinfo{pages}{243--249}.
\newblock
\showISBNx{9798400701405}
\urldef\tempurl%
\url{https://doi.org/10.1145/3587135.3592766}
\showDOI{\tempurl}


\bibitem[\protect\citeauthoryear{De, Shafique, and Corporaal}{De et~al\mbox{.}}{2023}]%
        {De2023hls}
\bibfield{author}{\bibinfo{person}{Sayandip De}, \bibinfo{person}{Muhammad Shafique}, {and} \bibinfo{person}{Henk Corporaal}.} \bibinfo{year}{2023}\natexlab{}.
\newblock \showarticletitle{Delay Prediction for ASIC HLS: Comparing Graph-Based and Nongraph-Based Learning Models}.
\newblock \bibinfo{journal}{\emph{IEEE Transactions on Computer-Aided Design of Integrated Circuits and Systems}}  \bibinfo{volume}{42}, \bibinfo{pages}{1133--1146}.
\newblock
Issue 4.
\showISSN{19374151}
\urldef\tempurl%
\url{https://doi.org/10.1109/TCAD.2022.3197977}
\showDOI{\tempurl}


\bibitem[\protect\citeauthoryear{de~Bruin, Vadivel, Wijtvliet, J{\"a}{\"a}skel{\"a}inen, and Corporaal}{de~Bruin et~al\mbox{.}}{2024}]%
        {debruin2024rblocks}
\bibfield{author}{\bibinfo{person}{Barry de Bruin}, \bibinfo{person}{Kanishkan Vadivel}, \bibinfo{person}{Mark Wijtvliet}, \bibinfo{person}{Pekka J{\"a}{\"a}skel{\"a}inen}, {and} \bibinfo{person}{Henk Corporaal}.} \bibinfo{year}{2024}\natexlab{}.
\newblock \showarticletitle{R-Blocks: an Energy-Efficient, Flexible, and Programmable CGRA}.
\newblock \bibinfo{journal}{\emph{ACM Transactions on Reconfigurable Technology and Systems}} (\bibinfo{year}{2024}).
\newblock


\bibitem[\protect\citeauthoryear{De~Sio, Azimi, Sterpone, Codinachs, and Decuzzi}{De~Sio et~al\mbox{.}}{2023}]%
        {DeSio2023PyXEL:FPGAs}
\bibfield{author}{\bibinfo{person}{Corrado De~Sio}, \bibinfo{person}{Sarah Azimi}, \bibinfo{person}{Luca Sterpone}, \bibinfo{person}{David~Merodio Codinachs}, {and} \bibinfo{person}{Filomena Decuzzi}.} \bibinfo{year}{2023}\natexlab{}.
\newblock \showarticletitle{{PyXEL: Exploring Bitstream Analysis to Assess and Enhance the Robustness of Designs on FPGAs}}.
\newblock \bibinfo{journal}{\emph{Proceedings - 2023 19th International Conference on Synthesis, Modeling, Analysis and Simulation Methods, and Applications to Circuit Design, SMACD 2023}} (\bibinfo{year}{2023}).
\newblock
\showISBNx{9798350332650}
\urldef\tempurl%
\url{https://doi.org/10.1109/SMACD58065.2023.10192116}
\showDOI{\tempurl}


\bibitem[\protect\citeauthoryear{Deng, Zhang, Pan, Gao, Mo, et~al\mbox{.}}{Deng et~al\mbox{.}}{2022}]%
        {Deng2022Physics-BasedImplementation}
\bibfield{author}{\bibinfo{person}{Xiong Deng}, \bibinfo{person}{Miao Zhang}, \bibinfo{person}{Wei Pan}, \bibinfo{person}{Ziqiang Gao}, \bibinfo{person}{Jundao Mo}, {et~al\mbox{.}}} \bibinfo{year}{2022}\natexlab{}.
\newblock \showarticletitle{{Physics-Based LED Modeling and Nonlinear Distortion Mitigating With Real-Time Implementation}}.
\newblock \bibinfo{journal}{\emph{IEEE Photonics Journal}} \bibinfo{volume}{14}, \bibinfo{number}{6} (\bibinfo{date}{12} \bibinfo{year}{2022}).
\newblock
\showISSN{19430655}
\urldef\tempurl%
\url{https://doi.org/10.1109/JPHOT.2022.3210021}
\showDOI{\tempurl}


\bibitem[\protect\citeauthoryear{Diamantopoulos, Ringlein, Purandare, Singh, and Hagleitner}{Diamantopoulos et~al\mbox{.}}{2020}]%
        {ml-agile-tuned-tpu}
\bibfield{author}{\bibinfo{person}{Dionysios Diamantopoulos}, \bibinfo{person}{Burkhard Ringlein}, \bibinfo{person}{Mitra Purandare}, \bibinfo{person}{Gagandeep Singh}, {and} \bibinfo{person}{Christoph Hagleitner}.} \bibinfo{year}{2020}\natexlab{}.
\newblock \showarticletitle{Agile Autotuning of a Transprecision Tensor Accelerator Overlay for TVM Compiler Stack}. In \bibinfo{booktitle}{\emph{2020 30th International Conference on Field-Programmable Logic and Applications (FPL)}}. \bibinfo{pages}{310--316}.
\newblock
\showISSN{1946-1488}
\urldef\tempurl%
\url{https://doi.org/10.1109/FPL50879.2020.00058}
\showDOI{\tempurl}


\bibitem[\protect\citeauthoryear{Du, Sterpone, Azimi, Codinachs, Ferlet-Cavrois, et~al\mbox{.}}{Du et~al\mbox{.}}{2019}]%
        {Du2019UltrahighFPGA}
\bibfield{author}{\bibinfo{person}{Boyang Du}, \bibinfo{person}{Luca Sterpone}, \bibinfo{person}{Sarah Azimi}, \bibinfo{person}{David~Merodio Codinachs}, \bibinfo{person}{Veronique Ferlet-Cavrois}, {et~al\mbox{.}}} \bibinfo{year}{2019}\natexlab{}.
\newblock \showarticletitle{{Ultrahigh Energy Heavy Ion Test Beam on Xilinx Kintex-7 SRAM-Based FPGA}}.
\newblock \bibinfo{journal}{\emph{IEEE Transactions on Nuclear Science}} \bibinfo{volume}{66}, \bibinfo{number}{7} (\bibinfo{date}{7} \bibinfo{year}{2019}), \bibinfo{pages}{1813--1819}.
\newblock
\showISSN{15581578}
\urldef\tempurl%
\url{https://doi.org/10.1109/TNS.2019.2915207}
\showDOI{\tempurl}


\bibitem[\protect\citeauthoryear{Ernst}{Ernst}{2020}]%
        {ernst2020competing}
\bibfield{author}{\bibinfo{person}{Dieter Ernst}.} \bibinfo{year}{2020}\natexlab{}.
\newblock \showarticletitle{Competing in artificial intelligence chips: China’s challenge amid technology war}.
\newblock  (\bibinfo{year}{2020}).
\newblock


\bibitem[\protect\citeauthoryear{Fang, Chen, Lee, Al-Ars, and Hofstee}{Fang et~al\mbox{.}}{2019}]%
        {Fang2019AModel}
\bibfield{author}{\bibinfo{person}{Jian Fang}, \bibinfo{person}{Jianyu Chen}, \bibinfo{person}{Jinho Lee}, \bibinfo{person}{Zaid Al-Ars}, {and} \bibinfo{person}{H.~Peter Hofstee}.} \bibinfo{year}{2019}\natexlab{}.
\newblock \showarticletitle{{A Fine-grained parallel snappy decompressor for FPGAS using a relaxed execution model}}.
\newblock \bibinfo{journal}{\emph{Proceedings - 27th IEEE International Symposium on Field-Programmable Custom Computing Machines, FCCM 2019}} (\bibinfo{date}{4} \bibinfo{year}{2019}), \bibinfo{pages}{335}.
\newblock
\showISBNx{9781728111315}
\urldef\tempurl%
\url{https://doi.org/10.1109/FCCM.2019.00076}
\showDOI{\tempurl}


\bibitem[\protect\citeauthoryear{Fang, Chen, Lee, Al-Ars, and Hofstee}{Fang et~al\mbox{.}}{2020a}]%
        {Fang2020AnLogic}
\bibfield{author}{\bibinfo{person}{Jian Fang}, \bibinfo{person}{Jianyu Chen}, \bibinfo{person}{Jinho Lee}, \bibinfo{person}{Zaid Al-Ars}, {and} \bibinfo{person}{H.~Peter Hofstee}.} \bibinfo{year}{2020}\natexlab{a}.
\newblock \showarticletitle{{An Efficient High-Throughput LZ77-Based Decompressor in Reconfigurable Logic}}.
\newblock \bibinfo{journal}{\emph{Journal of Signal Processing Systems}} \bibinfo{volume}{92}, \bibinfo{number}{9} (\bibinfo{date}{9} \bibinfo{year}{2020}), \bibinfo{pages}{931--947}.
\newblock
\showISSN{19398115}
\urldef\tempurl%
\url{https://doi.org/10.1007/S11265-020-01547-W/FIGURES/12}
\showDOI{\tempurl}


\bibitem[\protect\citeauthoryear{Fang, Mulder, Hidders, Lee, and Hofstee}{Fang et~al\mbox{.}}{2020b}]%
        {Fang2020In-memorySurvey}
\bibfield{author}{\bibinfo{person}{Jian Fang}, \bibinfo{person}{Yvo~T.B. Mulder}, \bibinfo{person}{Jan Hidders}, \bibinfo{person}{Jinho Lee}, {and} \bibinfo{person}{H.~Peter Hofstee}.} \bibinfo{year}{2020}\natexlab{b}.
\newblock \showarticletitle{{In-memory database acceleration on FPGAs: a survey}}.
\newblock \bibinfo{journal}{\emph{VLDB Journal}} \bibinfo{volume}{29}, \bibinfo{number}{1} (\bibinfo{date}{1} \bibinfo{year}{2020}), \bibinfo{pages}{33--59}.
\newblock
\showISSN{0949877X}
\urldef\tempurl%
\url{https://doi.org/10.1007/S00778-019-00581-W/FIGURES/9}
\showDOI{\tempurl}


\bibitem[\protect\citeauthoryear{Fernandez~Prieto, Regueiro, Hennessy, Buytaert, Van~Beuzekom, et~al\mbox{.}}{Fernandez~Prieto et~al\mbox{.}}{2020}]%
        {FernandezPrieto2020PhaseExperiment}
\bibfield{author}{\bibinfo{person}{Antonio Fernandez~Prieto}, \bibinfo{person}{Pablo~Vazquez Regueiro}, \bibinfo{person}{Karol Hennessy}, \bibinfo{person}{Jan Buytaert}, \bibinfo{person}{Martin Van~Beuzekom}, {et~al\mbox{.}}} \bibinfo{year}{2020}\natexlab{}.
\newblock \showarticletitle{{Phase I Upgrade of the Readout System of the Vertex Detector at the LHCb Experiment}}.
\newblock \bibinfo{journal}{\emph{IEEE Transactions on Nuclear Science}} \bibinfo{volume}{67}, \bibinfo{number}{4} (\bibinfo{date}{4} \bibinfo{year}{2020}), \bibinfo{pages}{732--739}.
\newblock
\showISSN{15581578}
\urldef\tempurl%
\url{https://doi.org/10.1109/TNS.2020.2970534}
\showDOI{\tempurl}


\bibitem[\protect\citeauthoryear{Firestone, Putnam, Mundkur, Chiou, Dabagh, et~al\mbox{.}}{Firestone et~al\mbox{.}}{2018}]%
        {firestone2018azure}
\bibfield{author}{\bibinfo{person}{Daniel Firestone}, \bibinfo{person}{Andrew Putnam}, \bibinfo{person}{Sambhrama Mundkur}, \bibinfo{person}{Derek Chiou}, \bibinfo{person}{Alireza Dabagh}, {et~al\mbox{.}}} \bibinfo{year}{2018}\natexlab{}.
\newblock \showarticletitle{Azure accelerated networking:$\{$SmartNICs$\}$ in the public cloud}. In \bibinfo{booktitle}{\emph{15th USENIX Symposium on Networked Systems Design and Implementation (NSDI 18)}}. \bibinfo{pages}{51--66}.
\newblock


\bibitem[\protect\citeauthoryear{Flich, Agosta, Ampletzer, Alonso, Brandolese, et~al\mbox{.}}{Flich et~al\mbox{.}}{2018}]%
        {Flich2018ExploringApproach}
\bibfield{author}{\bibinfo{person}{José Flich}, \bibinfo{person}{Giovanni Agosta}, \bibinfo{person}{Philipp Ampletzer}, \bibinfo{person}{David~Atienza Alonso}, \bibinfo{person}{Carlo Brandolese}, {et~al\mbox{.}}} \bibinfo{year}{2018}\natexlab{}.
\newblock \showarticletitle{{Exploring manycore architectures for next-generation HPC systems through the MANGO approach}}.
\newblock \bibinfo{journal}{\emph{Microprocessors and Microsystems}}  \bibinfo{volume}{61} (\bibinfo{date}{9} \bibinfo{year}{2018}), \bibinfo{pages}{154--170}.
\newblock
\showISSN{0141-9331}
\urldef\tempurl%
\url{https://doi.org/10.1016/J.MICPRO.2018.05.011}
\showDOI{\tempurl}


\bibitem[\protect\citeauthoryear{Forlin, Van~Huffelen, Cazzaniga, Rech, Alachiotis, et~al\mbox{.}}{Forlin et~al\mbox{.}}{2023}]%
        {Forlin2023AnSounds}
\bibfield{author}{\bibinfo{person}{Bruno~Endres Forlin}, \bibinfo{person}{Wouter Van~Huffelen}, \bibinfo{person}{Carlo Cazzaniga}, \bibinfo{person}{Paolo Rech}, \bibinfo{person}{Nikolaos Alachiotis}, {et~al\mbox{.}}} \bibinfo{year}{2023}\natexlab{}.
\newblock \showarticletitle{{An unprotected RISC-V Soft-core processor on an SRAM FPGA: Is it as bad as it sounds?}}
\newblock \bibinfo{journal}{\emph{Proceedings of the European Test Workshop}}  \bibinfo{volume}{2023-May} (\bibinfo{year}{2023}).
\newblock
\showISBNx{9798350336344}
\showISSN{15581780}
\urldef\tempurl%
\url{https://doi.org/10.1109/ETS56758.2023.10174076}
\showDOI{\tempurl}


\bibitem[\protect\citeauthoryear{Gambardella, Fraser, Zahid, Furano, and Blott}{Gambardella et~al\mbox{.}}{2022}]%
        {Gambardella2022AcceleratedTraining}
\bibfield{author}{\bibinfo{person}{Giulio Gambardella}, \bibinfo{person}{Nicholas~J. Fraser}, \bibinfo{person}{Ussama Zahid}, \bibinfo{person}{Gianluca Furano}, {and} \bibinfo{person}{Michaela Blott}.} \bibinfo{year}{2022}\natexlab{}.
\newblock \showarticletitle{{Accelerated Radiation Test on Quantized Neural Networks trained with Fault Aware Training}}.
\newblock \bibinfo{journal}{\emph{IEEE Aerospace Conference Proceedings}}  \bibinfo{volume}{2022-March} (\bibinfo{year}{2022}).
\newblock
\showISBNx{9781665437608}
\showISSN{1095323X}
\urldef\tempurl%
\url{https://doi.org/10.1109/AERO53065.2022.9843614}
\showDOI{\tempurl}


\bibitem[\protect\citeauthoryear{Garaffa, Aljuffri, Reinbrecht, Hamdioui, Taouil, et~al\mbox{.}}{Garaffa et~al\mbox{.}}{2021}]%
        {Garaffa2021RevealingNeuron}
\bibfield{author}{\bibinfo{person}{Luíza~C. Garaffa}, \bibinfo{person}{Abdullah Aljuffri}, \bibinfo{person}{Cezar Reinbrecht}, \bibinfo{person}{Said Hamdioui}, \bibinfo{person}{Mottaqiallah Taouil}, {et~al\mbox{.}}} \bibinfo{year}{2021}\natexlab{}.
\newblock \showarticletitle{{Revealing the secrets of spiking neural networks: The case of izhikevich neuron}}.
\newblock \bibinfo{journal}{\emph{Proceedings - 2021 24th Euromicro Conference on Digital System Design, DSD 2021}} (\bibinfo{year}{2021}), \bibinfo{pages}{514--518}.
\newblock
\showISBNx{9781665427036}
\urldef\tempurl%
\url{https://doi.org/10.1109/DSD53832.2021.00083}
\showDOI{\tempurl}


\bibitem[\protect\citeauthoryear{Gibney}{Gibney}{2014}]%
        {Gibney-nat-2014}
\bibfield{author}{\bibinfo{person}{Elizabeth Gibney}.} \bibinfo{year}{2014}\natexlab{}.
\newblock \showarticletitle{Physics: Quantum computer quest}.
\newblock \bibinfo{journal}{\emph{Nature}}  \bibinfo{volume}{516} (\bibinfo{year}{2014}), \bibinfo{pages}{24--26}.
\newblock


\bibitem[\protect\citeauthoryear{González, Nelissen, and Tovar}{González et~al\mbox{.}}{2022}]%
        {IPDeN}
\bibfield{author}{\bibinfo{person}{Yilian~Ribot González}, \bibinfo{person}{Geoffrey Nelissen}, {and} \bibinfo{person}{Eduardo Tovar}.} \bibinfo{year}{2022}\natexlab{}.
\newblock \showarticletitle{IPDeN: Real-Time deflection-based NoC with in-order flits delivery}. In \bibinfo{booktitle}{\emph{2022 IEEE 28th International Conference on Embedded and Real-Time Computing Systems and Applications (RTCSA)}}. \bibinfo{pages}{160--169}.
\newblock
\urldef\tempurl%
\url{https://doi.org/10.1109/RTCSA55878.2022.00023}
\showDOI{\tempurl}


\bibitem[\protect\citeauthoryear{Gottardo}{Gottardo}{2020}]%
        {Gottardo2020FEliXSystem}
\bibfield{author}{\bibinfo{person}{Carlo~Alberto Gottardo}.} \bibinfo{year}{2020}\natexlab{}.
\newblock \showarticletitle{{FEliX and SW Rod Commissioning of the New ATLAS Readout System}}.
\newblock \bibinfo{journal}{\emph{2020 IEEE Nuclear Science Symposium and Medical Imaging Conference, NSS/MIC 2020}} (\bibinfo{year}{2020}).
\newblock
\showISBNx{9781728176932}
\urldef\tempurl%
\url{https://doi.org/10.1109/NSS/MIC42677.2020.9507984}
\showDOI{\tempurl}


\bibitem[\protect\citeauthoryear{Groet, Hoozemans, Grapentin, Eberhardt, Al-Ars, et~al\mbox{.}}{Groet et~al\mbox{.}}{2024}]%
        {groet2024leveraging}
\bibfield{author}{\bibinfo{person}{Philip Groet}, \bibinfo{person}{Joost Hoozemans}, \bibinfo{person}{Andreas Grapentin}, \bibinfo{person}{Felix Eberhardt}, \bibinfo{person}{Zaid Al-Ars}, {et~al\mbox{.}}} \bibinfo{year}{2024}\natexlab{}.
\newblock \bibinfo{title}{Leveraging Apache Arrow for Zero-copy, Zero-serialization Cluster Shared Memory}.
\newblock
\newblock
\showeprint[arxiv]{cs.ET/2404.03030}


\bibitem[\protect\citeauthoryear{Group}{Group}{2015}]%
        {opencl}
\bibfield{author}{\bibinfo{person}{Khronos Group}.} \bibinfo{year}{2015}\natexlab{}.
\newblock \bibinfo{title}{The Open Standard for Parallel Programming of Heterogeneous Systems}.
\newblock \bibinfo{howpublished}{\url{https://www.khronos.org/opencl/}}.
\newblock


\bibitem[\protect\citeauthoryear{Guerrieri, Silva~Marques, Regazzoni, and Upegui}{Guerrieri et~al\mbox{.}}{2022}]%
        {Guerrieri2022OptimizingSynthesis}
\bibfield{author}{\bibinfo{person}{Andrea Guerrieri}, \bibinfo{person}{Gabriel~Da Silva~Marques}, \bibinfo{person}{Francesco Regazzoni}, {and} \bibinfo{person}{Andres Upegui}.} \bibinfo{year}{2022}\natexlab{}.
\newblock \showarticletitle{{Optimizing Lattice-based Post-Quantum Cryptography Codes for High-Level Synthesis}}.
\newblock \bibinfo{journal}{\emph{Proceedings - 2022 25th Euromicro Conference on Digital System Design, DSD 2022}} (\bibinfo{year}{2022}), \bibinfo{pages}{777--784}.
\newblock
\showISBNx{9781665474047}
\urldef\tempurl%
\url{https://doi.org/10.1109/DSD57027.2022.00109}
\showDOI{\tempurl}


\bibitem[\protect\citeauthoryear{Guo, Xue, Yan, Pan, Exarchakos, et~al\mbox{.}}{Guo et~al\mbox{.}}{2022}]%
        {Guo2022DACON:Invited}
\bibfield{author}{\bibinfo{person}{Xiaotao Guo}, \bibinfo{person}{Xuwei Xue}, \bibinfo{person}{Fulong Yan}, \bibinfo{person}{Bitao Pan}, \bibinfo{person}{Georgios Exarchakos}, {et~al\mbox{.}}} \bibinfo{year}{2022}\natexlab{}.
\newblock \showarticletitle{{DACON: A reconfigurable application-centric optical network for disaggregated data center infrastructures [Invited]}}.
\newblock \bibinfo{journal}{\emph{Journal of Optical Communications and Networking}} \bibinfo{volume}{14}, \bibinfo{number}{1} (\bibinfo{date}{1} \bibinfo{year}{2022}), \bibinfo{pages}{A69--A80}.
\newblock
\showISSN{19430639}
\urldef\tempurl%
\url{https://doi.org/10.1364/JOCN.438950}
\showDOI{\tempurl}


\bibitem[\protect\citeauthoryear{Haarman, Almeida, Heskes, Zwanenburg, and Alachiotis}{Haarman et~al\mbox{.}}{2023}]%
        {Hartman-2023}
\bibfield{author}{\bibinfo{person}{Timo Haarman}, \bibinfo{person}{Antonio Sousa~de Almeida}, \bibinfo{person}{Amber Heskes}, \bibinfo{person}{Floris Zwanenburg}, {and} \bibinfo{person}{Nikolaos Alachiotis}.} \bibinfo{year}{2023}\natexlab{}.
\newblock \showarticletitle{FPGA-accelerated Quantum Transport Measurements}. In \bibinfo{booktitle}{\emph{2023 International Conference on Field Programmable Technology (ICFPT)}}. \bibinfo{pages}{44--52}.
\newblock
\urldef\tempurl%
\url{https://doi.org/10.1109/ICFPT59805.2023.00010}
\showDOI{\tempurl}


\bibitem[\protect\citeauthoryear{Hall}{Hall}{2005}]%
        {book-SKA}
\bibfield{author}{\bibinfo{person}{Peter~J. Hall}.} \bibinfo{year}{2005}\natexlab{}.
\newblock \bibinfo{booktitle}{\emph{The Square Kilometre Array: An Engineering Perspective}}.
\newblock \bibinfo{publisher}{Springer}.
\newblock


\bibitem[\protect\citeauthoryear{Hampson, Bunton, Gunst, Baillie, and Vaate}{Hampson et~al\mbox{.}}{2016}]%
        {SKA-CBF}
\bibfield{author}{\bibinfo{person}{Grant Hampson}, \bibinfo{person}{John Bunton}, \bibinfo{person}{Andre Gunst}, \bibinfo{person}{P. Baillie}, {and} \bibinfo{person}{J.G. Vaate}.} \bibinfo{year}{2016}\natexlab{}.
\newblock \showarticletitle{Introduction to the SKA low correlator and beamformer system}. \bibinfo{pages}{99062S}.
\newblock
\urldef\tempurl%
\url{https://doi.org/10.1117/12.2231524}
\showDOI{\tempurl}


\bibitem[\protect\citeauthoryear{Hessels, Tai, Jansen, and Deuten}{Hessels et~al\mbox{.}}{[n. d.]}]%
        {rathenauEuropeseWetenschap}
\bibfield{author}{\bibinfo{person}{Laurens Hessels}, \bibinfo{person}{Sue-Yen Tjong~Tjin Tai}, \bibinfo{person}{Julia Jansen}, {and} \bibinfo{person}{Jasper Deuten}.} \bibinfo{year}{[n. d.]}\natexlab{}.
\newblock \bibinfo{title}{{E}uropese wetenschap en innovatie in een nieuw geopolitiek speelveld | {R}athenau {I}nstituut}.
\newblock \bibinfo{howpublished}{\url{https://www.rathenau.nl/nl/werking-van-het-wetenschapssysteem/europese-wetenschap-en-innovatie-een-nieuw-geopolitiek-speelveld}}.
\newblock
\newblock
\shownote{[Accessed 19-02-2024].}


\bibitem[\protect\citeauthoryear{Hirvonen, Lepp{\"{a}}nen, Hepola, Multanen, Hoozemans, et~al\mbox{.}}{Hirvonen et~al\mbox{.}}{2023}]%
        {Hirvonen2023AEx:Co-Processors}
\bibfield{author}{\bibinfo{person}{Alex Hirvonen}, \bibinfo{person}{Topi Lepp{\"{a}}nen}, \bibinfo{person}{Kari Hepola}, \bibinfo{person}{Joonas Multanen}, \bibinfo{person}{Joost Hoozemans}, {et~al\mbox{.}}} \bibinfo{year}{2023}\natexlab{}.
\newblock \showarticletitle{{AEx: Automated High-Level Synthesis of Compiler Programmable Co-Processors}}.
\newblock \bibinfo{journal}{\emph{Journal of Signal Processing Systems}} \bibinfo{volume}{95}, \bibinfo{number}{9} (\bibinfo{date}{9} \bibinfo{year}{2023}), \bibinfo{pages}{1051--1065}.
\newblock
\showISSN{19398115}
\urldef\tempurl%
\url{https://doi.org/10.1007/S11265-023-01841-3/TABLES/5}
\showDOI{\tempurl}


\bibitem[\protect\citeauthoryear{Homulle, Visser, Patra, Ferrari, Prati, et~al\mbox{.}}{Homulle et~al\mbox{.}}{2017}]%
        {Homulle-2017}
\bibfield{author}{\bibinfo{person}{Harald Homulle}, \bibinfo{person}{Stefan Visser}, \bibinfo{person}{Bishnu Patra}, \bibinfo{person}{Giorgio Ferrari}, \bibinfo{person}{Enrico Prati}, {et~al\mbox{.}}} \bibinfo{year}{2017}\natexlab{}.
\newblock \showarticletitle{{A reconfigurable cryogenic platform for the classical control of quantum processors}}.
\newblock \bibinfo{journal}{\emph{Review of Scientific Instruments}} \bibinfo{volume}{88}, \bibinfo{number}{4} (\bibinfo{date}{04} \bibinfo{year}{2017}), \bibinfo{pages}{045103}.
\newblock
\showISSN{0034-6748}
\urldef\tempurl%
\url{https://doi.org/10.1063/1.4979611}
\showDOI{\tempurl}
\showeprint{https://pubs.aip.org/aip/rsi/article-pdf/doi/10.1063/1.4979611/15975928/045103\_1\_online.pdf}


\bibitem[\protect\citeauthoryear{Hoozemans, Peltenburg, Nonnemacher, Hadnagy, Al-Ars, et~al\mbox{.}}{Hoozemans et~al\mbox{.}}{2021a}]%
        {Hoozemans2021FPGAOpportunities}
\bibfield{author}{\bibinfo{person}{Joost Hoozemans}, \bibinfo{person}{Johan Peltenburg}, \bibinfo{person}{Fabian Nonnemacher}, \bibinfo{person}{Akos Hadnagy}, \bibinfo{person}{Zaid Al-Ars}, {et~al\mbox{.}}} \bibinfo{year}{2021}\natexlab{a}.
\newblock \showarticletitle{{FPGA Acceleration for Big Data Analytics: Challenges and Opportunities}}.
\newblock \bibinfo{journal}{\emph{IEEE Circuits and Systems Magazine}} \bibinfo{volume}{21}, \bibinfo{number}{2} (\bibinfo{date}{4} \bibinfo{year}{2021}), \bibinfo{pages}{30--47}.
\newblock
\showISSN{15580830}
\urldef\tempurl%
\url{https://doi.org/10.1109/MCAS.2021.3071608}
\showDOI{\tempurl}


\bibitem[\protect\citeauthoryear{Hoozemans, Tervo, Jaaskelainen, and Al-Ars}{Hoozemans et~al\mbox{.}}{2021b}]%
        {Hoozemans2021EnergyASIP}
\bibfield{author}{\bibinfo{person}{Joost Hoozemans}, \bibinfo{person}{Kati Tervo}, \bibinfo{person}{Pekka Jaaskelainen}, {and} \bibinfo{person}{Zaid Al-Ars}.} \bibinfo{year}{2021}\natexlab{b}.
\newblock \showarticletitle{{Energy Efficient Multistandard Decompressor ASIP}}.
\newblock \bibinfo{journal}{\emph{ACM International Conference Proceeding Series}}  \bibinfo{volume}{Part F 174233} (\bibinfo{date}{1} \bibinfo{year}{2021}), \bibinfo{pages}{14--19}.
\newblock
\showISBNx{9781450388450}
\urldef\tempurl%
\url{https://doi.org/10.1145/3456172.3456218}
\showDOI{\tempurl}


\bibitem[\protect\citeauthoryear{Hoozemans, van Straten, and Wong}{Hoozemans et~al\mbox{.}}{2018}]%
        {Hoozemans2018IncreasingProcessor}
\bibfield{author}{\bibinfo{person}{Joost Hoozemans}, \bibinfo{person}{Jeroen van Straten}, {and} \bibinfo{person}{Stephan Wong}.} \bibinfo{year}{2018}\natexlab{}.
\newblock \showarticletitle{{Increasing resource utilization in mixed-criticality systems using a polymorphic VLIW processor}}.
\newblock \bibinfo{journal}{\emph{Journal of Systems Architecture}}  \bibinfo{volume}{84} (\bibinfo{date}{3} \bibinfo{year}{2018}), \bibinfo{pages}{2--11}.
\newblock
\showISSN{1383-7621}
\urldef\tempurl%
\url{https://doi.org/10.1016/J.SYSARC.2018.01.003}
\showDOI{\tempurl}


\bibitem[\protect\citeauthoryear{Hu, Chitnis, Monos, and Dinh}{Hu et~al\mbox{.}}{2021}]%
        {hu2021next}
\bibfield{author}{\bibinfo{person}{Taishan Hu}, \bibinfo{person}{Nilesh Chitnis}, \bibinfo{person}{Dimitri Monos}, {and} \bibinfo{person}{Anh Dinh}.} \bibinfo{year}{2021}\natexlab{}.
\newblock \showarticletitle{Next-generation sequencing technologies: An overview}.
\newblock \bibinfo{journal}{\emph{Human Immunology}} \bibinfo{volume}{82}, \bibinfo{number}{11} (\bibinfo{year}{2021}), \bibinfo{pages}{801--811}.
\newblock
\showISSN{0198-8859}
\urldef\tempurl%
\url{https://doi.org/10.1016/j.humimm.2021.02.012}
\showDOI{\tempurl}
\newblock
\shownote{Next Generation Sequencing and its Application to Medical Laboratory Immunology.}


\bibitem[\protect\citeauthoryear{Huang, Wu, Jeong, Wang, Chen, et~al\mbox{.}}{Huang et~al\mbox{.}}{2021}]%
        {pylog-2021}
\bibfield{author}{\bibinfo{person}{Sitao Huang}, \bibinfo{person}{Kun Wu}, \bibinfo{person}{Hyunmin Jeong}, \bibinfo{person}{Chengyue Wang}, \bibinfo{person}{Deming Chen}, {et~al\mbox{.}}} \bibinfo{year}{2021}\natexlab{}.
\newblock \showarticletitle{PyLog: An Algorithm-Centric Python-Based FPGA Programming and Synthesis Flow}.
\newblock \bibinfo{journal}{\emph{IEEE Trans. Comput.}} \bibinfo{volume}{70}, \bibinfo{number}{12} (\bibinfo{year}{2021}), \bibinfo{pages}{2015--2028}.
\newblock
\urldef\tempurl%
\url{https://doi.org/10.1109/TC.2021.3123465}
\showDOI{\tempurl}


\bibitem[\protect\citeauthoryear{Iakovidis, Levinson, Afik, Alexa, Alexopoulos, et~al\mbox{.}}{Iakovidis et~al\mbox{.}}{2023}]%
        {Iakovidis2023TheElectronics}
\bibfield{author}{\bibinfo{person}{G. Iakovidis}, \bibinfo{person}{L. Levinson}, \bibinfo{person}{Y. Afik}, \bibinfo{person}{C. Alexa}, \bibinfo{person}{T. Alexopoulos}, {et~al\mbox{.}}} \bibinfo{year}{2023}\natexlab{}.
\newblock \showarticletitle{{The New Small Wheel electronics}}.
\newblock \bibinfo{journal}{\emph{Journal of Instrumentation}} \bibinfo{volume}{18}, \bibinfo{number}{05} (\bibinfo{date}{5} \bibinfo{year}{2023}), \bibinfo{pages}{P05012}.
\newblock
\showISSN{1748-0221}
\urldef\tempurl%
\url{https://doi.org/10.1088/1748-0221/18/05/P05012}
\showDOI{\tempurl}


\bibitem[\protect\citeauthoryear{Iosup, Prodan, Varbanescu, Talluri, Magalhaes, et~al\mbox{.}}{Iosup et~al\mbox{.}}{2023}]%
        {Iosup2023GraphContinuum}
\bibfield{author}{\bibinfo{person}{Alexandru Iosup}, \bibinfo{person}{Radu Prodan}, \bibinfo{person}{Ana~Lucia Varbanescu}, \bibinfo{person}{Sacheendra Talluri}, \bibinfo{person}{Gilles Magalhaes}, {et~al\mbox{.}}} \bibinfo{year}{2023}\natexlab{}.
\newblock \showarticletitle{{Graph Greenifier: Towards Sustainable and Energy-Aware Massive Graph Processing in the Computing Continuum}}.
\newblock \bibinfo{journal}{\emph{ICPE 2023 - Companion of the 2023 ACM/SPEC International Conference on Performance Engineering}} (\bibinfo{date}{4} \bibinfo{year}{2023}), \bibinfo{pages}{209--214}.
\newblock
\showISBNx{9798400700729}
\urldef\tempurl%
\url{https://doi.org/10.1145/3578245.3585329}
\showDOI{\tempurl}


\bibitem[\protect\citeauthoryear{Ireland, Protheroe, Williams, Belcher, Dekker, et~al\mbox{.}}{Ireland et~al\mbox{.}}{2020}]%
        {Ireland-2020}
\bibfield{author}{\bibinfo{person}{Jane Ireland}, \bibinfo{person}{Stephen Protheroe}, \bibinfo{person}{Jonathan Williams}, \bibinfo{person}{Allan Belcher}, \bibinfo{person}{Ronald Dekker}, {et~al\mbox{.}}} \bibinfo{year}{2020}\natexlab{}.
\newblock \showarticletitle{Real-time quantum-accurate voltage waveform synthesis}. In \bibinfo{booktitle}{\emph{2020 Conference on Precision Electromagnetic Measurements (CPEM)}}. \bibinfo{pages}{1--2}.
\newblock
\urldef\tempurl%
\url{https://doi.org/10.1109/CPEM49742.2020.9191715}
\showDOI{\tempurl}


\bibitem[\protect\citeauthoryear{Irmak, Alachiotis, and Ziener}{Irmak et~al\mbox{.}}{2021a}]%
        {ml-energy-efficient-cnn}
\bibfield{author}{\bibinfo{person}{Hasan Irmak}, \bibinfo{person}{Nikolaos Alachiotis}, {and} \bibinfo{person}{Daniel Ziener}.} \bibinfo{year}{2021}\natexlab{a}.
\newblock \showarticletitle{An Energy-Efficient FPGA-based Convolutional Neural Network Implementation}. In \bibinfo{booktitle}{\emph{2021 29th Signal Processing and Communications Applications Conference (SIU)}}. \bibinfo{pages}{1--4}.
\newblock
\urldef\tempurl%
\url{https://doi.org/10.1109/SIU53274.2021.9477823}
\showDOI{\tempurl}


\bibitem[\protect\citeauthoryear{Irmak, Corradi, Detterer, Alachiotis, and Ziener}{Irmak et~al\mbox{.}}{2021b}]%
        {Irmak2021ADesigns}
\bibfield{author}{\bibinfo{person}{Hasan Irmak}, \bibinfo{person}{Federico Corradi}, \bibinfo{person}{Paul Detterer}, \bibinfo{person}{Nikolaos Alachiotis}, {and} \bibinfo{person}{Daniel Ziener}.} \bibinfo{year}{2021}\natexlab{b}.
\newblock \showarticletitle{{A Dynamic Reconfigurable Architecture for Hybrid Spiking and Convolutional FPGA-Based Neural Network Designs}}.
\newblock \bibinfo{journal}{\emph{Journal of Low Power Electronics and Applications 2021, Vol. 11, Page 32}} \bibinfo{volume}{11}, \bibinfo{number}{3} (\bibinfo{date}{8} \bibinfo{year}{2021}), \bibinfo{pages}{32}.
\newblock
\showISSN{2079-9268}
\urldef\tempurl%
\url{https://doi.org/10.3390/JLPEA11030032}
\showDOI{\tempurl}


\bibitem[\protect\citeauthoryear{Irmak, Ziener, and Alachiotis}{Irmak et~al\mbox{.}}{2021c}]%
        {ml-cnn-acclr-part-reconf}
\bibfield{author}{\bibinfo{person}{Hasan Irmak}, \bibinfo{person}{Daniel Ziener}, {and} \bibinfo{person}{Nikolaos Alachiotis}.} \bibinfo{year}{2021}\natexlab{c}.
\newblock \showarticletitle{Increasing Flexibility of FPGA-based CNN Accelerators with Dynamic Partial Reconfiguration}. In \bibinfo{booktitle}{\emph{2021 31st International Conference on Field-Programmable Logic and Applications (FPL)}}. \bibinfo{pages}{306--311}.
\newblock
\urldef\tempurl%
\url{https://doi.org/10.1109/FPL53798.2021.00061}
\showDOI{\tempurl}


\bibitem[\protect\citeauthoryear{Ji, Al-Ars, Hofstee, Chang, and Zhang}{Ji et~al\mbox{.}}{2023}]%
        {ml-FPQNet}
\bibfield{author}{\bibinfo{person}{Mengfei Ji}, \bibinfo{person}{Zaid Al-Ars}, \bibinfo{person}{Peter Hofstee}, \bibinfo{person}{Yuchun Chang}, {and} \bibinfo{person}{Baolin Zhang}.} \bibinfo{year}{2023}\natexlab{}.
\newblock \showarticletitle{FPQNet: Fully Pipelined and Quantized CNN for Ultra-Low Latency Image Classification on FPGAs Using OpenCAPI}.
\newblock \bibinfo{journal}{\emph{Electronics}} \bibinfo{volume}{12}, \bibinfo{number}{19} (\bibinfo{year}{2023}).
\newblock
\showISSN{2079-9292}
\urldef\tempurl%
\url{https://doi.org/10.3390/electronics12194085}
\showDOI{\tempurl}


\bibitem[\protect\citeauthoryear{Jin, Burleson, Van~Dijk, and R{\"{u}}hrmair}{Jin et~al\mbox{.}}{2020}]%
        {Jin2020ErasableDesign}
\bibfield{author}{\bibinfo{person}{Chenglu Jin}, \bibinfo{person}{Wayne Burleson}, \bibinfo{person}{Marten Van~Dijk}, {and} \bibinfo{person}{Ulrich R{\"{u}}hrmair}.} \bibinfo{year}{2020}\natexlab{}.
\newblock \showarticletitle{{Erasable PUFs: Formal Treatment and Generic Design}}.
\newblock \bibinfo{journal}{\emph{ASHES 2020 - Proceedings of the 4th ACM Workshop on Attacks and Solutions in Hardware Security}} (\bibinfo{date}{11} \bibinfo{year}{2020}), \bibinfo{pages}{21--33}.
\newblock
\showISBNx{9781450380904}
\urldef\tempurl%
\url{https://doi.org/10.1145/3411504.3421215}
\showDOI{\tempurl}


\bibitem[\protect\citeauthoryear{Jin, Burleson, van Dijk, and R{\"{u}}hrmair}{Jin et~al\mbox{.}}{2022}]%
        {Jin2022ProgrammableApplications}
\bibfield{author}{\bibinfo{person}{Chenglu Jin}, \bibinfo{person}{Wayne Burleson}, \bibinfo{person}{Marten van Dijk}, {and} \bibinfo{person}{Ulrich R{\"{u}}hrmair}.} \bibinfo{year}{2022}\natexlab{}.
\newblock \showarticletitle{{Programmable access-controlled and generic erasable PUF design and its applications}}.
\newblock \bibinfo{journal}{\emph{Journal of Cryptographic Engineering}} \bibinfo{volume}{12}, \bibinfo{number}{4} (\bibinfo{date}{11} \bibinfo{year}{2022}), \bibinfo{pages}{413--432}.
\newblock
\showISSN{21908516}
\urldef\tempurl%
\url{https://doi.org/10.1007/S13389-022-00284-Z/FIGURES/7}
\showDOI{\tempurl}


\bibitem[\protect\citeauthoryear{Karapakula, Brinkerink, Vecchio, Pourshaghaghi, Dolron, et~al\mbox{.}}{Karapakula et~al\mbox{.}}{2024}]%
        {karapakula2024ncle}
\bibfield{author}{\bibinfo{person}{Sukanth Karapakula}, \bibinfo{person}{Christiaan Brinkerink}, \bibinfo{person}{Antonio Vecchio}, \bibinfo{person}{Hamid~R. Pourshaghaghi}, \bibinfo{person}{Peter Dolron}, {et~al\mbox{.}}} \bibinfo{year}{2024}\natexlab{}.
\newblock \showarticletitle{Architecture design and ground performance of Netherlands-China low-frequency explorer}.
\newblock \bibinfo{journal}{\emph{Radio Science}} \bibinfo{volume}{59}, \bibinfo{number}{8} (\bibinfo{year}{2024}), \bibinfo{pages}{1--39}.
\newblock
\urldef\tempurl%
\url{https://doi.org/10.1029/2023RS007906}
\showDOI{\tempurl}


\bibitem[\protect\citeauthoryear{Katal, Dahiya, and Choudhury}{Katal et~al\mbox{.}}{2023}]%
        {enegry-efficiency-cloud-dc}
\bibfield{author}{\bibinfo{person}{Avita Katal}, \bibinfo{person}{Susheela Dahiya}, {and} \bibinfo{person}{Tanupriya Choudhury}.} \bibinfo{year}{2023}\natexlab{}.
\newblock \showarticletitle{Energy efficiency in cloud computing data centers: a survey on software technologies}.
\newblock \bibinfo{journal}{\emph{Cluster Computing}} \bibinfo{volume}{26}, \bibinfo{number}{3} (\bibinfo{date}{01 Jun} \bibinfo{year}{2023}), \bibinfo{pages}{1845--1875}.
\newblock
\showISSN{1573-7543}
\urldef\tempurl%
\url{https://doi.org/10.1007/s10586-022-03713-0}
\showDOI{\tempurl}


\bibitem[\protect\citeauthoryear{Katevenis, Ammendola, Biagioni, Cretaro, Frezza, et~al\mbox{.}}{Katevenis et~al\mbox{.}}{2018}]%
        {Katevenis2018NextDevelopment}
\bibfield{author}{\bibinfo{person}{Manolis Katevenis}, \bibinfo{person}{Roberto Ammendola}, \bibinfo{person}{Andrea Biagioni}, \bibinfo{person}{Paolo Cretaro}, \bibinfo{person}{Ottorino Frezza}, {et~al\mbox{.}}} \bibinfo{year}{2018}\natexlab{}.
\newblock \showarticletitle{{Next generation of Exascale-class systems: ExaNeSt project and the status of its interconnect and storage development}}.
\newblock \bibinfo{journal}{\emph{Microprocessors and Microsystems}}  \bibinfo{volume}{61} (\bibinfo{date}{9} \bibinfo{year}{2018}), \bibinfo{pages}{58--71}.
\newblock
\showISSN{0141-9331}
\urldef\tempurl%
\url{https://doi.org/10.1016/J.MICPRO.2018.05.009}
\showDOI{\tempurl}


\bibitem[\protect\citeauthoryear{Kim and Nielsen}{Kim and Nielsen}{2004}]%
        {kim2004linkage}
\bibfield{author}{\bibinfo{person}{Yuseob Kim} {and} \bibinfo{person}{Rasmus Nielsen}.} \bibinfo{year}{2004}\natexlab{}.
\newblock \showarticletitle{Linkage disequilibrium as a signature of selective sweeps}.
\newblock \bibinfo{journal}{\emph{Genetics}} \bibinfo{volume}{167}, \bibinfo{number}{3} (\bibinfo{year}{2004}), \bibinfo{pages}{1513--1524}.
\newblock


\bibitem[\protect\citeauthoryear{Koylu, Garaffa, Reinbrecht, Zahedi, Hamdioui, et~al\mbox{.}}{Koylu et~al\mbox{.}}{2022}]%
        {Koylu2022ExploitingAttacks}
\bibfield{author}{\bibinfo{person}{Troya Koylu}, \bibinfo{person}{Luiza Garaffa}, \bibinfo{person}{Cezar Reinbrecht}, \bibinfo{person}{Mahdi Zahedi}, \bibinfo{person}{Said Hamdioui}, {et~al\mbox{.}}} \bibinfo{year}{2022}\natexlab{}.
\newblock \showarticletitle{{Exploiting PUF Variation to Detect Fault Injection Attacks}}.
\newblock \bibinfo{journal}{\emph{Proceedings - 2022 25th International Symposium on Design and Diagnostics of Electronic Circuits and Systems, DDECS 2022}} (\bibinfo{year}{2022}), \bibinfo{pages}{74--79}.
\newblock
\showISBNx{9781665494311}
\urldef\tempurl%
\url{https://doi.org/10.1109/DDECS54261.2022.9770154}
\showDOI{\tempurl}


\bibitem[\protect\citeauthoryear{K{\"{o}}yl{\"{u}}, Wedig~Reinbrecht, Brandalero, Hamdioui, and Taouil}{K{\"{o}}yl{\"{u}} et~al\mbox{.}}{2022}]%
        {Koylu2022InstructionAttacks}
\bibfield{author}{\bibinfo{person}{Troya~Çağıl K{\"{o}}yl{\"{u}}}, \bibinfo{person}{Cezar~Rodolfo Wedig~Reinbrecht}, \bibinfo{person}{Marcelo Brandalero}, \bibinfo{person}{Said Hamdioui}, {and} \bibinfo{person}{Mottaqiallah Taouil}.} \bibinfo{year}{2022}\natexlab{}.
\newblock \showarticletitle{{Instruction flow-based detectors against fault injection attacks}}.
\newblock \bibinfo{journal}{\emph{Microprocessors and Microsystems}}  \bibinfo{volume}{94} (\bibinfo{date}{10} \bibinfo{year}{2022}), \bibinfo{pages}{104638}.
\newblock
\showISSN{0141-9331}
\urldef\tempurl%
\url{https://doi.org/10.1016/J.MICPRO.2022.104638}
\showDOI{\tempurl}


\bibitem[\protect\citeauthoryear{Kundel, Nobach, Blendin, Maas, Zimber, et~al\mbox{.}}{Kundel et~al\mbox{.}}{2021}]%
        {Kundel2021OpenBNG:Hardware}
\bibfield{author}{\bibinfo{person}{Ralf Kundel}, \bibinfo{person}{Leonhard Nobach}, \bibinfo{person}{Jeremias Blendin}, \bibinfo{person}{Wilfried Maas}, \bibinfo{person}{Andreas Zimber}, {et~al\mbox{.}}} \bibinfo{year}{2021}\natexlab{}.
\newblock \showarticletitle{{OpenBNG: Central office network functions on programmable data plane hardware}}.
\newblock \bibinfo{journal}{\emph{International Journal of Network Management}} \bibinfo{volume}{31}, \bibinfo{number}{1} (\bibinfo{date}{1} \bibinfo{year}{2021}), \bibinfo{pages}{e2134}.
\newblock
\showISSN{1099-1190}
\urldef\tempurl%
\url{https://doi.org/10.1002/NEM.2134}
\showDOI{\tempurl}


\bibitem[\protect\citeauthoryear{Labafniya, Picek, Etemadi~Borujeni, and Mentens}{Labafniya et~al\mbox{.}}{2020}]%
        {Labafniya2020OnPrevention}
\bibfield{author}{\bibinfo{person}{Mansoureh Labafniya}, \bibinfo{person}{Stjepan Picek}, \bibinfo{person}{Shahram Etemadi~Borujeni}, {and} \bibinfo{person}{Nele Mentens}.} \bibinfo{year}{2020}\natexlab{}.
\newblock \showarticletitle{{On the feasibility of using evolvable hardware for hardware Trojan detection and prevention}}.
\newblock \bibinfo{journal}{\emph{Applied Soft Computing}}  \bibinfo{volume}{91} (\bibinfo{date}{6} \bibinfo{year}{2020}), \bibinfo{pages}{106247}.
\newblock
\showISSN{1568-4946}
\urldef\tempurl%
\url{https://doi.org/10.1016/J.ASOC.2020.106247}
\showDOI{\tempurl}


\bibitem[\protect\citeauthoryear{Lahr, Niederhagen, Petri, and Samardjiska}{Lahr et~al\mbox{.}}{2020}]%
        {Lahr2020SideImplementation}
\bibfield{author}{\bibinfo{person}{Norman Lahr}, \bibinfo{person}{Ruben Niederhagen}, \bibinfo{person}{Richard Petri}, {and} \bibinfo{person}{Simona Samardjiska}.} \bibinfo{year}{2020}\natexlab{}.
\newblock \showarticletitle{{Side Channel Information Set Decoding Using Iterative Chunking: Plaintext Recovery from the “Classic McEliece” Hardware Reference Implementation}}.
\newblock \bibinfo{journal}{\emph{Lecture Notes in Computer Science (including subseries Lecture Notes in Artificial Intelligence and Lecture Notes in Bioinformatics)}}  \bibinfo{volume}{12491 LNCS} (\bibinfo{year}{2020}), \bibinfo{pages}{881--910}.
\newblock
\showISBNx{9783030648367}
\showISSN{16113349}
\urldef\tempurl%
\url{https://doi.org/10.1007/978-3-030-64837-4{\_}29/TABLES/5}
\showDOI{\tempurl}


\bibitem[\protect\citeauthoryear{Lahti, Sjövall, Vanne, and Hämäläinen}{Lahti et~al\mbox{.}}{2019}]%
        {Lahti-2019}
\bibfield{author}{\bibinfo{person}{Sakari Lahti}, \bibinfo{person}{Panu Sjövall}, \bibinfo{person}{Jarno Vanne}, {and} \bibinfo{person}{Timo~D. Hämäläinen}.} \bibinfo{year}{2019}\natexlab{}.
\newblock \showarticletitle{Are We There Yet? A Study on the State of High-Level Synthesis}.
\newblock \bibinfo{journal}{\emph{IEEE Transactions on Computer-Aided Design of Integrated Circuits and Systems}} \bibinfo{volume}{38}, \bibinfo{number}{5} (\bibinfo{year}{2019}), \bibinfo{pages}{898--911}.
\newblock
\urldef\tempurl%
\url{https://doi.org/10.1109/TCAD.2018.2834439}
\showDOI{\tempurl}


\bibitem[\protect\citeauthoryear{Lai, Chi, Hu, Wang, Yu, et~al\mbox{.}}{Lai et~al\mbox{.}}{2019}]%
        {heterocl-2019}
\bibfield{author}{\bibinfo{person}{Yi-Hsiang Lai}, \bibinfo{person}{Yuze Chi}, \bibinfo{person}{Yuwei Hu}, \bibinfo{person}{Jie Wang}, \bibinfo{person}{Cody~Hao Yu}, {et~al\mbox{.}}} \bibinfo{year}{2019}\natexlab{}.
\newblock \showarticletitle{{HeteroCL}: A multi-paradigm programming infrastructure for software-defined reconfigurable computing}. In \bibinfo{booktitle}{\emph{Proceedings of the 2019 ACM/SIGDA International Symposium on Field-Programmable Gate Arrays (FPGA'19)}}. \bibinfo{pages}{242--251}.
\newblock


\bibitem[\protect\citeauthoryear{Lattner and Adve}{Lattner and Adve}{2004}]%
        {lattner2004llvm}
\bibfield{author}{\bibinfo{person}{Chris Lattner} {and} \bibinfo{person}{Vikram Adve}.} \bibinfo{year}{2004}\natexlab{}.
\newblock \showarticletitle{LLVM: A compilation framework for lifelong program analysis \& transformation}. In \bibinfo{booktitle}{\emph{International symposium on code generation and optimization, 2004. CGO 2004.}} IEEE, \bibinfo{pages}{75--86}.
\newblock


\bibitem[\protect\citeauthoryear{Lenkiewicz, Broekema, and Metzler}{Lenkiewicz et~al\mbox{.}}{2018}]%
        {big-data-radio-astronomy}
\bibfield{author}{\bibinfo{person}{Przemyslaw Lenkiewicz}, \bibinfo{person}{P.~Chris Broekema}, {and} \bibinfo{person}{Bernard Metzler}.} \bibinfo{year}{2018}\natexlab{}.
\newblock \showarticletitle{Energy-efficient data transfers in radio astronomy with software UDP RDMA}.
\newblock \bibinfo{journal}{\emph{Future Generation Computer Systems}}  \bibinfo{volume}{79} (\bibinfo{year}{2018}), \bibinfo{pages}{215--224}.
\newblock
\showISSN{0167-739X}
\urldef\tempurl%
\url{https://doi.org/10.1016/j.future.2017.03.027}
\showDOI{\tempurl}


\bibitem[\protect\citeauthoryear{Leon, Bezaitis, Lentaris, Soudris, Reisis, et~al\mbox{.}}{Leon et~al\mbox{.}}{2021a}]%
        {Leon2021FPGABenchmarks}
\bibfield{author}{\bibinfo{person}{Vasileios Leon}, \bibinfo{person}{Charalampos Bezaitis}, \bibinfo{person}{George Lentaris}, \bibinfo{person}{Dimitrios Soudris}, \bibinfo{person}{Dionysios Reisis}, {et~al\mbox{.}}} \bibinfo{year}{2021}\natexlab{a}.
\newblock \showarticletitle{{FPGA VPU Co-Processing in Space Applications: Development and Testing with DSP/AI Benchmarks}}.
\newblock \bibinfo{journal}{\emph{2021 28th IEEE International Conference on Electronics, Circuits, and Systems, ICECS 2021 - Proceedings}} (\bibinfo{year}{2021}).
\newblock
\showISBNx{9781728182810}
\urldef\tempurl%
\url{https://doi.org/10.1109/ICECS53924.2021.9665462}
\showDOI{\tempurl}


\bibitem[\protect\citeauthoryear{Leon, Lentaris, Petrongonas, Soudris, Furano, et~al\mbox{.}}{Leon et~al\mbox{.}}{2021b}]%
        {Leon2021ImprovingSoC}
\bibfield{author}{\bibinfo{person}{Vasileios Leon}, \bibinfo{person}{George Lentaris}, \bibinfo{person}{Evangelos Petrongonas}, \bibinfo{person}{Dimitrios Soudris}, \bibinfo{person}{Gianluca Furano}, {et~al\mbox{.}}} \bibinfo{year}{2021}\natexlab{b}.
\newblock \showarticletitle{{Improving Performance-Power-Programmability in Space Avionics with Edge Devices: VBN on Myriad2 SoC}}.
\newblock \bibinfo{journal}{\emph{ACM Transactions on Embedded Computing Systems (TECS)}} \bibinfo{volume}{20}, \bibinfo{number}{3} (\bibinfo{date}{3} \bibinfo{year}{2021}).
\newblock
\showISBNx{10.1145/3440885}
\showISSN{15583465}
\urldef\tempurl%
\url{https://doi.org/10.1145/3440885}
\showDOI{\tempurl}


\bibitem[\protect\citeauthoryear{Leon, Stamoulias, Lentaris, Soudris, Gonzalez-Arjona, et~al\mbox{.}}{Leon et~al\mbox{.}}{2021c}]%
        {Leon2021DevelopmentBenchmarks}
\bibfield{author}{\bibinfo{person}{Vasileios Leon}, \bibinfo{person}{Ioannis Stamoulias}, \bibinfo{person}{George Lentaris}, \bibinfo{person}{Dimitrios Soudris}, \bibinfo{person}{David Gonzalez-Arjona}, {et~al\mbox{.}}} \bibinfo{year}{2021}\natexlab{c}.
\newblock \showarticletitle{{Development and Testing on the European Space-Grade BRAVE FPGAs: Evaluation of NG-Large Using High-Performance DSP Benchmarks}}.
\newblock \bibinfo{journal}{\emph{IEEE Access}}  \bibinfo{volume}{9} (\bibinfo{year}{2021}), \bibinfo{pages}{131877--131892}.
\newblock
\showISSN{21693536}
\urldef\tempurl%
\url{https://doi.org/10.1109/ACCESS.2021.3114502}
\showDOI{\tempurl}


\bibitem[\protect\citeauthoryear{Licht, De~Matteis, Ben-Nun, Kuster, Rausch, et~al\mbox{.}}{Licht et~al\mbox{.}}{2022}]%
        {Licht2022PythonDesign}
\bibfield{author}{\bibinfo{person}{Johannes de~Fine Licht}, \bibinfo{person}{Tiziano De~Matteis}, \bibinfo{person}{Tal Ben-Nun}, \bibinfo{person}{Andreas Kuster}, \bibinfo{person}{Oliver Rausch}, {et~al\mbox{.}}} \bibinfo{year}{2022}\natexlab{}.
\newblock \showarticletitle{{Python FPGA Programming with Data-Centric Multi-Level Design}}.
\newblock  (\bibinfo{date}{12} \bibinfo{year}{2022}).
\newblock
\urldef\tempurl%
\url{http://arxiv.org/abs/2212.13768}
\showURL{%
\tempurl}


\bibitem[\protect\citeauthoryear{Malakonakis, Brokalakis, Alachiotis, Sotiriades, and Dollas}{Malakonakis et~al\mbox{.}}{2020}]%
        {Malakonakis2020ExploringRAxML}
\bibfield{author}{\bibinfo{person}{Pavlos Malakonakis}, \bibinfo{person}{Andreas Brokalakis}, \bibinfo{person}{Nikolaos Alachiotis}, \bibinfo{person}{Evripides Sotiriades}, {and} \bibinfo{person}{Apostolos Dollas}.} \bibinfo{year}{2020}\natexlab{}.
\newblock \showarticletitle{{Exploring Modern FPGA Platforms for Faster Phylogeny Reconstruction with RAxML}}.
\newblock \bibinfo{journal}{\emph{Proceedings - IEEE 20th International Conference on Bioinformatics and Bioengineering, BIBE 2020}} (\bibinfo{date}{10} \bibinfo{year}{2020}), \bibinfo{pages}{97--104}.
\newblock
\showISBNx{9781728195742}
\urldef\tempurl%
\url{https://doi.org/10.1109/BIBE50027.2020.00024}
\showDOI{\tempurl}


\bibitem[\protect\citeauthoryear{Massolino, Longa, Renes, and Batina}{Massolino et~al\mbox{.}}{2020}]%
        {Massolino2020ASIKE}
\bibfield{author}{\bibinfo{person}{Pedro Maat~C. Massolino}, \bibinfo{person}{Patrick Longa}, \bibinfo{person}{Joost Renes}, {and} \bibinfo{person}{Lejla Batina}.} \bibinfo{year}{2020}\natexlab{}.
\newblock \showarticletitle{{A Compact and Scalable Hardware/Software Co-design of SIKE}}.
\newblock \bibinfo{journal}{\emph{Cryptology ePrint Archive}} \bibinfo{volume}{2020}, \bibinfo{number}{2} (\bibinfo{year}{2020}), \bibinfo{pages}{245--271}.
\newblock
\showISSN{25692925}
\urldef\tempurl%
\url{https://doi.org/10.13154/TCHES.V2020.I2.245-271}
\showDOI{\tempurl}


\bibitem[\protect\citeauthoryear{Meyer, Sancho, Mrdakovic, Miao, and Calabretta}{Meyer et~al\mbox{.}}{2018}]%
        {Meyer2018OpticalPerformance}
\bibfield{author}{\bibinfo{person}{Hugo Meyer}, \bibinfo{person}{Jose~Carlos Sancho}, \bibinfo{person}{Milica Mrdakovic}, \bibinfo{person}{Wang Miao}, {and} \bibinfo{person}{Nicola Calabretta}.} \bibinfo{year}{2018}\natexlab{}.
\newblock \showarticletitle{{Optical packet switching in HPC. An analysis of applications performance}}.
\newblock \bibinfo{journal}{\emph{Future Generation Computer Systems}}  \bibinfo{volume}{82} (\bibinfo{date}{5} \bibinfo{year}{2018}), \bibinfo{pages}{606--616}.
\newblock
\showISSN{0167-739X}
\urldef\tempurl%
\url{https://doi.org/10.1016/J.FUTURE.2017.02.041}
\showDOI{\tempurl}


\bibitem[\protect\citeauthoryear{Miteloudi, Batina, Daemen, and Mentens}{Miteloudi et~al\mbox{.}}{2022}]%
        {Miteloudi2022ROCKY:Data}
\bibfield{author}{\bibinfo{person}{Konstantina Miteloudi}, \bibinfo{person}{Lejla Batina}, \bibinfo{person}{Joan Daemen}, {and} \bibinfo{person}{Nele Mentens}.} \bibinfo{year}{2022}\natexlab{}.
\newblock \showarticletitle{{ROCKY: Rotation Countermeasure for the Protection of Keys and Other Sensitive Data}}.
\newblock \bibinfo{journal}{\emph{Lecture Notes in Computer Science (including subseries Lecture Notes in Artificial Intelligence and Lecture Notes in Bioinformatics)}}  \bibinfo{volume}{13227 LNCS} (\bibinfo{year}{2022}), \bibinfo{pages}{288--299}.
\newblock
\showISBNx{9783031045790}
\showISSN{16113349}
\urldef\tempurl%
\url{https://doi.org/10.1007/978-3-031-04580-6{\_}19/TABLES/3}
\showDOI{\tempurl}


\bibitem[\protect\citeauthoryear{Miteloudi, Chmielewski, Batina, and Mentens}{Miteloudi et~al\mbox{.}}{2021}]%
        {Miteloudi2021EvaluatingLeakage}
\bibfield{author}{\bibinfo{person}{Konstantina Miteloudi}, \bibinfo{person}{Lukasz Chmielewski}, \bibinfo{person}{Lejla Batina}, {and} \bibinfo{person}{Nele Mentens}.} \bibinfo{year}{2021}\natexlab{}.
\newblock \showarticletitle{{Evaluating the ROCKY Countermeasure for Side-Channel Leakage}}.
\newblock \bibinfo{journal}{\emph{IEEE/IFIP International Conference on VLSI and System-on-Chip, VLSI-SoC}}  \bibinfo{volume}{2021-October} (\bibinfo{year}{2021}).
\newblock
\showISBNx{9781665426145}
\showISSN{23248440}
\urldef\tempurl%
\url{https://doi.org/10.1109/VLSI-SOC53125.2021.9606973}
\showDOI{\tempurl}


\bibitem[\protect\citeauthoryear{Moonen, Vogt-Ardatjew, and Leferink}{Moonen et~al\mbox{.}}{2021}]%
        {Moonen2021Simulink-BasedSystems}
\bibfield{author}{\bibinfo{person}{Niek Moonen}, \bibinfo{person}{Robert Vogt-Ardatjew}, {and} \bibinfo{person}{Frank Leferink}.} \bibinfo{year}{2021}\natexlab{}.
\newblock \showarticletitle{{Simulink-Based FPGA Control for EMI Investigations of Power Electronic Systems}}.
\newblock \bibinfo{journal}{\emph{IEEE Transactions on Electromagnetic Compatibility}} \bibinfo{volume}{63}, \bibinfo{number}{4} (\bibinfo{date}{8} \bibinfo{year}{2021}), \bibinfo{pages}{1266--1273}.
\newblock
\showISSN{1558187X}
\urldef\tempurl%
\url{https://doi.org/10.1109/TEMC.2020.3042301}
\showDOI{\tempurl}


\bibitem[\protect\citeauthoryear{Mousavi, Pourshaghaghi, Kumar, and Corporaal}{Mousavi et~al\mbox{.}}{2023}]%
        {Mousavi2023MTTRSensitivity}
\bibfield{author}{\bibinfo{person}{Mahsa Mousavi}, \bibinfo{person}{Hamid~Reza Pourshaghaghi}, \bibinfo{person}{Akash Kumar}, {and} \bibinfo{person}{Henk Corporaal}.} \bibinfo{year}{2023}\natexlab{}.
\newblock \showarticletitle{{MTTR reduction of FPGA scrubbing: Exploring SEU sensitivity}}.
\newblock \bibinfo{journal}{\emph{Microprocessors and Microsystems}}  \bibinfo{volume}{101} (\bibinfo{date}{9} \bibinfo{year}{2023}), \bibinfo{pages}{104841}.
\newblock
\showISSN{0141-9331}
\urldef\tempurl%
\url{https://doi.org/10.1016/J.MICPRO.2023.104841}
\showDOI{\tempurl}


\bibitem[\protect\citeauthoryear{Mystkowska and Steenari}{Mystkowska and Steenari}{[n. d.]}]%
        {MystkowskaSimulationSpaceFibre}
\bibfield{author}{\bibinfo{person}{Gabriela Mystkowska} {and} \bibinfo{person}{David Steenari}.} \bibinfo{year}{[n. d.]}\natexlab{}.
\newblock \showarticletitle{{Simulation and hardware validation of SerDes links for SpaceFibre}}.
\newblock  (\bibinfo{year}{[n. d.]}).
\newblock


\bibitem[\protect\citeauthoryear{Nechi, Groth, Mulhem, Merchant, Buchty, et~al\mbox{.}}{Nechi et~al\mbox{.}}{2023}]%
        {10.1145/3613963}
\bibfield{author}{\bibinfo{person}{Anouar Nechi}, \bibinfo{person}{Lukas Groth}, \bibinfo{person}{Saleh Mulhem}, \bibinfo{person}{Farhad Merchant}, \bibinfo{person}{Rainer Buchty}, {et~al\mbox{.}}} \bibinfo{year}{2023}\natexlab{}.
\newblock \showarticletitle{FPGA-based Deep Learning Inference Accelerators: Where Are We Standing?}
\newblock \bibinfo{journal}{\emph{ACM Trans. Reconfigurable Technol. Syst.}} \bibinfo{volume}{16}, \bibinfo{number}{4}, Article \bibinfo{articleno}{60} (\bibinfo{date}{oct} \bibinfo{year}{2023}), \bibinfo{numpages}{32}~pages.
\newblock
\showISSN{1936-7406}
\urldef\tempurl%
\url{https://doi.org/10.1145/3613963}
\showDOI{\tempurl}


\bibitem[\protect\citeauthoryear{Nikiema, Palumbo, Aasma, Cassano, Kritikakou, et~al\mbox{.}}{Nikiema et~al\mbox{.}}{2023}]%
        {Nikiema2023TowardsDevices}
\bibfield{author}{\bibinfo{person}{Pegdwende~Romaric Nikiema}, \bibinfo{person}{Alessandro Palumbo}, \bibinfo{person}{Allan Aasma}, \bibinfo{person}{Luca Cassano}, \bibinfo{person}{Angeliki Kritikakou}, {et~al\mbox{.}}} \bibinfo{year}{2023}\natexlab{}.
\newblock \showarticletitle{{Towards Dependable RISC-V Cores for Edge Computing Devices}}.
\newblock \bibinfo{journal}{\emph{Proceedings - 2023 IEEE 29th International Symposium on On-Line Testing and Robust System Design, IOLTS 2023}} (\bibinfo{year}{2023}).
\newblock
\showISBNx{9798350341355}
\urldef\tempurl%
\url{https://doi.org/10.1109/IOLTS59296.2023.10224862}
\showDOI{\tempurl}


\bibitem[\protect\citeauthoryear{Niu, Deng, Fan, Mo, Chen, et~al\mbox{.}}{Niu et~al\mbox{.}}{2021}]%
        {Niu2021LEDCommunications}
\bibfield{author}{\bibinfo{person}{Yinan Niu}, \bibinfo{person}{Xiong Deng}, \bibinfo{person}{Wenxiang Fan}, \bibinfo{person}{Jundao Mo}, \bibinfo{person}{Chen Chen}, {et~al\mbox{.}}} \bibinfo{year}{2021}\natexlab{}.
\newblock \showarticletitle{{LED Nonlinearity Post-compensator with Legendre polynomials in Visible Light Communications}}.
\newblock \bibinfo{journal}{\emph{Proceedings of the 16th IEEE Conference on Industrial Electronics and Applications, ICIEA 2021}} (\bibinfo{date}{8} \bibinfo{year}{2021}), \bibinfo{pages}{1322--1327}.
\newblock
\showISBNx{9781665422482}
\urldef\tempurl%
\url{https://doi.org/10.1109/ICIEA51954.2021.9516222}
\showDOI{\tempurl}


\bibitem[\protect\citeauthoryear{Nurvitadhi, Sheffield, Sim, Mishra, Venkatesh, et~al\mbox{.}}{Nurvitadhi et~al\mbox{.}}{2016}]%
        {hw-efficiency-compare}
\bibfield{author}{\bibinfo{person}{Eriko Nurvitadhi}, \bibinfo{person}{David Sheffield}, \bibinfo{person}{Jaewoong Sim}, \bibinfo{person}{Asit Mishra}, \bibinfo{person}{Ganesh Venkatesh}, {et~al\mbox{.}}} \bibinfo{year}{2016}\natexlab{}.
\newblock \showarticletitle{Accelerating Binarized Neural Networks: Comparison of FPGA, CPU, GPU, and ASIC}. In \bibinfo{booktitle}{\emph{2016 International Conference on Field-Programmable Technology (FPT)}}. \bibinfo{pages}{77--84}.
\newblock
\urldef\tempurl%
\url{https://doi.org/10.1109/FPT.2016.7929192}
\showDOI{\tempurl}


\bibitem[\protect\citeauthoryear{Ofek, Petrenko, Heeres, Reinhold, Leghtas, et~al\mbox{.}}{Ofek et~al\mbox{.}}{2016}]%
        {Ofek-nat-2016}
\bibfield{author}{\bibinfo{person}{Nissim Ofek}, \bibinfo{person}{Andrei Petrenko}, \bibinfo{person}{Reinier Heeres}, \bibinfo{person}{Philip Reinhold}, \bibinfo{person}{Zaki Leghtas}, {et~al\mbox{.}}} \bibinfo{year}{2016}\natexlab{}.
\newblock \showarticletitle{Extending the lifetime of a quantum bit with error correction in superconducting circuits}.
\newblock \bibinfo{journal}{\emph{Nature}}  \bibinfo{volume}{536} (\bibinfo{year}{2016}), \bibinfo{pages}{441--445}.
\newblock


\bibitem[\protect\citeauthoryear{O'Neal, Liu, Tang, Kalantar, DeRenard, et~al\mbox{.}}{O'Neal et~al\mbox{.}}{2018}]%
        {oneal-2018}
\bibfield{author}{\bibinfo{person}{Kenneth O'Neal}, \bibinfo{person}{Mitch Liu}, \bibinfo{person}{Hans Tang}, \bibinfo{person}{Amin Kalantar}, \bibinfo{person}{Kennen DeRenard}, {et~al\mbox{.}}} \bibinfo{year}{2018}\natexlab{}.
\newblock \showarticletitle{HLSPredict: cross platform performance prediction for FPGA high-level synthesis}. In \bibinfo{booktitle}{\emph{Proceedings of the International Conference on Computer-Aided Design}} \emph{(\bibinfo{series}{ICCAD '18})}. \bibinfo{publisher}{Association for Computing Machinery}, \bibinfo{address}{New York, NY, USA}, Article \bibinfo{articleno}{104}, \bibinfo{numpages}{8}~pages.
\newblock
\showISBNx{9781450359504}
\urldef\tempurl%
\url{https://doi.org/10.1145/3240765.3264635}
\showDOI{\tempurl}


\bibitem[\protect\citeauthoryear{Overwater, Babaie, and Sebastiano}{Overwater et~al\mbox{.}}{2022}]%
        {Overwater-2022}
\bibfield{author}{\bibinfo{person}{Ramon W.~J. Overwater}, \bibinfo{person}{Masoud Babaie}, {and} \bibinfo{person}{Fabio Sebastiano}.} \bibinfo{year}{2022}\natexlab{}.
\newblock \showarticletitle{Neural-Network Decoders for Quantum Error Correction Using Surface Codes: A Space Exploration of the Hardware Cost-Performance Tradeoffs}.
\newblock \bibinfo{journal}{\emph{IEEE Transactions on Quantum Engineering}}  \bibinfo{volume}{3} (\bibinfo{year}{2022}), \bibinfo{pages}{1--19}.
\newblock
\urldef\tempurl%
\url{https://doi.org/10.1109/TQE.2022.3174017}
\showDOI{\tempurl}


\bibitem[\protect\citeauthoryear{Owens, Houston, Luebke, Green, Stone, et~al\mbox{.}}{Owens et~al\mbox{.}}{2008}]%
        {gpuComputing}
\bibfield{author}{\bibinfo{person}{John~D. Owens}, \bibinfo{person}{Mike Houston}, \bibinfo{person}{David Luebke}, \bibinfo{person}{Simon Green}, \bibinfo{person}{John~E. Stone}, {et~al\mbox{.}}} \bibinfo{year}{2008}\natexlab{}.
\newblock \showarticletitle{GPU Computing}.
\newblock \bibinfo{journal}{\emph{Proc. IEEE}} \bibinfo{volume}{96}, \bibinfo{number}{5} (\bibinfo{year}{2008}), \bibinfo{pages}{879--899}.
\newblock
\urldef\tempurl%
\url{https://doi.org/10.1109/JPROC.2008.917757}
\showDOI{\tempurl}


\bibitem[\protect\citeauthoryear{Palumbo, Cassano, Luzzi, Hern{\'{a}}ndez, Reviriego, et~al\mbox{.}}{Palumbo et~al\mbox{.}}{2022}]%
        {Palumbo2022IsAnswer}
\bibfield{author}{\bibinfo{person}{Alessandro Palumbo}, \bibinfo{person}{Luca Cassano}, \bibinfo{person}{Bruno Luzzi}, \bibinfo{person}{José~Alberto Hern{\'{a}}ndez}, \bibinfo{person}{Pedro Reviriego}, {et~al\mbox{.}}} \bibinfo{year}{2022}\natexlab{}.
\newblock \showarticletitle{{Is your FPGA bitstream Hardware Trojan-free? Machine learning can provide an answer}}.
\newblock \bibinfo{journal}{\emph{Journal of Systems Architecture}}  \bibinfo{volume}{128} (\bibinfo{date}{7} \bibinfo{year}{2022}), \bibinfo{pages}{102543}.
\newblock
\showISSN{1383-7621}
\urldef\tempurl%
\url{https://doi.org/10.1016/J.SYSARC.2022.102543}
\showDOI{\tempurl}


\bibitem[\protect\citeauthoryear{Peltenburg, Hadnagy, Brobbel, Morrow, and Al-Ars}{Peltenburg et~al\mbox{.}}{2021a}]%
        {Peltenburg2021TensAccelerators}
\bibfield{author}{\bibinfo{person}{Johan Peltenburg}, \bibinfo{person}{Akos Hadnagy}, \bibinfo{person}{Matthijs Brobbel}, \bibinfo{person}{Robert Morrow}, {and} \bibinfo{person}{Zaid Al-Ars}.} \bibinfo{year}{2021}\natexlab{a}.
\newblock \showarticletitle{{Tens of gigabytes per second JSON-to-Arrow conversion with FPGA accelerators}}.
\newblock \bibinfo{journal}{\emph{2021 International Conference on Field-Programmable Technology, ICFPT 2021}} (\bibinfo{year}{2021}).
\newblock
\showISBNx{9781665420105}
\urldef\tempurl%
\url{https://doi.org/10.1109/ICFPT52863.2021.9609833}
\showDOI{\tempurl}


\bibitem[\protect\citeauthoryear{Peltenburg, Van~Leeuwen, Hoozemans, Fang, Al-Ars, et~al\mbox{.}}{Peltenburg et~al\mbox{.}}{2020a}]%
        {Peltenburg2020BattlingFPGA}
\bibfield{author}{\bibinfo{person}{Johan Peltenburg}, \bibinfo{person}{Lars~T.J. Van~Leeuwen}, \bibinfo{person}{Joost Hoozemans}, \bibinfo{person}{Jian Fang}, \bibinfo{person}{Zaid Al-Ars}, {et~al\mbox{.}}} \bibinfo{year}{2020}\natexlab{a}.
\newblock \showarticletitle{{Battling the CPU Bottleneck in Apache Parquet to Arrow Conversion Using FPGA}}.
\newblock \bibinfo{journal}{\emph{Proceedings - 2020 International Conference on Field-Programmable Technology, ICFPT 2020}} (\bibinfo{date}{12} \bibinfo{year}{2020}), \bibinfo{pages}{281--286}.
\newblock
\showISBNx{9780738105185}
\urldef\tempurl%
\url{https://doi.org/10.1109/ICFPT51103.2020.00048}
\showDOI{\tempurl}


\bibitem[\protect\citeauthoryear{Peltenburg, Van~Straten, Brobbel, Al-Ars, and Hofstee}{Peltenburg et~al\mbox{.}}{2020b}]%
        {Peltenburg2020Tydi:Streams}
\bibfield{author}{\bibinfo{person}{Johan Peltenburg}, \bibinfo{person}{Jeroen Van~Straten}, \bibinfo{person}{Matthijs Brobbel}, \bibinfo{person}{Zaid Al-Ars}, {and} \bibinfo{person}{H.~Peter Hofstee}.} \bibinfo{year}{2020}\natexlab{b}.
\newblock \showarticletitle{{Tydi: An Open Specification for Complex Data Structures over Hardware Streams}}.
\newblock \bibinfo{journal}{\emph{IEEE Micro}} \bibinfo{volume}{40}, \bibinfo{number}{4} (\bibinfo{date}{7} \bibinfo{year}{2020}), \bibinfo{pages}{120--130}.
\newblock
\showISSN{19374143}
\urldef\tempurl%
\url{https://doi.org/10.1109/MM.2020.2996373}
\showDOI{\tempurl}


\bibitem[\protect\citeauthoryear{Peltenburg, van Straten, Brobbel, Al-Ars, and Hofstee}{Peltenburg et~al\mbox{.}}{2021b}]%
        {Peltenburg2021GeneratingArrow}
\bibfield{author}{\bibinfo{person}{Johan Peltenburg}, \bibinfo{person}{Jeroen van Straten}, \bibinfo{person}{Matthijs Brobbel}, \bibinfo{person}{Zaid Al-Ars}, {and} \bibinfo{person}{H.~Peter Hofstee}.} \bibinfo{year}{2021}\natexlab{b}.
\newblock \showarticletitle{{Generating High-Performance FPGA Accelerator Designs for Big Data Analytics with Fletcher and Apache Arrow}}.
\newblock \bibinfo{journal}{\emph{Journal of Signal Processing Systems}} \bibinfo{volume}{93}, \bibinfo{number}{5} (\bibinfo{date}{5} \bibinfo{year}{2021}), \bibinfo{pages}{565--586}.
\newblock
\showISSN{19398115}
\urldef\tempurl%
\url{https://doi.org/10.1007/S11265-021-01650-6/FIGURES/15}
\showDOI{\tempurl}


\bibitem[\protect\citeauthoryear{Peltenburg, Van~Straten, Wijtemans, Van~Leeuwen, Al-Ars, et~al\mbox{.}}{Peltenburg et~al\mbox{.}}{2019}]%
        {Peltenburg2019Fletcher:Arrow}
\bibfield{author}{\bibinfo{person}{Johan Peltenburg}, \bibinfo{person}{Jeroen Van~Straten}, \bibinfo{person}{Lars Wijtemans}, \bibinfo{person}{Lars Van~Leeuwen}, \bibinfo{person}{Zaid Al-Ars}, {et~al\mbox{.}}} \bibinfo{year}{2019}\natexlab{}.
\newblock \showarticletitle{{Fletcher: A framework to efficiently integrate FPGA accelerators with apache arrow}}.
\newblock \bibinfo{journal}{\emph{Proceedings - 29th International Conference on Field-Programmable Logic and Applications, FPL 2019}} (\bibinfo{date}{9} \bibinfo{year}{2019}), \bibinfo{pages}{270--277}.
\newblock
\showISBNx{9781728148847}
\urldef\tempurl%
\url{https://doi.org/10.1109/FPL.2019.00051}
\showDOI{\tempurl}


\bibitem[\protect\citeauthoryear{Pes, Luiken, Corradi, and Frenkel}{Pes et~al\mbox{.}}{2024}]%
        {Pes2024ActiveNetworks}
\bibfield{author}{\bibinfo{person}{Lorenzo Pes}, \bibinfo{person}{Rick Luiken}, \bibinfo{person}{Federico Corradi}, {and} \bibinfo{person}{Charlotte Frenkel}.} \bibinfo{year}{2024}\natexlab{}.
\newblock \showarticletitle{{Active Dendrites Enable Efficient Continual Learning in Time-To-First-Spike Neural Networks}}.
\newblock \bibinfo{journal}{\emph{International Conference on Artificial Intelligence Circuits and Systems (AICAS)}} (\bibinfo{date}{4} \bibinfo{year}{2024}).
\newblock
\urldef\tempurl%
\url{http://arxiv.org/abs/2404.19419}
\showURL{%
\tempurl}


\bibitem[\protect\citeauthoryear{Petrica, Alonso, Kroes, Fraser, Cotofana, et~al\mbox{.}}{Petrica et~al\mbox{.}}{2020}]%
        {ml-mem-efficient-df-inf}
\bibfield{author}{\bibinfo{person}{Lucian Petrica}, \bibinfo{person}{Tobias Alonso}, \bibinfo{person}{Mairin Kroes}, \bibinfo{person}{Nicholas Fraser}, \bibinfo{person}{Sorin Cotofana}, {et~al\mbox{.}}} \bibinfo{year}{2020}\natexlab{}.
\newblock \showarticletitle{Memory-Efficient Dataflow Inference for Deep CNNs on FPGA}. In \bibinfo{booktitle}{\emph{2020 International Conference on Field-Programmable Technology (ICFPT)}}. \bibinfo{pages}{48--55}.
\newblock
\urldef\tempurl%
\url{https://doi.org/10.1109/ICFPT51103.2020.00016}
\showDOI{\tempurl}


\bibitem[\protect\citeauthoryear{Philips, Madzik, Amitonov, de~Snoo, Russ, et~al\mbox{.}}{Philips et~al\mbox{.}}{2022}]%
        {Philips-nat-2022}
\bibfield{author}{\bibinfo{person}{Stephan G.~J. Philips}, \bibinfo{person}{Mateusz~T. Madzik}, \bibinfo{person}{Sergey~V. Amitonov}, \bibinfo{person}{Sander~L. de Snoo}, \bibinfo{person}{Maximilian Russ}, {et~al\mbox{.}}} \bibinfo{year}{2022}\natexlab{}.
\newblock \showarticletitle{Universal control of a six-qubit quantum processor in silicon}.
\newblock \bibinfo{journal}{\emph{Nature}}  \bibinfo{volume}{609} (\bibinfo{year}{2022}), \bibinfo{pages}{919–924}.
\newblock


\bibitem[\protect\citeauthoryear{Pinto, Syrivelis, Gazzetti, Koutsovasilis, Reale, et~al\mbox{.}}{Pinto et~al\mbox{.}}{2020}]%
        {9252003}
\bibfield{author}{\bibinfo{person}{Christian Pinto}, \bibinfo{person}{Dimitris Syrivelis}, \bibinfo{person}{Michele Gazzetti}, \bibinfo{person}{Panos Koutsovasilis}, \bibinfo{person}{Andrea Reale}, {et~al\mbox{.}}} \bibinfo{year}{2020}\natexlab{}.
\newblock \showarticletitle{ThymesisFlow: A Software-Defined, HW/SW co-Designed Interconnect Stack for Rack-Scale Memory Disaggregation}. In \bibinfo{booktitle}{\emph{2020 53rd Annual IEEE/ACM International Symposium on Microarchitecture (MICRO)}}. \bibinfo{pages}{868--880}.
\newblock
\urldef\tempurl%
\url{https://doi.org/10.1109/MICRO50266.2020.00075}
\showDOI{\tempurl}


\bibitem[\protect\citeauthoryear{Plimpton}{Plimpton}{1995}]%
        {plimpton1995fast}
\bibfield{author}{\bibinfo{person}{Steve Plimpton}.} \bibinfo{year}{1995}\natexlab{}.
\newblock \showarticletitle{Fast parallel algorithms for short-range molecular dynamics}.
\newblock \bibinfo{journal}{\emph{Journal of computational physics}} \bibinfo{volume}{117}, \bibinfo{number}{1} (\bibinfo{year}{1995}), \bibinfo{pages}{1--19}.
\newblock


\bibitem[\protect\citeauthoryear{Portaluri, Azimi, De~Sio, Sterpone, and Codinachs}{Portaluri et~al\mbox{.}}{2022}]%
        {Portaluri2022Radiation-inducedDevices}
\bibfield{author}{\bibinfo{person}{Andrea Portaluri}, \bibinfo{person}{Sarah Azimi}, \bibinfo{person}{Corrado De~Sio}, \bibinfo{person}{Luca Sterpone}, {and} \bibinfo{person}{David~Merodio Codinachs}.} \bibinfo{year}{2022}\natexlab{}.
\newblock \showarticletitle{{Radiation-induced Effects on DMA Data Transfer in Reconfigurable Devices}}.
\newblock \bibinfo{journal}{\emph{Proceedings - 2022 IEEE 28th International Symposium on On-Line Testing and Robust System Design, IOLTS 2022}} (\bibinfo{year}{2022}).
\newblock
\showISBNx{9781665473552}
\urldef\tempurl%
\url{https://doi.org/10.1109/IOLTS56730.2022.9897262}
\showDOI{\tempurl}


\bibitem[\protect\citeauthoryear{Prodan, Kimovski, Bartolini, Cochez, Iosup, et~al\mbox{.}}{Prodan et~al\mbox{.}}{2022}]%
        {Prodan2022TowardsEurope}
\bibfield{author}{\bibinfo{person}{Radu Prodan}, \bibinfo{person}{Dragi Kimovski}, \bibinfo{person}{Andrea Bartolini}, \bibinfo{person}{Michael Cochez}, \bibinfo{person}{Alexandru Iosup}, {et~al\mbox{.}}} \bibinfo{year}{2022}\natexlab{}.
\newblock \showarticletitle{{Towards Extreme and Sustainable Graph Processing for Urgent Societal Challenges in Europe}}.
\newblock \bibinfo{journal}{\emph{Proceedings - 2022 IEEE Cloud Summit, Cloud Summit 2022}} (\bibinfo{year}{2022}), \bibinfo{pages}{23--30}.
\newblock
\showISBNx{9781665451130}
\urldef\tempurl%
\url{https://doi.org/10.1109/CLOUDSUMMIT54781.2022.00010}
\showDOI{\tempurl}


\bibitem[\protect\citeauthoryear{Putnam, Caulfield, Chung, Chiou, Constantinides, et~al\mbox{.}}{Putnam et~al\mbox{.}}{[n. d.]}]%
        {PutnamAServices}
\bibfield{author}{\bibinfo{person}{Andrew Putnam}, \bibinfo{person}{Adrian~M Caulfield}, \bibinfo{person}{Eric~S Chung}, \bibinfo{person}{Derek Chiou}, \bibinfo{person}{Kypros Constantinides}, {et~al\mbox{.}}} \bibinfo{year}{[n. d.]}\natexlab{}.
\newblock \bibinfo{booktitle}{\emph{{A Reconfigurable Fabric for Accelerating Large-Scale Datacenter Services}}}.
\newblock \bibinfo{type}{{T}echnical {R}eport}.
\newblock


\bibitem[\protect\citeauthoryear{Rapuano, Meoni, Pacini, Dinelli, Furano, et~al\mbox{.}}{Rapuano et~al\mbox{.}}{2021}]%
        {ml-myriad-2-space-cnn}
\bibfield{author}{\bibinfo{person}{Emilio Rapuano}, \bibinfo{person}{Gabriele Meoni}, \bibinfo{person}{Tommaso Pacini}, \bibinfo{person}{Gianmarco Dinelli}, \bibinfo{person}{Gianluca Furano}, {et~al\mbox{.}}} \bibinfo{year}{2021}\natexlab{}.
\newblock \showarticletitle{An FPGA-Based Hardware Accelerator for CNNs Inference on Board Satellites: Benchmarking with Myriad 2-Based Solution for the CloudScout Case Study}.
\newblock \bibinfo{journal}{\emph{Remote Sensing}} \bibinfo{volume}{13}, \bibinfo{number}{8} (\bibinfo{year}{2021}).
\newblock
\showISSN{2072-4292}
\urldef\tempurl%
\url{https://doi.org/10.3390/rs13081518}
\showDOI{\tempurl}


\bibitem[\protect\citeauthoryear{Rellermeyer, Omranian~Khorasani, Graur, and Parthasarathy}{Rellermeyer et~al\mbox{.}}{2019}]%
        {Rellermeyer2019TheProcessing}
\bibfield{author}{\bibinfo{person}{Jan~S. Rellermeyer}, \bibinfo{person}{Sobhan Omranian~Khorasani}, \bibinfo{person}{Dan Graur}, {and} \bibinfo{person}{Apourva Parthasarathy}.} \bibinfo{year}{2019}\natexlab{}.
\newblock \showarticletitle{{The coming age of pervasive data processing}}.
\newblock \bibinfo{journal}{\emph{Proceedings - 2019 18th International Symposium on Parallel and Distributed Computing, ISPDC 2019}} (\bibinfo{date}{6} \bibinfo{year}{2019}), \bibinfo{pages}{58--65}.
\newblock
\showISBNx{9781728138008}
\urldef\tempurl%
\url{https://doi.org/10.1109/ISPDC.2019.00011}
\showDOI{\tempurl}


\bibitem[\protect\citeauthoryear{Reukers, Tian, Al-Ars, Hofstee, Brobbel, et~al\mbox{.}}{Reukers et~al\mbox{.}}{2023}]%
        {Reukers2023AnIR}
\bibfield{author}{\bibinfo{person}{Matthijs~A. Reukers}, \bibinfo{person}{Yongding Tian}, \bibinfo{person}{Zaid Al-Ars}, \bibinfo{person}{Peter Hofstee}, \bibinfo{person}{Matthijs Brobbel}, {et~al\mbox{.}}} \bibinfo{year}{2023}\natexlab{}.
\newblock \showarticletitle{An Intermediate Representation for Composable Typed Streaming Dataflow Designs}.
\newblock \bibinfo{journal}{\emph{ArXiv}}  \bibinfo{volume}{abs/2308.13436} (\bibinfo{year}{2023}).
\newblock
\urldef\tempurl%
\url{https://api.semanticscholar.org/CorpusID:261125186}
\showURL{%
\tempurl}


\bibitem[\protect\citeauthoryear{Ribot~Gonzalez and Nelissen}{Ribot~Gonzalez and Nelissen}{2020}]%
        {RibotGonzalez2020HopliteRT:FPGA}
\bibfield{author}{\bibinfo{person}{Yilian Ribot~Gonzalez} {and} \bibinfo{person}{Geoffrey Nelissen}.} \bibinfo{year}{2020}\natexlab{}.
\newblock \showarticletitle{{HopliteRT*: Real-Time NoC for FPGA}}.
\newblock \bibinfo{journal}{\emph{IEEE Transactions on Computer-Aided Design of Integrated Circuits and Systems}} \bibinfo{volume}{39}, \bibinfo{number}{11} (\bibinfo{date}{11} \bibinfo{year}{2020}), \bibinfo{pages}{3650--3661}.
\newblock
\showISSN{19374151}
\urldef\tempurl%
\url{https://doi.org/10.1109/TCAD.2020.3012748}
\showDOI{\tempurl}


\bibitem[\protect\citeauthoryear{Ribot~Gonz\'{a}lez, Nelissen, and Tovar}{Ribot~Gonz\'{a}lez et~al\mbox{.}}{2021}]%
        {nDimNoC}
\bibfield{author}{\bibinfo{person}{Yilian Ribot~Gonz\'{a}lez}, \bibinfo{person}{Geoffrey Nelissen}, {and} \bibinfo{person}{Eduardo Tovar}.} \bibinfo{year}{2021}\natexlab{}.
\newblock \showarticletitle{{nDimNoC: Real-Time D-dimensional NoC}}. In \bibinfo{booktitle}{\emph{33rd Euromicro Conference on Real-Time Systems (ECRTS 2021)}} \emph{(\bibinfo{series}{Leibniz International Proceedings in Informatics (LIPIcs)})}, \bibfield{editor}{\bibinfo{person}{Bj\"{o}rn~B. Brandenburg}} (Ed.), Vol.~\bibinfo{volume}{196}. \bibinfo{publisher}{Schloss Dagstuhl -- Leibniz-Zentrum f{\"u}r Informatik}, \bibinfo{address}{Dagstuhl, Germany}, \bibinfo{pages}{5:1--5:22}.
\newblock
\showISBNx{978-3-95977-192-4}
\showISSN{1868-8969}
\urldef\tempurl%
\url{https://doi.org/10.4230/LIPIcs.ECRTS.2021.5}
\showDOI{\tempurl}


\bibitem[\protect\citeauthoryear{Ribot~González and Nelissen}{Ribot~González and Nelissen}{2020}]%
        {HopliteRT*}
\bibfield{author}{\bibinfo{person}{Yilian Ribot~González} {and} \bibinfo{person}{Geoffrey Nelissen}.} \bibinfo{year}{2020}\natexlab{}.
\newblock \showarticletitle{HopliteRT*: Real-Time NoC for FPGA}.
\newblock \bibinfo{journal}{\emph{IEEE Transactions on Computer-Aided Design of Integrated Circuits and Systems}} \bibinfo{volume}{39}, \bibinfo{number}{11} (\bibinfo{year}{2020}), \bibinfo{pages}{3650--3661}.
\newblock
\urldef\tempurl%
\url{https://doi.org/10.1109/TCAD.2020.3012748}
\showDOI{\tempurl}


\bibitem[\protect\citeauthoryear{Rocha, Biswas, Verhoef, Bampi, Van~Hoof, et~al\mbox{.}}{Rocha et~al\mbox{.}}{2020}]%
        {Rocha2020BinaryWrist-PPG}
\bibfield{author}{\bibinfo{person}{Leandro~Giacomini Rocha}, \bibinfo{person}{Dwaipayan Biswas}, \bibinfo{person}{Bram~Ernst Verhoef}, \bibinfo{person}{Sergio Bampi}, \bibinfo{person}{Chris Van~Hoof}, {et~al\mbox{.}}} \bibinfo{year}{2020}\natexlab{}.
\newblock \showarticletitle{{Binary CorNET: Accelerator for HR Estimation from Wrist-PPG}}.
\newblock \bibinfo{journal}{\emph{IEEE Transactions on Biomedical Circuits and Systems}} \bibinfo{volume}{14}, \bibinfo{number}{4} (\bibinfo{date}{8} \bibinfo{year}{2020}), \bibinfo{pages}{715--726}.
\newblock
\showISSN{19409990}
\urldef\tempurl%
\url{https://doi.org/10.1109/TBCAS.2020.3001675}
\showDOI{\tempurl}


\bibitem[\protect\citeauthoryear{Rodriguez, Santos, Sarmiento, and Torre}{Rodriguez et~al\mbox{.}}{2019}]%
        {Rodriguez2019ScalableCompression}
\bibfield{author}{\bibinfo{person}{Alfonso Rodriguez}, \bibinfo{person}{Lucana Santos}, \bibinfo{person}{Roberto Sarmiento}, {and} \bibinfo{person}{Eduardo De~La Torre}.} \bibinfo{year}{2019}\natexlab{}.
\newblock \showarticletitle{{Scalable hardware-based on-board processing for run-time adaptive lossless hyperspectral compression}}.
\newblock \bibinfo{journal}{\emph{IEEE Access}}  \bibinfo{volume}{7} (\bibinfo{year}{2019}), \bibinfo{pages}{10644--10652}.
\newblock
\showISSN{21693536}
\urldef\tempurl%
\url{https://doi.org/10.1109/ACCESS.2019.2892308}
\showDOI{\tempurl}


\bibitem[\protect\citeauthoryear{Rowe, Tapia, Barry, Karkare, Papageorgiou, et~al\mbox{.}}{Rowe et~al\mbox{.}}{2023}]%
        {Rowe2023ThePerformance}
\bibfield{author}{\bibinfo{person}{S. Rowe}, \bibinfo{person}{M. Tapia}, \bibinfo{person}{P.~S. Barry}, \bibinfo{person}{K.~S. Karkare}, \bibinfo{person}{A. Papageorgiou}, {et~al\mbox{.}}} \bibinfo{year}{2023}\natexlab{}.
\newblock \showarticletitle{{The MUSCAT Readout Electronics Backend: Design and Pre-deployment Performance}}.
\newblock \bibinfo{journal}{\emph{Journal of Low Temperature Physics}} \bibinfo{volume}{211}, \bibinfo{number}{5-6} (\bibinfo{date}{6} \bibinfo{year}{2023}), \bibinfo{pages}{289--301}.
\newblock
\showISSN{15737357}
\urldef\tempurl%
\url{https://doi.org/10.1007/S10909-022-02868-9/FIGURES/4}
\showDOI{\tempurl}


\bibitem[\protect\citeauthoryear{Rubattu, Errico, and Marinis}{Rubattu et~al\mbox{.}}{[n. d.]}]%
        {RubattuASystems}
\bibfield{author}{\bibinfo{person}{Claudio Rubattu}, \bibinfo{person}{Walter Errico}, {and} \bibinfo{person}{Kostas Marinis}.} \bibinfo{year}{[n. d.]}\natexlab{}.
\newblock \showarticletitle{{A low-overhead AXI bridge on SpaceWire for multi-device spacecraft systems}}.
\newblock  (\bibinfo{year}{[n. d.]}).
\newblock


\bibitem[\protect\citeauthoryear{Sadasivam, Thompto, Kalla, and Starke}{Sadasivam et~al\mbox{.}}{2017}]%
        {7924241}
\bibfield{author}{\bibinfo{person}{Satish~Kumar Sadasivam}, \bibinfo{person}{Brian~W. Thompto}, \bibinfo{person}{Ron Kalla}, {and} \bibinfo{person}{William~J. Starke}.} \bibinfo{year}{2017}\natexlab{}.
\newblock \showarticletitle{IBM Power9 Processor Architecture}.
\newblock \bibinfo{journal}{\emph{IEEE Micro}} \bibinfo{volume}{37}, \bibinfo{number}{2} (\bibinfo{year}{2017}), \bibinfo{pages}{40--51}.
\newblock
\urldef\tempurl%
\url{https://doi.org/10.1109/MM.2017.40}
\showDOI{\tempurl}


\bibitem[\protect\citeauthoryear{Sahebi, Barbone, Procaccini, Luk, Gaydadjiev, et~al\mbox{.}}{Sahebi et~al\mbox{.}}{2023}]%
        {Sahebi2023DistributedFPGAs}
\bibfield{author}{\bibinfo{person}{Amin Sahebi}, \bibinfo{person}{Marco Barbone}, \bibinfo{person}{Marco Procaccini}, \bibinfo{person}{Wayne Luk}, \bibinfo{person}{Georgi Gaydadjiev}, {et~al\mbox{.}}} \bibinfo{year}{2023}\natexlab{}.
\newblock \showarticletitle{{Distributed large-scale graph processing on FPGAs}}.
\newblock \bibinfo{journal}{\emph{Journal of Big Data}} \bibinfo{volume}{10}, \bibinfo{number}{1} (\bibinfo{date}{12} \bibinfo{year}{2023}), \bibinfo{pages}{1--28}.
\newblock
\showISSN{21961115}
\urldef\tempurl%
\url{https://doi.org/10.1186/S40537-023-00756-X/FIGURES/10}
\showDOI{\tempurl}


\bibitem[\protect\citeauthoryear{Salazar-Garcia, Chacon-Rodriguez, Rfmolo-Donadfo, Garcia-Ramirez, Solorzano-Pacheco, et~al\mbox{.}}{Salazar-Garcia et~al\mbox{.}}{2022}]%
        {Salazar-Garcia2022AApplications}
\bibfield{author}{\bibinfo{person}{Carlos Salazar-Garcia}, \bibinfo{person}{Alfonso Chacon-Rodriguez}, \bibinfo{person}{Renato Rfmolo-Donadfo}, \bibinfo{person}{Ronny Garcia-Ramirez}, \bibinfo{person}{David Solorzano-Pacheco}, {et~al\mbox{.}}} \bibinfo{year}{2022}\natexlab{}.
\newblock \showarticletitle{{A custom interconnection multi-FPGA framework for distributed processing applications}}.
\newblock \bibinfo{journal}{\emph{35th SBC/SBMicro/IEEE/ACM Symposium on Integrated Circuits and Systems Design, SBCCI 2022 - Proceedings}} (\bibinfo{year}{2022}).
\newblock
\showISBNx{9781665481281}
\urldef\tempurl%
\url{https://doi.org/10.1109/SBCCI55532.2022.9893238}
\showDOI{\tempurl}


\bibitem[\protect\citeauthoryear{Salazar-Garc{\'{i}}a, Garc{\'{i}}a-Ram{\'{i}}rez, R{\'{i}}molo-Donad{\'{i}}o, Strydis, and Chac{\'{o}}n-Rodr{\'{i}}guez}{Salazar-Garc{\'{i}}a et~al\mbox{.}}{2021}]%
        {Salazar-Garcia2021PlasticNet+:Transceivers}
\bibfield{author}{\bibinfo{person}{Carlos Salazar-Garc{\'{i}}a}, \bibinfo{person}{Ronny Garc{\'{i}}a-Ram{\'{i}}rez}, \bibinfo{person}{Renato R{\'{i}}molo-Donad{\'{i}}o}, \bibinfo{person}{Christos Strydis}, {and} \bibinfo{person}{Alfonso Chac{\'{o}}n-Rodr{\'{i}}guez}.} \bibinfo{year}{2021}\natexlab{}.
\newblock \showarticletitle{{PlasticNet+: Extending multi-FPGA interconnect architecture via Gigabit transceivers}}.
\newblock \bibinfo{journal}{\emph{Proceedings - IEEE International Symposium on Circuits and Systems}}  \bibinfo{volume}{2021-May} (\bibinfo{year}{2021}).
\newblock
\showISBNx{9781728192017}
\showISSN{02714310}
\urldef\tempurl%
\url{https://doi.org/10.1109/ISCAS51556.2021.9401058}
\showDOI{\tempurl}


\bibitem[\protect\citeauthoryear{Salazar-Garc{\'\i}a, Gonz{\'a}lez-G{\'o}mez, Alfaro-Badilla, Garc{\'\i}a-Ram{\'\i}rez, R{\'\i}molo-Donad{\'\i}o, et~al\mbox{.}}{Salazar-Garc{\'\i}a et~al\mbox{.}}{2020}]%
        {salazar2020plasticnet}
\bibfield{author}{\bibinfo{person}{Carlos Salazar-Garc{\'\i}a}, \bibinfo{person}{Jeferson Gonz{\'a}lez-G{\'o}mez}, \bibinfo{person}{Kaleb Alfaro-Badilla}, \bibinfo{person}{Ronny Garc{\'\i}a-Ram{\'\i}rez}, \bibinfo{person}{Renato R{\'\i}molo-Donad{\'\i}o}, {et~al\mbox{.}}} \bibinfo{year}{2020}\natexlab{}.
\newblock \showarticletitle{PlasticNet: A low latency flexible network architecture for interconnected multi-FPGA systems}. In \bibinfo{booktitle}{\emph{2020 IEEE 3rd Conference on PhD Research in Microelectronics and Electronics in Latin America (PRIME-LA)}}. IEEE, \bibinfo{pages}{1--4}.
\newblock


\bibitem[\protect\citeauthoryear{Sanchez, Barrios, Ventura, Santos, and Sarmiento}{Sanchez et~al\mbox{.}}{2022}]%
        {Sanchez2022AbetoManagement}
\bibfield{author}{\bibinfo{person}{Antonio~J. Sanchez}, \bibinfo{person}{Yubal Barrios}, \bibinfo{person}{Diego Ventura}, \bibinfo{person}{Lucana Santos}, {and} \bibinfo{person}{Roberto Sarmiento}.} \bibinfo{year}{2022}\natexlab{}.
\newblock \showarticletitle{{Abeto framework: a Solution for Heterogeneous IP Management}}.
\newblock \bibinfo{journal}{\emph{Proceedings - 2022 25th Euromicro Conference on Digital System Design, DSD 2022}} (\bibinfo{year}{2022}), \bibinfo{pages}{720--725}.
\newblock
\showISBNx{9781665474047}
\urldef\tempurl%
\url{https://doi.org/10.1109/DSD57027.2022.00101}
\showDOI{\tempurl}


\bibitem[\protect\citeauthoryear{Sanchez-Garrido, Jurado, Jimenez-Lopez, Balzer, Prokoph, et~al\mbox{.}}{Sanchez-Garrido et~al\mbox{.}}{2021}]%
        {Sanchez-Garrido2021AArray}
\bibfield{author}{\bibinfo{person}{Jorge Sanchez-Garrido}, \bibinfo{person}{Antonio Jurado}, \bibinfo{person}{Miguel Jimenez-Lopez}, \bibinfo{person}{Arnim Balzer}, \bibinfo{person}{Heike Prokoph}, {et~al\mbox{.}}} \bibinfo{year}{2021}\natexlab{}.
\newblock \showarticletitle{{A White Rabbit-Synchronized Accurate Time-Stamping Solution for the Small-Sized Cameras of the Cherenkov Telescope Array}}.
\newblock \bibinfo{journal}{\emph{IEEE Transactions on Instrumentation and Measurement}}  \bibinfo{volume}{70} (\bibinfo{year}{2021}).
\newblock
\showISSN{15579662}
\urldef\tempurl%
\url{https://doi.org/10.1109/TIM.2020.3013343}
\showDOI{\tempurl}


\bibitem[\protect\citeauthoryear{Sankaran, Detterer, Nl, Kannan, Alachiotis, et~al\mbox{.}}{Sankaran et~al\mbox{.}}{[n. d.]}]%
        {SankaranAnInference}
\bibfield{author}{\bibinfo{person}{Anand Sankaran}, \bibinfo{person}{Paul Detterer}, \bibinfo{person}{Detterer@imec Nl}, \bibinfo{person}{Kalpana Kannan}, \bibinfo{person}{Nikolaos Alachiotis}, {et~al\mbox{.}}} \bibinfo{year}{[n. d.]}\natexlab{}.
\newblock \showarticletitle{{An Event-driven Recurrent Spiking Neural Network Architecture for Efficient Inference on FPGA Architecture for Efficient Inference}}.
\newblock  (\bibinfo{year}{[n. d.]}).
\newblock
\showISBNx{9781450397896}
\urldef\tempurl%
\url{https://doi.org/10.1145/3546790.3546802}
\showDOI{\tempurl}


\bibitem[\protect\citeauthoryear{Sano, Koshiba, Miyajima, and Ueno}{Sano et~al\mbox{.}}{2023}]%
        {Sano2023ESSPER:Fugaku}
\bibfield{author}{\bibinfo{person}{Kentaro Sano}, \bibinfo{person}{Atsushi Koshiba}, \bibinfo{person}{Takaaki Miyajima}, {and} \bibinfo{person}{Tomohiro Ueno}.} \bibinfo{year}{2023}\natexlab{}.
\newblock \showarticletitle{{ESSPER: Elastic and Scalable FPGA-Cluster System for High-Performance Reconfigurable Computing with Supercomputer Fugaku}}. In \bibinfo{booktitle}{\emph{ACM International Conference Proceeding Series}}. \bibinfo{publisher}{Association for Computing Machinery}, \bibinfo{pages}{140--150}.
\newblock
\showISBNx{9781450398060}
\urldef\tempurl%
\url{https://doi.org/10.1145/3578178.3579341}
\showDOI{\tempurl}


\bibitem[\protect\citeauthoryear{Santana, Mefleh, and Calabretta}{Santana et~al\mbox{.}}{2023}]%
        {Santana2023SOA-BasedApplications}
\bibfield{author}{\bibinfo{person}{Henrique Santana}, \bibinfo{person}{Ali Mefleh}, {and} \bibinfo{person}{Nicola Calabretta}.} \bibinfo{year}{2023}\natexlab{}.
\newblock \showarticletitle{{SOA-Based Optical Networks with Sub-Microsecond Control Plane for Low-Latency Applications}}.
\newblock \bibinfo{journal}{\emph{International Conference on Transparent Optical Networks}}  \bibinfo{volume}{2023-July} (\bibinfo{year}{2023}).
\newblock
\showISBNx{9798350303032}
\showISSN{21627339}
\urldef\tempurl%
\url{https://doi.org/10.1109/ICTON59386.2023.10207247}
\showDOI{\tempurl}


\bibitem[\protect\citeauthoryear{Santana, Pan, Kraemer, Prifti, Mefleh, et~al\mbox{.}}{Santana et~al\mbox{.}}{2022}]%
        {Santana2022TransparentApplications}
\bibfield{author}{\bibinfo{person}{Henrique Santana}, \bibinfo{person}{Bitao Pan}, \bibinfo{person}{Rafael Kraemer}, \bibinfo{person}{Krisitf Prifti}, \bibinfo{person}{Ali Mefleh}, {et~al\mbox{.}}} \bibinfo{year}{2022}\natexlab{}.
\newblock \showarticletitle{{Transparent and Fast Reconfigurable Optical Network with Edge Computing Nodes for Beyond 5G applications}}.
\newblock \bibinfo{journal}{\emph{2022 13th International Symposium on Communication Systems, Networks and Digital Signal Processing, CSNDSP 2022}} (\bibinfo{year}{2022}), \bibinfo{pages}{820--825}.
\newblock
\showISBNx{9781665410441}
\urldef\tempurl%
\url{https://doi.org/10.1109/CSNDSP54353.2022.9907904}
\showDOI{\tempurl}


\bibitem[\protect\citeauthoryear{Schmidt and Hildebrandt}{Schmidt and Hildebrandt}{2017}]%
        {genetic-big-data}
\bibfield{author}{\bibinfo{person}{Bertil Schmidt} {and} \bibinfo{person}{Andreas Hildebrandt}.} \bibinfo{year}{2017}\natexlab{}.
\newblock \showarticletitle{Next-generation sequencing: big data meets high performance computing}.
\newblock \bibinfo{journal}{\emph{Drug Discovery Today}} \bibinfo{volume}{22}, \bibinfo{number}{4} (\bibinfo{year}{2017}), \bibinfo{pages}{712--717}.
\newblock
\showISSN{1359-6446}
\urldef\tempurl%
\url{https://doi.org/10.1016/j.drudis.2017.01.014}
\showDOI{\tempurl}


\bibitem[\protect\citeauthoryear{Schoonderbeek, Hut, Kooistra, {Mika}, Pepping, et~al\mbox{.}}{Schoonderbeek et~al\mbox{.}}{2020}]%
        {Schoonderbeek2020ImplementationBeam-Weights}
\bibfield{author}{\bibinfo{person}{Gijs Schoonderbeek}, \bibinfo{person}{Boudewijn Hut}, \bibinfo{person}{E. Kooistra}, \bibinfo{person}{{Mika}}, \bibinfo{person}{H.~J. Pepping}, {et~al\mbox{.}}} \bibinfo{year}{2020}\natexlab{}.
\newblock \showarticletitle{{Implementation of a Correlator onto a Hardware Beam-Former to Calculate Beam-Weights}}.
\newblock \bibinfo{journal}{\emph{Journal of Astronomical Instrumentation}} \bibinfo{volume}{9}, \bibinfo{number}{2} (\bibinfo{date}{6} \bibinfo{year}{2020}).
\newblock
\showISSN{22511725}
\urldef\tempurl%
\url{https://doi.org/10.1142/S2251171720500075/ASSET/IMAGES/LARGE/S2251171720500075FIGF12.JPEG}
\showDOI{\tempurl}


\bibitem[\protect\citeauthoryear{Schoonderbeek, Szomoru, Gunst, Hiemstra, and Hargreaves}{Schoonderbeek et~al\mbox{.}}{2019}]%
        {doi:10.1142/S225117171950003X}
\bibfield{author}{\bibinfo{person}{G.~W. Schoonderbeek}, \bibinfo{person}{A. Szomoru}, \bibinfo{person}{A.~W. Gunst}, \bibinfo{person}{L. Hiemstra}, {and} \bibinfo{person}{J. Hargreaves}.} \bibinfo{year}{2019}\natexlab{}.
\newblock \showarticletitle{UniBoard2, A Generic Scalable High-Performance Computing Platform for Radio Astronomy}.
\newblock \bibinfo{journal}{\emph{Journal of Astronomical Instrumentation}} \bibinfo{volume}{08}, \bibinfo{number}{02} (\bibinfo{year}{2019}), \bibinfo{pages}{1950003}.
\newblock
\urldef\tempurl%
\url{https://doi.org/10.1142/S225117171950003X}
\showDOI{\tempurl}
\showeprint{https://doi.org/10.1142/S225117171950003X}


\bibitem[\protect\citeauthoryear{Shahroodi, Zahedi, Singh, Wong, and Hamdioui}{Shahroodi et~al\mbox{.}}{2022}]%
        {Shahroodi2022KrakenOnMem:Profiling}
\bibfield{author}{\bibinfo{person}{Taha Shahroodi}, \bibinfo{person}{Mahdi Zahedi}, \bibinfo{person}{Abhairaj Singh}, \bibinfo{person}{Stephan Wong}, {and} \bibinfo{person}{Said Hamdioui}.} \bibinfo{year}{2022}\natexlab{}.
\newblock \showarticletitle{{KrakenOnMem: A Memristor-Augmented HW/SW Framework for Taxonomic Profiling}}.
\newblock \bibinfo{journal}{\emph{Proceedings of the International Conference on Supercomputing}} \bibinfo{volume}{14}, \bibinfo{number}{22} (\bibinfo{date}{6} \bibinfo{year}{2022}), \bibinfo{pages}{2022}.
\newblock
\showISBNx{9781450392815}
\urldef\tempurl%
\url{https://doi.org/10.1145/3524059.3532367}
\showDOI{\tempurl}


\bibitem[\protect\citeauthoryear{Sidiropoulos, Chatzikonstantis, Soudris, and Strydis}{Sidiropoulos et~al\mbox{.}}{2018}]%
        {Sidiropoulos2018vineyard}
\bibfield{author}{\bibinfo{person}{Harry Sidiropoulos}, \bibinfo{person}{George Chatzikonstantis}, \bibinfo{person}{Dimitrios Soudris}, {and} \bibinfo{person}{Christos Strydis}.} \bibinfo{year}{2018}\natexlab{}.
\newblock \showarticletitle{The VINEYARD Framework for Heterogeneous Cloud Applications: The BrainFrame Case}.
\newblock \bibinfo{journal}{\emph{Conference on Design and Architectures for Signal and Image Processing, DASIP}}  \bibinfo{volume}{2018-October}, \bibinfo{pages}{70--75}.
\newblock
\showISBNx{9781538682371}
\showISSN{21649766}
\urldef\tempurl%
\url{https://doi.org/10.1109/DASIP.2018.8597119}
\showDOI{\tempurl}


\bibitem[\protect\citeauthoryear{Singh, Alser, Cali, DIamantopoulos, Gomez-Luna, et~al\mbox{.}}{Singh et~al\mbox{.}}{2021a}]%
        {Singh2021FPGA-BasedApplications}
\bibfield{author}{\bibinfo{person}{Gagandeep Singh}, \bibinfo{person}{Mohammed Alser}, \bibinfo{person}{Damla~Senol Cali}, \bibinfo{person}{DIonysios DIamantopoulos}, \bibinfo{person}{Juan Gomez-Luna}, {et~al\mbox{.}}} \bibinfo{year}{2021}\natexlab{a}.
\newblock \showarticletitle{{FPGA-Based Near-Memory Acceleration of Modern Data-Intensive Applications}}.
\newblock \bibinfo{journal}{\emph{IEEE Micro}} \bibinfo{volume}{41}, \bibinfo{number}{4} (\bibinfo{date}{7} \bibinfo{year}{2021}), \bibinfo{pages}{39--48}.
\newblock
\showISSN{19374143}
\urldef\tempurl%
\url{https://doi.org/10.1109/MM.2021.3088396}
\showDOI{\tempurl}


\bibitem[\protect\citeauthoryear{Singh, Chelini, Corda, Awan, Stuijk, et~al\mbox{.}}{Singh et~al\mbox{.}}{2019}]%
        {Singh2019Near-memoryFuture}
\bibfield{author}{\bibinfo{person}{Gagandeep Singh}, \bibinfo{person}{Lorenzo Chelini}, \bibinfo{person}{Stefano Corda}, \bibinfo{person}{Ahsan~Javed Awan}, \bibinfo{person}{Sander Stuijk}, {et~al\mbox{.}}} \bibinfo{year}{2019}\natexlab{}.
\newblock \showarticletitle{{Near-memory computing: Past, present, and future}}.
\newblock \bibinfo{journal}{\emph{Microprocessors and Microsystems}}  \bibinfo{volume}{71} (\bibinfo{date}{11} \bibinfo{year}{2019}), \bibinfo{pages}{102868}.
\newblock
\showISSN{0141-9331}
\urldef\tempurl%
\url{https://doi.org/10.1016/J.MICPRO.2019.102868}
\showDOI{\tempurl}


\bibitem[\protect\citeauthoryear{Singh, Diamantopolous, G{\'{o}}mez-Luna, Stuijk, Mutlu, et~al\mbox{.}}{Singh et~al\mbox{.}}{2021b}]%
        {Singh2021ModelingLearning}
\bibfield{author}{\bibinfo{person}{Gagandeep Singh}, \bibinfo{person}{Dionysios Diamantopolous}, \bibinfo{person}{Juan G{\'{o}}mez-Luna}, \bibinfo{person}{Sander Stuijk}, \bibinfo{person}{Onur Mutlu}, {et~al\mbox{.}}} \bibinfo{year}{2021}\natexlab{b}.
\newblock \showarticletitle{{Modeling FPGA-Based Systems via Few-Shot Learning}}.
\newblock  (\bibinfo{date}{2} \bibinfo{year}{2021}), \bibinfo{pages}{146--146}.
\newblock
\urldef\tempurl%
\url{https://doi.org/10.1145/3431920.3439460}
\showDOI{\tempurl}


\bibitem[\protect\citeauthoryear{Singh, Diamantopoulos, Hagleitner, G{\'{o}}mez-Luna, Stuijk, et~al\mbox{.}}{Singh et~al\mbox{.}}{2020}]%
        {Singh2020NERO:Modeling}
\bibfield{author}{\bibinfo{person}{Gagandeep Singh}, \bibinfo{person}{Dionysios Diamantopoulos}, \bibinfo{person}{Christoph Hagleitner}, \bibinfo{person}{Juan G{\'{o}}mez-Luna}, \bibinfo{person}{Sander Stuijk}, {et~al\mbox{.}}} \bibinfo{year}{2020}\natexlab{}.
\newblock \showarticletitle{{NERO: A near High-Bandwidth Memory Stencil Accelerator for Weather Prediction Modeling}}.
\newblock \bibinfo{journal}{\emph{Proceedings - 30th International Conference on Field-Programmable Logic and Applications, FPL 2020}} (\bibinfo{date}{8} \bibinfo{year}{2020}), \bibinfo{pages}{9--17}.
\newblock
\showISBNx{9781728199023}
\urldef\tempurl%
\url{https://doi.org/10.1109/FPL50879.2020.00014}
\showDOI{\tempurl}


\bibitem[\protect\citeauthoryear{Singh, Z{\"{u}}rich, G{\'{o}}mez-luna, Christoph~Hagleitner, Diamantopoulos, et~al\mbox{.}}{Singh et~al\mbox{.}}{2022}]%
        {Singh2022AcceleratingFabric}
\bibfield{author}{\bibinfo{person}{Gagandeep Singh}, \bibinfo{person}{Eth Z{\"{u}}rich}, \bibinfo{person}{Juan G{\'{o}}mez-luna}, \bibinfo{person}{Switzerland Christoph~Hagleitner}, \bibinfo{person}{D Diamantopoulos}, {et~al\mbox{.}}} \bibinfo{year}{2022}\natexlab{}.
\newblock \showarticletitle{{Accelerating Weather Prediction Using Near-Memory Reconfigurable Fabric}}.
\newblock \bibinfo{journal}{\emph{ACM Transactions on Reconfigurable Technology and Systems (TRETS)}} \bibinfo{volume}{15}, \bibinfo{number}{4} (\bibinfo{date}{6} \bibinfo{year}{2022}).
\newblock
\showISSN{19367414}
\urldef\tempurl%
\url{https://doi.org/10.1145/3501804}
\showDOI{\tempurl}


\bibitem[\protect\citeauthoryear{Singha, Diamantopoulosb, Gomez-Lunaa, Stuijkc, Corporaalc, et~al\mbox{.}}{Singha et~al\mbox{.}}{2022}]%
        {Singha2022leaper}
\bibfield{author}{\bibinfo{person}{Gagandeep Singha}, \bibinfo{person}{Dionysios Diamantopoulosb}, \bibinfo{person}{Juan Gomez-Lunaa}, \bibinfo{person}{Sander Stuijkc}, \bibinfo{person}{Henk Corporaalc}, {et~al\mbox{.}}} \bibinfo{year}{2022}\natexlab{}.
\newblock \showarticletitle{LEAPER: Fast and Accurate FPGA-based System Performance Prediction via Transfer Learning}.
\newblock \bibinfo{journal}{\emph{Proceedings - IEEE International Conference on Computer Design: VLSI in Computers and Processors}}  \bibinfo{volume}{2022-October} (\bibinfo{year}{2022}), \bibinfo{pages}{499--508}.
\newblock
\showISBNx{9781665461863}
\showISSN{10636404}
\urldef\tempurl%
\url{https://doi.org/10.1109/ICCD56317.2022.00080}
\showDOI{\tempurl}


\bibitem[\protect\citeauthoryear{Smith, Aguilar, Allison, Beatty, Bernhoff, et~al\mbox{.}}{Smith et~al\mbox{.}}{2022}]%
        {Smith2022HardwareRNO-G}
\bibfield{author}{\bibinfo{person}{Daniel Smith}, \bibinfo{person}{J.~A. Aguilar}, \bibinfo{person}{P. Allison}, \bibinfo{person}{J.~J. Beatty}, \bibinfo{person}{H. Bernhoff}, {et~al\mbox{.}}} \bibinfo{year}{2022}\natexlab{}.
\newblock \showarticletitle{{Hardware Development for the Radio Neutrino Observatory in Greenland (RNO-G)}}.
\newblock \bibinfo{journal}{\emph{Proceedings of Science}}  \bibinfo{volume}{395} (\bibinfo{date}{3} \bibinfo{year}{2022}).
\newblock
\showISSN{18248039}
\urldef\tempurl%
\url{https://doi.org/10.22323/1.395.1058}
\showDOI{\tempurl}


\bibitem[\protect\citeauthoryear{Socha, Brejn{\'{i}}k, Balasch, Novotn{\'{y}}, and Mentens}{Socha et~al\mbox{.}}{2020}]%
        {Socha2020Side-channelHardware}
\bibfield{author}{\bibinfo{person}{Petr Socha}, \bibinfo{person}{Jan Brejn{\'{i}}k}, \bibinfo{person}{Josep Balasch}, \bibinfo{person}{Martin Novotn{\'{y}}}, {and} \bibinfo{person}{Nele Mentens}.} \bibinfo{year}{2020}\natexlab{}.
\newblock \showarticletitle{{Side-channel countermeasures utilizing dynamic logic reconfiguration: Protecting AES/Rijndael and Serpent encryption in hardware}}.
\newblock \bibinfo{journal}{\emph{Microprocessors and Microsystems}}  \bibinfo{volume}{78} (\bibinfo{date}{10} \bibinfo{year}{2020}), \bibinfo{pages}{103208}.
\newblock
\showISSN{0141-9331}
\urldef\tempurl%
\url{https://doi.org/10.1016/J.MICPRO.2020.103208}
\showDOI{\tempurl}


\bibitem[\protect\citeauthoryear{Stamatakis}{Stamatakis}{2014}]%
        {stamatakis2014raxml}
\bibfield{author}{\bibinfo{person}{Alexandros Stamatakis}.} \bibinfo{year}{2014}\natexlab{}.
\newblock \showarticletitle{RAxML version 8: a tool for phylogenetic analysis and post-analysis of large phylogenies}.
\newblock \bibinfo{journal}{\emph{Bioinformatics}} \bibinfo{volume}{30}, \bibinfo{number}{9} (\bibinfo{year}{2014}), \bibinfo{pages}{1312--1313}.
\newblock


\bibitem[\protect\citeauthoryear{Strubell, Ganesh, and McCallum}{Strubell et~al\mbox{.}}{2019}]%
        {energy-llm}
\bibfield{author}{\bibinfo{person}{Emma Strubell}, \bibinfo{person}{Ananya Ganesh}, {and} \bibinfo{person}{Andrew McCallum}.} \bibinfo{year}{2019}\natexlab{}.
\newblock \bibinfo{title}{Energy and Policy Considerations for Deep Learning in NLP}.
\newblock
\newblock
\showeprint[arxiv]{cs.CL/1906.02243}


\bibitem[\protect\citeauthoryear{Sun, Chen, Liu, Miao, Chen, et~al\mbox{.}}{Sun et~al\mbox{.}}{2021}]%
        {Sun-2021}
\bibfield{author}{\bibinfo{person}{Qi Sun}, \bibinfo{person}{Tinghuan Chen}, \bibinfo{person}{Siting Liu}, \bibinfo{person}{Jin Miao}, \bibinfo{person}{Jianli Chen}, {et~al\mbox{.}}} \bibinfo{year}{2021}\natexlab{}.
\newblock \showarticletitle{Correlated Multi-objective Multi-fidelity Optimization for HLS Directives Design}. In \bibinfo{booktitle}{\emph{2021 Design, Automation \& Test in Europe Conference \& Exhibition (DATE)}}. \bibinfo{pages}{46--51}.
\newblock
\urldef\tempurl%
\url{https://doi.org/10.23919/DATE51398.2021.9474241}
\showDOI{\tempurl}


\bibitem[\protect\citeauthoryear{Theodoropoulos, Michel, Malakonakis, Georgopoulos, Isotton, et~al\mbox{.}}{Theodoropoulos et~al\mbox{.}}{2023}]%
        {Theodoropoulos2023optima}
\bibfield{author}{\bibinfo{person}{Dimitris Theodoropoulos}, \bibinfo{person}{Olivier Michel}, \bibinfo{person}{Pavlos Malakonakis}, \bibinfo{person}{Konstantinos Georgopoulos}, \bibinfo{person}{Giovanni Isotton}, {et~al\mbox{.}}} \bibinfo{year}{2023}\natexlab{}.
\newblock \showarticletitle{Optimizing Industrial Applications for Heterogeneous HPC Systems: The OPTIMA Project Intermediate stage}.
\newblock \bibinfo{journal}{\emph{Proceedings -Design, Automation and Test in Europe, DATE}}  \bibinfo{volume}{2023-April}.
\newblock
\showISBNx{9783981926378}
\showISSN{15301591}
\urldef\tempurl%
\url{https://doi.org/10.23919/DATE56975.2023.10136980}
\showDOI{\tempurl}


\bibitem[\protect\citeauthoryear{Tian, Reukers, Al-Ars, Hofstee, Brobbel, et~al\mbox{.}}{Tian et~al\mbox{.}}{2022}]%
        {Tian2022TydilangAL}
\bibfield{author}{\bibinfo{person}{Yongding Tian}, \bibinfo{person}{Matthijs~A. Reukers}, \bibinfo{person}{Zaid Al-Ars}, \bibinfo{person}{Peter Hofstee}, \bibinfo{person}{Matthijs Brobbel}, {et~al\mbox{.}}} \bibinfo{year}{2022}\natexlab{}.
\newblock \showarticletitle{Tydi-lang: A Language for Typed Streaming Hardware}.
\newblock \bibinfo{journal}{\emph{Proceedings of the SC '23 Workshops of The International Conference on High Performance Computing, Network, Storage, and Analysis}} (\bibinfo{year}{2022}).
\newblock
\urldef\tempurl%
\url{https://api.semanticscholar.org/CorpusID:254591636}
\showURL{%
\tempurl}


\bibitem[\protect\citeauthoryear{Umuroglu, Fraser, Gambardella, Blott, Leong, et~al\mbox{.}}{Umuroglu et~al\mbox{.}}{2017}]%
        {umuroglu2017finn}
\bibfield{author}{\bibinfo{person}{Yaman Umuroglu}, \bibinfo{person}{Nicholas~J Fraser}, \bibinfo{person}{Giulio Gambardella}, \bibinfo{person}{Michaela Blott}, \bibinfo{person}{Philip Leong}, {et~al\mbox{.}}} \bibinfo{year}{2017}\natexlab{}.
\newblock \showarticletitle{Finn: A framework for fast, scalable binarized neural network inference}. In \bibinfo{booktitle}{\emph{Proceedings of the 2017 ACM/SIGDA international symposium on field-programmable gate arrays}}. \bibinfo{pages}{65--74}.
\newblock


\bibitem[\protect\citeauthoryear{Ustun, Deng, Pal, Li, and Zhang}{Ustun et~al\mbox{.}}{2020}]%
        {ustun-2020}
\bibfield{author}{\bibinfo{person}{Ecenur Ustun}, \bibinfo{person}{Chenhui Deng}, \bibinfo{person}{Debjit Pal}, \bibinfo{person}{Zhijing Li}, {and} \bibinfo{person}{Zhiru Zhang}.} \bibinfo{year}{2020}\natexlab{}.
\newblock \showarticletitle{Accurate operation delay prediction for FPGA HLS using graph neural networks}. In \bibinfo{booktitle}{\emph{Proceedings of the 39th International Conference on Computer-Aided Design}} \emph{(\bibinfo{series}{ICCAD '20})}. \bibinfo{publisher}{Association for Computing Machinery}, \bibinfo{address}{New York, NY, USA}, Article \bibinfo{articleno}{87}, \bibinfo{numpages}{9}~pages.
\newblock
\showISBNx{9781450380263}
\urldef\tempurl%
\url{https://doi.org/10.1145/3400302.3415657}
\showDOI{\tempurl}


\bibitem[\protect\citeauthoryear{van Cappellen, Oosterloo, Verheijen, Adams, Adebahr, et~al\mbox{.}}{van Cappellen et~al\mbox{.}}{2022}]%
        {van_Cappellen_2022}
\bibfield{author}{\bibinfo{person}{W.~A. van Cappellen}, \bibinfo{person}{T.~A. Oosterloo}, \bibinfo{person}{M.~A.~W. Verheijen}, \bibinfo{person}{E.~A.~K. Adams}, \bibinfo{person}{B. Adebahr}, {et~al\mbox{.}}} \bibinfo{year}{2022}\natexlab{}.
\newblock \showarticletitle{Apertif: Phased array feeds for the Westerbork Synthesis Radio Telescope: System overview and performance characteristics}.
\newblock \bibinfo{journal}{\emph{Astronomy \& Astrophysics}}  \bibinfo{volume}{658} (\bibinfo{date}{Feb.} \bibinfo{year}{2022}), \bibinfo{pages}{A146}.
\newblock
\showISSN{1432-0746}
\urldef\tempurl%
\url{https://doi.org/10.1051/0004-6361/202141739}
\showDOI{\tempurl}


\bibitem[\protect\citeauthoryear{{van der Byl}, {Smith}, {Martens}, {Manley}, {van Balla}, et~al\mbox{.}}{{van der Byl} et~al\mbox{.}}{2022}]%
        {2022JATIS...8a1006V}
\bibfield{author}{\bibinfo{person}{Andrew {van der Byl}}, \bibinfo{person}{James {Smith}}, \bibinfo{person}{Andrew {Martens}}, \bibinfo{person}{Jason {Manley}}, \bibinfo{person}{Tyrone {van Balla}}, {et~al\mbox{.}}} \bibinfo{year}{2022}\natexlab{}.
\newblock \showarticletitle{{MeerKAT correlator-beamformer: a real-time processing back-end for astronomical observations}}.
\newblock \bibinfo{journal}{\emph{Journal of Astronomical Telescopes, Instruments, and Systems}}  \bibinfo{volume}{8}, Article \bibinfo{articleno}{011006} (\bibinfo{date}{Jan.} \bibinfo{year}{2022}), \bibinfo{numpages}{011006}~pages.
\newblock
\urldef\tempurl%
\url{https://doi.org/10.1117/1.JATIS.8.1.011006}
\showDOI{\tempurl}


\bibitem[\protect\citeauthoryear{van Haarlem, Wise, Gunst, Heald, McKean, et~al\mbox{.}}{van Haarlem et~al\mbox{.}}{2013}]%
        {van_Haarlem_2013}
\bibfield{author}{\bibinfo{person}{M.~P. van Haarlem}, \bibinfo{person}{M.~W. Wise}, \bibinfo{person}{A.~W. Gunst}, \bibinfo{person}{G. Heald}, \bibinfo{person}{J.~P. McKean}, {et~al\mbox{.}}} \bibinfo{year}{2013}\natexlab{}.
\newblock \showarticletitle{LOFAR: The LOw-Frequency ARray}.
\newblock \bibinfo{journal}{\emph{Astronomy \& Astrophysics}}  \bibinfo{volume}{556} (\bibinfo{date}{July} \bibinfo{year}{2013}), \bibinfo{pages}{A2}.
\newblock
\showISSN{1432-0746}
\urldef\tempurl%
\url{https://doi.org/10.1051/0004-6361/201220873}
\showDOI{\tempurl}


\bibitem[\protect\citeauthoryear{van Meter and Horsman}{van Meter and Horsman}{2013}]%
        {Meter-2013}
\bibfield{author}{\bibinfo{person}{Rodney van Meter} {and} \bibinfo{person}{Dominic Horsman}.} \bibinfo{year}{2013}\natexlab{}.
\newblock \showarticletitle{A blueprint for building a quantum computer}.
\newblock \bibinfo{journal}{\emph{Commun. ACM}} \bibinfo{volume}{56}, \bibinfo{number}{10} (\bibinfo{year}{2013}), \bibinfo{pages}{84--93}.
\newblock


\bibitem[\protect\citeauthoryear{van Wijhe, Sprave, Passaretti, Alachiotis, Grutzeck, et~al\mbox{.}}{van Wijhe et~al\mbox{.}}{2024}]%
        {Versal-ACAP}
\bibfield{author}{\bibinfo{person}{Victor van Wijhe}, \bibinfo{person}{Vincent Sprave}, \bibinfo{person}{Daniele Passaretti}, \bibinfo{person}{Nikolaos Alachiotis}, \bibinfo{person}{Gerrit Grutzeck}, {et~al\mbox{.}}} \bibinfo{year}{2024}\natexlab{}.
\newblock \showarticletitle{Exploring the Versal AI Engines for Signal Processing in Radio Astronomy}. In \bibinfo{booktitle}{\emph{2024 34rd International Conference on Field-Programmable Logic and Applications (FPL)}}.
\newblock


\bibitem[\protect\citeauthoryear{Vasile, Martoiu, Boukadida, Antonescu, Ulmamei, et~al\mbox{.}}{Vasile et~al\mbox{.}}{2022}]%
        {Vasile2022FPGALHC}
\bibfield{author}{\bibinfo{person}{M. Vasile}, \bibinfo{person}{S. Martoiu}, \bibinfo{person}{N. Boukadida}, \bibinfo{person}{M. Antonescu}, \bibinfo{person}{A. Ulmamei}, {et~al\mbox{.}}} \bibinfo{year}{2022}\natexlab{}.
\newblock \showarticletitle{{FPGA implementation of RDMA for ATLAS readout with FELIX at high luminosity LHC}}.
\newblock \bibinfo{journal}{\emph{Journal of Instrumentation}} \bibinfo{volume}{17}, \bibinfo{number}{05} (\bibinfo{date}{5} \bibinfo{year}{2022}), \bibinfo{pages}{C05022}.
\newblock
\showISSN{1748-0221}
\urldef\tempurl%
\url{https://doi.org/10.1088/1748-0221/17/05/C05022}
\showDOI{\tempurl}


\bibitem[\protect\citeauthoryear{Vasile, Martoiu, Boukhadida, Stoicea, Micu, et~al\mbox{.}}{Vasile et~al\mbox{.}}{2023}]%
        {Vasile2023IntegrationLHC}
\bibfield{author}{\bibinfo{person}{M. Vasile}, \bibinfo{person}{S. Martoiu}, \bibinfo{person}{N. Boukhadida}, \bibinfo{person}{G. Stoicea}, \bibinfo{person}{P. Micu}, {et~al\mbox{.}}} \bibinfo{year}{2023}\natexlab{}.
\newblock \showarticletitle{{Integration of FPGA RDMA into the ATLAS readout with FELIX in High Luminosity LHC}}.
\newblock \bibinfo{journal}{\emph{Journal of Instrumentation}} \bibinfo{volume}{18}, \bibinfo{number}{01} (\bibinfo{date}{1} \bibinfo{year}{2023}), \bibinfo{pages}{C01025}.
\newblock
\showISBNx{9781665488723}
\showISSN{1748-0221}
\urldef\tempurl%
\url{https://doi.org/10.1088/1748-0221/18/01/C01025}
\showDOI{\tempurl}


\bibitem[\protect\citeauthoryear{Veenboer and Romein}{Veenboer and Romein}{2019}]%
        {10.1007/978-3-030-29400-7_36}
\bibfield{author}{\bibinfo{person}{Bram Veenboer} {and} \bibinfo{person}{John~W. Romein}.} \bibinfo{year}{2019}\natexlab{}.
\newblock \showarticletitle{Radio-Astronomical Imaging: FPGAs vs GPUs}. In \bibinfo{booktitle}{\emph{Euro-Par 2019: Parallel Processing}}, \bibfield{editor}{\bibinfo{person}{Ramin Yahyapour}} (Ed.). \bibinfo{publisher}{Springer International Publishing}, \bibinfo{address}{Cham}, \bibinfo{pages}{509--521}.
\newblock
\showISBNx{978-3-030-29400-7}


\bibitem[\protect\citeauthoryear{Viel, Gouveia, Costa, Oliveira, Boing, et~al\mbox{.}}{Viel et~al\mbox{.}}{2023}]%
        {Viel2023Payload-XL:FPGA}
\bibfield{author}{\bibinfo{person}{Felipe Viel}, \bibinfo{person}{Kleber~R. Gouveia}, \bibinfo{person}{Edilberto Costa}, \bibinfo{person}{Marcio Oliveira}, \bibinfo{person}{Miguel Boing}, {et~al\mbox{.}}} \bibinfo{year}{2023}\natexlab{}.
\newblock \showarticletitle{{Payload-XL: A Platform for the In-Orbit Validation of the BRAVE FPGA}}.
\newblock \bibinfo{journal}{\emph{IEEE Embedded Systems Letters}} \bibinfo{volume}{15}, \bibinfo{number}{2} (\bibinfo{date}{6} \bibinfo{year}{2023}), \bibinfo{pages}{93--96}.
\newblock
\showISSN{19430671}
\urldef\tempurl%
\url{https://doi.org/10.1109/LES.2022.3191638}
\showDOI{\tempurl}


\bibitem[\protect\citeauthoryear{Vlagkoulis, Sari, Antonopoulos, Psarakis, Tavoularis, et~al\mbox{.}}{Vlagkoulis et~al\mbox{.}}{2022}]%
        {Vlagkoulis2022ConfigurationTechnique}
\bibfield{author}{\bibinfo{person}{Vasileios Vlagkoulis}, \bibinfo{person}{Aitzan Sari}, \bibinfo{person}{Georgios Antonopoulos}, \bibinfo{person}{Mihalis Psarakis}, \bibinfo{person}{Antonios Tavoularis}, {et~al\mbox{.}}} \bibinfo{year}{2022}\natexlab{}.
\newblock \showarticletitle{{Configuration Memory Scrubbing of SRAM-Based FPGAs Using a Mixed 2-D Coding Technique}}.
\newblock \bibinfo{journal}{\emph{IEEE Transactions on Nuclear Science}} \bibinfo{volume}{69}, \bibinfo{number}{4} (\bibinfo{date}{4} \bibinfo{year}{2022}), \bibinfo{pages}{871--882}.
\newblock
\showISSN{15581578}
\urldef\tempurl%
\url{https://doi.org/10.1109/TNS.2022.3151977}
\showDOI{\tempurl}


\bibitem[\protect\citeauthoryear{Vlagkoulis, Sari, Vrachnis, Antonopoulos, Segkos, et~al\mbox{.}}{Vlagkoulis et~al\mbox{.}}{2021}]%
        {Vlagkoulis2021SingleIons}
\bibfield{author}{\bibinfo{person}{Vasileios Vlagkoulis}, \bibinfo{person}{Aitzan Sari}, \bibinfo{person}{John Vrachnis}, \bibinfo{person}{Georgios Antonopoulos}, \bibinfo{person}{Nikolaos Segkos}, {et~al\mbox{.}}} \bibinfo{year}{2021}\natexlab{}.
\newblock \showarticletitle{{Single Event Effects Characterization of the Programmable Logic of Xilinx Zynq-7000 FPGA Using Very/Ultra High-Energy Heavy Ions}}.
\newblock \bibinfo{journal}{\emph{IEEE Transactions on Nuclear Science}} \bibinfo{volume}{68}, \bibinfo{number}{1} (\bibinfo{date}{1} \bibinfo{year}{2021}), \bibinfo{pages}{36--45}.
\newblock
\showISSN{15581578}
\urldef\tempurl%
\url{https://doi.org/10.1109/TNS.2020.3033188}
\showDOI{\tempurl}


\bibitem[\protect\citeauthoryear{Voicu and Al-Ars}{Voicu and Al-Ars}{2019}]%
        {Voicu2019SparkJNI:Spark}
\bibfield{author}{\bibinfo{person}{Tudor~Alexandru Voicu} {and} \bibinfo{person}{Zaid Al-Ars}.} \bibinfo{year}{2019}\natexlab{}.
\newblock \showarticletitle{{SparkJNI: A Toolchain for Hardware Accelerated Big Data Apache Spark}}.
\newblock \bibinfo{journal}{\emph{2019 4th IEEE International Conference on Big Data Analytics, ICBDA 2019}} (\bibinfo{date}{5} \bibinfo{year}{2019}), \bibinfo{pages}{152--157}.
\newblock
\showISBNx{9781728112824}
\urldef\tempurl%
\url{https://doi.org/10.1109/ICBDA.2019.8713201}
\showDOI{\tempurl}


\bibitem[\protect\citeauthoryear{Vos, Kirchhoff, and Ziener}{Vos et~al\mbox{.}}{2020}]%
        {vos2020gowin}
\bibfield{author}{\bibinfo{person}{Pepijn~De Vos}, \bibinfo{person}{Michael Kirchhoff}, {and} \bibinfo{person}{Daniel Ziener}.} \bibinfo{year}{2020}\natexlab{}.
\newblock \showarticletitle{A Complete Open Source Design Flow for Gowin FPGAs}.
\newblock \bibinfo{journal}{\emph{Proceedings - 2020 International Conference on Field-Programmable Technology, ICFPT 2020}}, \bibinfo{pages}{182--189}.
\newblock
\showISBNx{9780738105185}
\urldef\tempurl%
\url{https://doi.org/10.1109/ICFPT51103.2020.00033}
\showDOI{\tempurl}


\bibitem[\protect\citeauthoryear{Wijerathne, Li, Pathania, Mitra, and Thiele}{Wijerathne et~al\mbox{.}}{2022}]%
        {Wijerathne2022HiMap:Abstraction}
\bibfield{author}{\bibinfo{person}{Dhananjaya Wijerathne}, \bibinfo{person}{Zhaoying Li}, \bibinfo{person}{Anuj Pathania}, \bibinfo{person}{Tulika Mitra}, {and} \bibinfo{person}{Lothar Thiele}.} \bibinfo{year}{2022}\natexlab{}.
\newblock \showarticletitle{{HiMap: Fast and Scalable High-Quality Mapping on CGRA via Hierarchical Abstraction}}.
\newblock \bibinfo{journal}{\emph{IEEE Transactions on Computer-Aided Design of Integrated Circuits and Systems}} \bibinfo{volume}{41}, \bibinfo{number}{10} (\bibinfo{date}{10} \bibinfo{year}{2022}), \bibinfo{pages}{3290--3303}.
\newblock
\showISSN{19374151}
\urldef\tempurl%
\url{https://doi.org/10.1109/TCAD.2021.3132551}
\showDOI{\tempurl}


\bibitem[\protect\citeauthoryear{Wijtvliet, Corporaal, and Kumar}{Wijtvliet et~al\mbox{.}}{2021}]%
        {Wijtvliet2021cgra}
\bibfield{author}{\bibinfo{person}{Mark Wijtvliet}, \bibinfo{person}{Henk Corporaal}, {and} \bibinfo{person}{Akash Kumar}.} \bibinfo{year}{2021}\natexlab{}.
\newblock \showarticletitle{CGRA-EAM—Rapid Energy and Area Estimation for Coarse-grained Reconfigurable Architectures}.
\newblock \bibinfo{journal}{\emph{ACM Transactions on Reconfigurable Technology and Systems (TRETS)}}  \bibinfo{volume}{14} (\bibinfo{date}{9} \bibinfo{year}{2021}).
\newblock
Issue 4.
\showISSN{19367414}
\urldef\tempurl%
\url{https://doi.org/10.1145/3468874}
\showDOI{\tempurl}


\bibitem[\protect\citeauthoryear{Wijtvliet, Huisken, Waeijen, and Corporaal}{Wijtvliet et~al\mbox{.}}{2019}]%
        {Wijtvliet2019Blocks:Efficiency}
\bibfield{author}{\bibinfo{person}{Mark Wijtvliet}, \bibinfo{person}{Jos Huisken}, \bibinfo{person}{Luc Waeijen}, {and} \bibinfo{person}{Henk Corporaal}.} \bibinfo{year}{2019}\natexlab{}.
\newblock \showarticletitle{{Blocks: Redesigning coarse grained reconfigurable architectures for energy efficiency}}.
\newblock \bibinfo{journal}{\emph{Proceedings - 29th International Conference on Field-Programmable Logic and Applications, FPL 2019}} (\bibinfo{date}{9} \bibinfo{year}{2019}), \bibinfo{pages}{17--23}.
\newblock
\showISBNx{9781728148847}
\urldef\tempurl%
\url{https://doi.org/10.1109/FPL.2019.00013}
\showDOI{\tempurl}


\bibitem[\protect\citeauthoryear{Wood, Lu, and Langmead}{Wood et~al\mbox{.}}{2019}]%
        {wood2019improved}
\bibfield{author}{\bibinfo{person}{Derrick~E Wood}, \bibinfo{person}{Jennifer Lu}, {and} \bibinfo{person}{Ben Langmead}.} \bibinfo{year}{2019}\natexlab{}.
\newblock \showarticletitle{Improved metagenomic analysis with Kraken 2}.
\newblock \bibinfo{journal}{\emph{Genome biology}}  \bibinfo{volume}{20} (\bibinfo{year}{2019}), \bibinfo{pages}{1--13}.
\newblock


\bibitem[\protect\citeauthoryear{Xue and Calabretta}{Xue and Calabretta}{2022}]%
        {Xue2022NanosecondNetworks}
\bibfield{author}{\bibinfo{person}{Xuwei Xue} {and} \bibinfo{person}{Nicola Calabretta}.} \bibinfo{year}{2022}\natexlab{}.
\newblock \showarticletitle{{Nanosecond optical switching and control system for data center networks}}.
\newblock \bibinfo{journal}{\emph{Nature Communications 2022 13:1}} \bibinfo{volume}{13}, \bibinfo{number}{1} (\bibinfo{date}{4} \bibinfo{year}{2022}), \bibinfo{pages}{1--8}.
\newblock
\showISSN{2041-1723}
\urldef\tempurl%
\url{https://doi.org/10.1038/s41467-022-29913-1}
\showDOI{\tempurl}


\bibitem[\protect\citeauthoryear{Xue, Patra, van Dijk, Samkharadze, Subramanian, et~al\mbox{.}}{Xue et~al\mbox{.}}{2021}]%
        {Xue-nat-2021}
\bibfield{author}{\bibinfo{person}{Xiao Xue}, \bibinfo{person}{Bishnu Patra}, \bibinfo{person}{Jeroen P.~G. van Dijk}, \bibinfo{person}{Nodar Samkharadze}, \bibinfo{person}{Sushil Subramanian}, {et~al\mbox{.}}} \bibinfo{year}{2021}\natexlab{}.
\newblock \showarticletitle{CMOS-based cryogenic control of silicon quantum circuits}.
\newblock \bibinfo{journal}{\emph{Nature}}  \bibinfo{volume}{593} (\bibinfo{year}{2021}), \bibinfo{pages}{205–210}.
\newblock


\bibitem[\protect\citeauthoryear{Yan, Xue, and Calabretta}{Yan et~al\mbox{.}}{2018}]%
        {Yan2018HiFOST:Switches}
\bibfield{author}{\bibinfo{person}{Fulong Yan}, \bibinfo{person}{Xuwei Xue}, {and} \bibinfo{person}{Nicola Calabretta}.} \bibinfo{year}{2018}\natexlab{}.
\newblock \showarticletitle{{HiFOST: A scalable and low-latency hybrid data center network architecture based on flow-controlled fast optical switches}}.
\newblock \bibinfo{journal}{\emph{Journal of Optical Communications and Networking}} \bibinfo{volume}{10}, \bibinfo{number}{7} (\bibinfo{date}{7} \bibinfo{year}{2018}), \bibinfo{pages}{B1--B14}.
\newblock
\showISSN{19430620}
\urldef\tempurl%
\url{https://doi.org/10.1364/JOCN.10.0000B1}
\showDOI{\tempurl}


\bibitem[\protect\citeauthoryear{Yang, Ji, Chen, Zhuang, Zhang, et~al\mbox{.}}{Yang et~al\mbox{.}}{2023}]%
        {asic-challenges}
\bibfield{author}{\bibinfo{person}{Zhuoping Yang}, \bibinfo{person}{Shixin Ji}, \bibinfo{person}{Xingzhen Chen}, \bibinfo{person}{Jinming Zhuang}, \bibinfo{person}{Weifeng Zhang}, {et~al\mbox{.}}} \bibinfo{year}{2023}\natexlab{}.
\newblock \bibinfo{title}{Challenges and Opportunities to Enable Large-Scale Computing via Heterogeneous Chiplets}.
\newblock
\newblock
\showeprint[arxiv]{cs.AR/2311.16417}


\bibitem[\protect\citeauthoryear{Yasudo, Coutinho, Varbanescu, Luk, Amano, et~al\mbox{.}}{Yasudo et~al\mbox{.}}{2018}]%
        {Yasudo2018PerformancePlatforms}
\bibfield{author}{\bibinfo{person}{Ryota Yasudo}, \bibinfo{person}{Jose Coutinho}, \bibinfo{person}{Ana Varbanescu}, \bibinfo{person}{Wayne Luk}, \bibinfo{person}{Hideharu Amano}, {et~al\mbox{.}}} \bibinfo{year}{2018}\natexlab{}.
\newblock \showarticletitle{{Performance Estimation for Exascale Reconfigurable Dataflow Platforms}}.
\newblock \bibinfo{journal}{\emph{Proceedings - 2018 International Conference on Field-Programmable Technology, FPT 2018}} (\bibinfo{date}{12} \bibinfo{year}{2018}), \bibinfo{pages}{317--320}.
\newblock
\showISBNx{9781728102139}
\urldef\tempurl%
\url{https://doi.org/10.1109/FPT.2018.00062}
\showDOI{\tempurl}


\bibitem[\protect\citeauthoryear{Yasudo, Coutinho, Varbanescu, Luk, Amano, et~al\mbox{.}}{Yasudo et~al\mbox{.}}{2021}]%
        {Yasudo2021AnalyticalPlatforms}
\bibfield{author}{\bibinfo{person}{Ryota Yasudo}, \bibinfo{person}{José~G.F. Coutinho}, \bibinfo{person}{Ana~Lucia Varbanescu}, \bibinfo{person}{Wayne Luk}, \bibinfo{person}{Hideharu Amano}, {et~al\mbox{.}}} \bibinfo{year}{2021}\natexlab{}.
\newblock \showarticletitle{{Analytical Performance Estimation for Large-Scale Reconfigurable Dataflow Platforms}}.
\newblock \bibinfo{journal}{\emph{ACM Transactions on Reconfigurable Technology and Systems (TRETS)}} \bibinfo{volume}{14}, \bibinfo{number}{3} (\bibinfo{date}{8} \bibinfo{year}{2021}).
\newblock
\showISSN{19367414}
\urldef\tempurl%
\url{https://doi.org/10.1145/3452742}
\showDOI{\tempurl}


\bibitem[\protect\citeauthoryear{Zeitouni, Vliegen, Frassetto, Koch, Sadeghi, et~al\mbox{.}}{Zeitouni et~al\mbox{.}}{2021}]%
        {Zeitouni2021TrustedFPGAs}
\bibfield{author}{\bibinfo{person}{Shaza Zeitouni}, \bibinfo{person}{Jo Vliegen}, \bibinfo{person}{Tommaso Frassetto}, \bibinfo{person}{Dirk Koch}, \bibinfo{person}{Ahmad~Reza Sadeghi}, {et~al\mbox{.}}} \bibinfo{year}{2021}\natexlab{}.
\newblock \showarticletitle{{Trusted Configuration in Cloud FPGAs}}.
\newblock \bibinfo{journal}{\emph{Proceedings - 29th IEEE International Symposium on Field-Programmable Custom Computing Machines, FCCM 2021}} (\bibinfo{date}{5} \bibinfo{year}{2021}), \bibinfo{pages}{233--241}.
\newblock
\showISBNx{9780738126739}
\urldef\tempurl%
\url{https://doi.org/10.1109/FCCM51124.2021.00036}
\showDOI{\tempurl}


\bibitem[\protect\citeauthoryear{Zhang and Qu}{Zhang and Qu}{2014}]%
        {Zhang2014ASystems}
\bibfield{author}{\bibinfo{person}{Jiliang Zhang} {and} \bibinfo{person}{Gang Qu}.} \bibinfo{year}{2014}\natexlab{}.
\newblock \showarticletitle{{A survey on security and trust of FPGA-based systems}}. In \bibinfo{booktitle}{\emph{2014 International Conference on Field-Programmable Technology (FPT)}}. \bibinfo{publisher}{IEEE}, \bibinfo{pages}{147--152}.
\newblock
\showISBNx{978-1-4799-6245-7}
\urldef\tempurl%
\url{https://doi.org/10.1109/FPT.2014.7082768}
\showDOI{\tempurl}


\bibitem[\protect\citeauthoryear{Zhang, Xue, Tangdiongga, and Calabretta}{Zhang et~al\mbox{.}}{2022}]%
        {Zhang2022Low-LatencyRouter}
\bibfield{author}{\bibinfo{person}{Shaojuan Zhang}, \bibinfo{person}{Xuwei Xue}, \bibinfo{person}{Eduward Tangdiongga}, {and} \bibinfo{person}{Nicola Calabretta}.} \bibinfo{year}{2022}\natexlab{}.
\newblock \showarticletitle{{Low-Latency Optical Wireless Data-Center Networks Using Nanoseconds Semiconductor-Based Wavelength Selectors and Arrayed Waveguide Grating Router}}.
\newblock \bibinfo{journal}{\emph{Photonics 2022, Vol. 9, Page 203}} \bibinfo{volume}{9}, \bibinfo{number}{3} (\bibinfo{date}{3} \bibinfo{year}{2022}), \bibinfo{pages}{203}.
\newblock
\showISSN{2304-6732}
\urldef\tempurl%
\url{https://doi.org/10.3390/PHOTONICS9030203}
\showDOI{\tempurl}


\bibitem[\protect\citeauthoryear{Zhao, Feng, Sinha, Zhang, Liang, et~al\mbox{.}}{Zhao et~al\mbox{.}}{2020}]%
        {comba-2020}
\bibfield{author}{\bibinfo{person}{Jieru Zhao}, \bibinfo{person}{Liang Feng}, \bibinfo{person}{Sharad Sinha}, \bibinfo{person}{Wei Zhang}, \bibinfo{person}{Yun Liang}, {et~al\mbox{.}}} \bibinfo{year}{2020}\natexlab{}.
\newblock \showarticletitle{Performance Modeling and Directives Optimization for High-Level Synthesis on FPGA}.
\newblock \bibinfo{journal}{\emph{IEEE Transactions on Computer-Aided Design of Integrated Circuits and Systems}} \bibinfo{volume}{39}, \bibinfo{number}{7} (\bibinfo{year}{2020}), \bibinfo{pages}{1428--1441}.
\newblock
\urldef\tempurl%
\url{https://doi.org/10.1109/TCAD.2019.2912916}
\showDOI{\tempurl}


\bibitem[\protect\citeauthoryear{Ziogas, Schneider, Ben-Nun, Calotoiu, De~Matteis, et~al\mbox{.}}{Ziogas et~al\mbox{.}}{2021}]%
        {dace-2021}
\bibfield{author}{\bibinfo{person}{Alexandros~Nikolaos Ziogas}, \bibinfo{person}{Timo Schneider}, \bibinfo{person}{Tal Ben-Nun}, \bibinfo{person}{Alexandru Calotoiu}, \bibinfo{person}{Tiziano De~Matteis}, {et~al\mbox{.}}} \bibinfo{year}{2021}\natexlab{}.
\newblock \showarticletitle{Productivity, Portability, Performance: Data-Centric Python}. In \bibinfo{booktitle}{\emph{Proceedings of the International Conference for High Performance Computing, Networking, Storage and Analysis}} \emph{(\bibinfo{series}{SC '21})}. \bibinfo{publisher}{Association for Computing Machinery}, \bibinfo{address}{New York, NY, USA}, Article \bibinfo{articleno}{95}, \bibinfo{numpages}{13}~pages.
\newblock
\showISBNx{9781450384421}
\urldef\tempurl%
\url{https://doi.org/10.1145/3458817.3476176}
\showDOI{\tempurl}


\end{thebibliography}

\end{document}